\documentclass[12pt]{iopart}
\expandafter\let\csname equation*\endcsname\relax
\expandafter\let\csname endequation*\endcsname\relax
\usepackage{mathtools}
\usepackage{amssymb}
\usepackage{graphicx}
\usepackage{subfigure}
\usepackage[justification=raggedright]{caption}
\usepackage{tabularx}
\usepackage[colorlinks=true,linkcolor=blue,
     urlcolor = blue,
     citecolor=blue,]{hyperref}
\usepackage{enumerate}
\usepackage{tensor}
\usepackage{braket}
\usepackage[doi=true,isbn=false,url=false,eprint=false,natbib=true,maxcitenames=1]{biblatex}
\DefineBibliographyStrings{english}{%
  andothers = {\textit{et~al\adddot}}
}
\usepackage{appendix}
\usepackage{dcolumn}
\usepackage{bm}

\newcommand{\bt}{\bm{\theta}}
\newcommand{\dt}{\Delta t_c}
\newcommand{\mms}{\, {\rm ms}}
\newcommand{\cac}{{\cal A}^c}
\newcommand{\cas}{{\cal A}^s}

\newcommand{\uf}{^{(1)}}
\newcommand{\ud}{^{(2)}}
\newcommand{\etall}{{\it et al.}}

\newcommand{\refereeA}[1]{#1}
\newcommand{\refereeB}[1]{#1}

\begin{document}

\title[Anatomy of parameter-estimation biases in overlapping gravitational-wave
signals]{Anatomy of parameter-estimation biases in overlapping gravitational-wave
signals}

\author[Ziming Wang et al.]{Ziming~Wang$^{1,2}$, Dicong~Liang$^{2}$, Junjie~Zhao$^{3,4}$, Chang~Liu$^{1,2,5}$, Lijing~Shao$^{2,6}$}

\address{$^{1}$ Department of Astronomy, School of Physics, Peking University,
Beijing 100871, China} 
\address{$^{2}$ Kavli Institute for Astronomy and Astrophysics, Peking University,
Beijing 100871, China}
\address{$^{3}$ Institute for Frontiers in Astronomy and Astrophysics, Beijing Normal University, Beijing 102206, China}
\address{$^{4}$ Department of Astronomy, Beijing Normal University, Beijing 100875,
China}
\address{$^{5}$ Laboratoire des 2 Infinis - Toulouse (L2IT-IN2P3), Universit\'e de
Toulouse, CNRS, UPS, F-31062 Toulouse Cedex 9, France} 
\address{$^{6}$ National Astronomical Observatories, Chinese Academy of Sciences,
Beijing 100012, China}

\ead{lshao@pku.edu.cn {\rm (LS)}}

\begin{abstract}
    In future gravitational-wave (GW) detections, a large number of overlapping GW
    signals will appear in the data stream of detectors.  When extracting
    information from one signal, the presence of other signals can cause large
    parameter estimation biases.  Using the Fisher matrix (FM), we develop a bias
    analysis procedure to investigate how each parameter of other signals affects
    the inference biases. Taking two-signal overlapping as an example, we show
    detailedly and quantitatively that the biases essentially originate from the
    overlapping of the frequency evolution. Furthermore, we find that the behaviors
    of the correlation coefficients between the parameters of the two signals are
    similar to the biases. Both of them can be used as characterization of the
    influence between signals.  We also corroborate the bias results of the FM
    method with full Bayesian analysis. Our results can provide guidance for the development of new PE algorithms on overlapping signals, and the analysis methodology has the potential to generalize.
\end{abstract}
\maketitle
\section{Introduction}\label{sec1:intro}

Since the first gravitational-wave (GW) event was detected in
2015~\cite{Abbott:2016blz}, nowadays GW observatories LIGO, Virgo, and KAGRA
have detected nearly 90 GW events, which greatly enhanced our understanding of
fundamental physics, astrophysics, and cosmology~\cite{LIGOScientific:2018mvr,
LIGOScientific:2020ibl, LIGOScientific:2021djp}.  In the future, the
next-generation (XG) ground-based GW detectors, such as Cosmic Explorer
(CE)~\cite{Reitze:2019iox, Reitze:2019dyk} and Einstein Telescope
(ET)~\cite{Hild:2010id, Sathyaprakash:2012jk} will be constructed. Compared with
the present GW detectors, XG detectors improve the sensitivity by an order of
magnitude, and have a lower cut-off frequency~\cite{Kalogera:2021bya}, which
leads to more detectable GW events with longer signal duration.  Therefore, in
the data stream of XG detectors, there will be a large number of overlapping
signals (OSs)~\cite{Regimbau:2009rk, Samajdar:2021egv, Pizzati:2021apa,
Relton:2021cax, Himemoto:2021ukb}.  For space-based GW detectors, such as Laser
Interferometer Space Antenna (LISA)~\cite{LISA:2017pwj}, Taiji~\cite{Hu:2017mde}
and TianQin \cite{TianQin:2015yph, Gong:2021gvw}, a large number of OSs will
also be present~\cite{Antonelli:2021vwg}.

Searching for GW signals from detector data and extracting information of GW
sources are the basis for GW science. 
The impact of OSs is mainly reflected in two
aspects. On the one hand, every OS comes from the superposition of several
independent signals, so the dimension of the parameter space is  high. 
Conducting parameter estimation (PE) on the whole OS requires considerable computational time and
resources. On the other hand, if one deals with every single signal in the OS
one by one, the existence of other signals will have an impact on the PE of the
target, usually reflecting as a systematic error, or the PE bias more quantitatively.
Janquart \etall~\cite{Janquart:2022nyz} referred to these two PE cases as the joint PE (JPE)
and the single PE (SPE) respectively. 
If the SPE bias of each signal is small, then we do
not need to conduct complex JPE.  
\refereeA{On the contrary, when the number of GW events is known, JPE is a better choice when SPE biases are
large.}  Compared to JPE, computing SPE is faster, and the results of SPE can
tell us whether we need JPE. Therefore, as the basis of studying the whole PE
technology of OS, it is necessary and useful to investigate the behaviors of SPE
first. 

A number of publications~\cite{Regimbau:2009rk, Samajdar:2021egv,
Pizzati:2021apa,Relton:2021cax, Himemoto:2021ukb, Janquart:2022nyz,
Smith:2021bqc} have studied how overlapping will affect SPE. They consider
different detectors, such as Advanced LIGO (AdvLIGO), LIGO Voyager, ground-based
XG detectors (CE and ET), and space-based detector LISA~\cite{Antonelli:2021vwg}
with different compact binary coalescences, including binary black holes (BBHs)
\cite{Relton:2021cax, Janquart:2022nyz}, binary neutron stars (BNSs)
\cite{Pizzati:2021apa, Smith:2021bqc}, and neutron star--black hole (NSBH)
systems \cite{Samajdar:2021egv}.  The GW parameters used in these works are also
different. In addition to the necessary parameters---binary masses, luminosity
distance, and merger time---Himemoto \etall~\cite{Himemoto:2021ukb} considered the spin-orbit
coupling and spin-spin coupling effects, and Hu \etall~\cite{Hu:2022bji} considered the
parameterized post-Newtonian coefficients to test general relativity and
calculated the accumulating errors in the PE of OSs.

Qualitatively, when the merging times of different signals are close to each
other, the mutual influence between the signals is expected to be more
significant. In this case, the SPE results will have obvious biases, and we call
the corresponding OS the ``biased OS''. In other words, the biased OSs are {\it
real} OSs in the sense of PE.  On the contrary, if the OS can be decomposed into
individual signals and analyzed separately, we call it ``unbiased OS''.
Quantitatively,
Relton \etall~\cite{Relton:2021cax} and Himemoto \etall~\cite{Himemoto:2021ukb}
stated that for an OS generated by two BBHs, it can be regarded as unbiased as
long as the merger time difference $\Delta t_c$ is larger than $0.1\,\rm{s}$. 
The critical $\Delta t_c$ given by Pizzati \etall~\cite{Pizzati:2021apa} is more conservative,
which is $\Delta t_c > 0.5\,\rm{s}$. 
Samajdar \etall~\cite{Samajdar:2021egv} argued that in
most cases, for the OS of two compact binaries the intrinsic parameters of each
signal can be extracted with negligible biases. These inconsistencies are mainly
caused by the different parameter choices. 
In
addition, these works only pointed out whether the simulated OSs are biased OSs or
unbiased OSs. They summarized the simulation results but did not systematically
investigate the dependence of bias on the parameters. 

In this work, we analyze the SPE biases of OSs generated by two BBHs, which is
the basis to conduct PE for more complex OSs. Specifically, our work answers the
following questions:
\begin{enumerate}[(i)]
	\item What is the origin of biases? In other words, why could there be
	biases when conducting SPE for OSs?
	\item How do the biases depend on the merger time of each component in the
	OS? Will large biases arise when the merger times are close?
	\item Given a target signal, what kind of signals are more likely to
	generate a biased OS when overlapping with it?
\end{enumerate}
As mentioned above, it is difficult to directly describe the behaviors of bias
in the whole parameter space.
Therefore, we use
the Fisher matrix (FM) method to quickly forecast the SPE biases for numerous
OSs. The FM method can be regarded as an approximation of a full Bayesian
analysis, and has been widely used to forecast and estimate parameters in GW
researches~\cite{kay1993statistical, Cutler:1997ta, Berti:2004bd, Isoyama:2018rjb,  Zhao:2021bjw,
Shuman:2021ruh, Liu:2021dcr, Wang:2022yxb}. In particular, the FM method has
also been applied in OS analysis~\cite{Pizzati:2021apa, Himemoto:2021ukb,
Antonelli:2021vwg}.

As answers to the above questions, here we summarize our main conclusions:
\begin{enumerate}[(i)]
	\item Biases originate from the overlapping of the frequency evolution of
	signals. In the frequency domain, this is equivalent to a nearly constant
	phase difference between signals in a certain frequency band. 
	For unbiased OSs, their component
	signals only overlap in the data stream in the time domain, but can be
	decomposed into independent signals and analyzed separately in the frequency
	domain.
	\item A small merger time difference is only a necessary condition for large
	biases. Although the bias is indeed larger when $\dt \sim 0$, there is a
	significant asymmetry in the order of the mergers. When the binary with lighter
	masses merges after the heavier one, the frequency evolution is easier to
	overlap, leading to a large bias. 
	In
	addition, when $|\Delta t_c|$ increases, biases decay to zero in an
	oscillating way rather than monotonically. Moreover, the bias dependence on
	$\dt$ is different for different parameters. The parameters can be divided
	into two categories according to whether they enter in amplitude or phase in
	the waveform. Varying $\dt$, when the bias for one category is close to
	extreme values, the bias for  the other category is almost zero. 
	\item Biased OSs are more likely to arise when the component masses, especially
	the chirp masses, of individual signals are close. In this case, the
	frequency evolution of the two signals is so similar that large biases will
	occur in a wide range of $\dt$. In addition, the louder the other signal is,
	the larger the SPE bias of the target signal is.
\end{enumerate}
It is worth noting that all the masses in the above conclusions are defined in
the detector frame, since signals are overlapping in the detector data stream.
When considering a detector network, the biases will depend on the relative locations between GW sources and detectors~\cite{Relton:2021cax}. However,
in each detector, the biases' behaviors are the same as we summarized above.

These conclusions are consistent with the intuitive understanding of
the GW PE process, but they are analyzed and justified in detail for the first
time in this work. In the attempt to obtain these conclusions, we
establish an explicit expression [see Eq.~(\ref{eq12:})] and its corresponding
analyzing pipeline for SPE bias.  Benefiting from the rapidity of the FM method,
this analyzing pipeline allows one to discuss and analyze the dependence of bias
on various parameters easily.

In this paper, we also calculated the correlation coefficients between the
corresponding parameters of two signals using the FM method. 
\refereeB{In this work, we detailedly calculate and compare the behaviors of correlation coefficients and biases. Their similar explicit expressions [see Eq.~(\ref{eq12:}) and Eq.~(\ref{eq18:})] lead to their similar dependence on parameters. This is consistent with the qualitative results by Himemoto \etall~\cite{Himemoto:2021ukb}. Large correlation coefficients mean a strong indistinguishability and lead to a large bias in
the SPE.} Finally, to verify the validity of FM, we conduct a full Bayesian
analysis for some representative OSs. For most OSs, the FM results are
consistent with the full Bayesian SPE results.  In the large-bias cases, there are slight differences between the FM method and the full
Bayesian analysis. We justify the use of the FM method to forecast biases, which will provide a 
reference and guidance in the simulation of SPE and JPE for OSs. 

The paper is organized as follows. Section~\ref{sec2:method} introduces the
basic PE methods for GWs, including the full Bayesian inference and the
FM approximation. In Section~\ref{sec3:bias}, we investigate the bias behaviors and analyze the origin of biases in depth using the FM
method. Section~\ref{sec4:other} calculates the correlation coefficients of two
signals and then compares them with the FM results. The applicability of the FM method in analyzing biases is verified using full Bayesian inference. Section~\ref{sec5:}
gives the summary and conclusion.
\section{Parameter Estimation Methods}
\label{sec2:method}

\subsection{Basic concepts of Bayesian inference and FM approximation}
\label{sec2.1}

The data $g(t)$ in a GW detector are the superposition of the GW strain
$h(t;\bt)$ and the detector noise $n(t)$. In PE, we want to extract the GW
parameters $\bt$ from data $g(t)$. Using Bayes' theorem, one has the posterior distribution of $\bt$,
\begin{equation}\label{eq1:Bayes}
	P(\bt|g)\propto p(\bt)P(g|\bt)\,,
\end{equation}
where $p(\bt)$ is the prior, and $P(g|\bt)$ is called likelihood, which
describes the conditional probability of collecting the data $g(t)$ given the
parameters $\bt$. Suppose that $n(t)$ is stationary and follows a Gaussian
distribution with a mean of zero, the likelihood can be further expressed
as~\cite{Finn:1992wt}
\begin{equation}\label{eq2:full likelihood}
	P(g| \bt) \propto e^{-\frac{1}{2}(g-h,\, g-h)}\,,
\end{equation}
where the inner product $(u\,,v)$ is defined for any two data streams, $u(t)$
and $v(t)$, as
\begin{align}\label{eq3:inner product}
	(u\,,v):=  2 \Re \int_{-\infty}^{\infty} \frac{u^{*}(f) v(f)}{S_{n}(|f|)} {\rm d} f=4 \Re \int_{0}^{\infty} \frac{u^{*}(f) v(f)}{S_{n}(f)} {\rm d} f\,,
\end{align}
where $S_n(f)$ is the one-sided power spectral density of the noise; $u(f)$
and $v(f)$ are the Fourier transforms of $u(t)$ and $v(t)$, respectively.
Using $u(f)$ as example, the Fourier transform is defined by,
\begin{equation}\label{eq4:fft}
	u(f) := \int_{-\infty}^{\infty} u(t) e^{-2 \pi i f t} {\rm d} t\,.
\end{equation}
In addition, we define the whitened data stream $u_{\rm w}(f):= 2
u(f)/\sqrt{S_n}$, which can be interpreted as the signal strength with respect
to the noise in the frequency domain.  Once the GW data $g(t)$ is obtained, one
can construct the posterior distribution of the parameters $\bt$ by the above
formulas. After that, we can use some
sampling techniques, such as the Markov-Chain Monte Carlo (MCMC) method
\cite{Christensen:1998gf, Christensen:2004jm, Sharma:2017wfu} and Nested
Sampling \cite{2004AIPC..735..395S, Skilling:2006gxv}, to obtain the posterior
distributions of $\bt$.

In principle, Bayesian inference gives the full posterior at the desired
precision, but it can also take considerable computational time and resources,
especially when the parameter space dimension is high. \refereeA{In contrast, the FM method is
one of the fastest methods to give the approximate posterior, and it is widely
used in the GW data analysis~\cite{Vallisneri:2007ev}, where it reads}
\begin{equation}\label{eq5:FM}
	F_{\alpha\beta} = \left(h_{,\alpha}\,,h_{,\beta}\right)\,,
\end{equation}
where $h_{,\alpha}\equiv \partial h/\partial \theta^{\alpha}$ is the derivative
of the waveform with respect to the $\alpha$-th parameter, computed at the truth
point $\tilde{\bt}$ of the parameters.  FM is a symmetric, positive
semi-definite matrix, whose inverse exists in most cases in GW
analysis~\cite{Vallisneri:2007ev,Wang:2022kia}. Denoting the inverse of FM as
$\Sigma \equiv F^{-1}$, one obtain
\begin{equation}\label{eq6:Sigmas}
	\sigma_{\alpha} =\sqrt{\Sigma^{\alpha \alpha}}\,,\quad
c_{\alpha \beta} =\frac{\Sigma^{\alpha \beta}}{\sigma_{\alpha} \sigma_{\beta}}=\frac{\Sigma^{\alpha \beta}}{\sqrt{\Sigma^{\alpha \alpha} \Sigma^{ \beta \beta}}}\, .
\end{equation}
Here, $\sigma_{\alpha}$ is the standard deviation of $\theta^\alpha$, $c_{\alpha \beta}$ is the correlation coefficient between $\theta^\alpha$ and $\theta^\beta$~\cite{kay1993statistical}. The FM approximation, which uses
Eqs.~(\ref{eq5:FM}) and (\ref{eq6:Sigmas}) to obtain the errors and correlations
of the parameters, is very fast to calculate, as it requires the derivative of
the waveform only at one point.  FM approximation is equivalent to
the linearized-signal approximation~\cite{Finn:1992wt,Vallisneri:2007ev},
that is, the dependence of the waveform template on the parameters is linear,
which holds if and only if the GW signal is loud enough.  In addition, when
using the FM approximation result (\ref{eq6:Sigmas}), we have defaulted that the
priors of all parameters are uniform. \refereeB{In the frequentist view, the inverse matrix $\Sigma$ is the covariance matrix of the maximum likelihood estimate (MLE) values
$\hat{\bt}$ under linearized-signal approximation, and $\sigma_\alpha$ can be interpreted as (the lower limit of) measurement uncertainty due to the
presence of noise~\cite{Vallisneri:2007ev}.}

\subsection{PE methods of OSs}
\label{sec2.2}

In the case of OSs, $h(t)$ should be the superposition of individual independent
GW signals, then
\begin{equation}\label{eq7:OS}
	g(t) = h(t;\tilde{\bt})+n(t) = \sum_{j=1}^m h^{(j)}\big(t;\tilde{\bt}^{(j)}\big)+n(t)\,,
\end{equation}
where $m$ is the number of GWs overlapping together, $h^{(j)}\big(t;\tilde{\bt}^{(j)}\big)$ is the strain of the $j$-th signal, and $\tilde{\bt}^{(j)}$ is the corresponding true parameters. It should be noted that in practice the components of $g(t)$ are unknown; an OS can contain any number of GWs and the template corresponding to each signal can also be different~\cite{Antonelli:2021vwg} .

When conducting PE, one can use
different models to fit the GW strain. If the number of individual signals is known, and the parameters
of all GWs are investigated simultaneously, one should take a model $h=h(t;\bt)
= \sum_{j=1}^m h^{(j)}\big(t;\bt^{(j)}\big)$, with  $\bt =
\{\bt^{(1)},\cdots,\bt ^{(m)}\}$. This is called JPE~\cite{Janquart:2022nyz}.
In the SPE case, only one signal, $h^{(1)}$, and its corresponding parameters
are considered.  However, when conducting SPE, the strain data are still
determined by Eq.~(\ref{eq7:OS}). The presence of
$\tilde{h}^{(2)},\cdots,\tilde{h}^{(m)}$ inevitably has an impact on the
estimation of $\bt\uf$, which leads to a bias in the SPE results. 

For simplicity, we consider the likelihood function of SPE with the inclusion
of two GW signals overlapping, 
\begin{align}\label{eq8:real likelihood}
	P\big(g(t)|\bt\uf\big) \propto \exp & \left\{-\frac{1}{2}\Big(\tilde{h}\uf+\tilde{h}\ud+n-h\uf\big(\bt\uf\big)\,,\notag \right.  \\ &\left. \tilde{h}\uf+\tilde{h}\ud+n-h\uf\big(\bt\uf\big)\Big)\right\}\,.
\end{align}
With Eq.~(\ref{eq8:real likelihood}), one can exhaustively discuss how one
signal in the OS affects the SPE of another signal. In principle, this requires
exploring throughout the parameter space and comparing the SPE result with the
truth $\tilde{\bt}\uf$ and $\tilde{\bt}\ud$.  However, due to the limited
computational resources, one in practice can only select some representative
points for full Bayesian analysis, as was done in
Refs.~\cite{Samajdar:2021egv,Pizzati:2021apa,Relton:2021cax,Himemoto:2021ukb}.
It will be further discussed in Sec.~\ref{sec4:other}.

Now, we use the FM method to derive the explicit expression of biases in SPE.
Using linearized-signal approximation and expanding the waveform at the truth point, we have
$h\uf\big(\bt\uf\big) \approx \tilde{h}\uf+\tilde{h} \uf_{,\alpha}\big(
\theta^{(1)\alpha} - \tilde{\theta}^{(1)\alpha}\big)$, where
$\tilde{h}\uf_{,\alpha} \equiv \partial h\uf\big(\bt\uf\big)/\partial
\theta^{(1)\alpha}$. Note that all quantities are computed at $\tilde{\bt}\uf$,
and we have used the Einstein summation convention.  Substituting it into
Eq.~(\ref{eq8:real likelihood}) and using MLE, we obtain 
\begin{equation}\label{eq9:}
	\hat{\theta}^\alpha-\tilde{\theta}^\alpha  = \Sigma^{\alpha\beta}\big(\tilde{h}\uf_{,\beta}\,,n\big)+\Sigma^{\alpha\beta}\big(\tilde{h}\uf_{,\beta}\,,\tilde{h}\ud\big)\,.
\end{equation}
Since the SPE is only for $\bt\uf$, the superscript ``(1)'' of $\bt$ is omitted
for conciseness. 
To measure the
deviation of the MLE results from the true value in the statistical sense, we
use the most common approach, namely, the mean square error
(MSE)~\cite{casella2021statistical}, which is defined as the expected value of
the square of the error $\hat{\theta}^\alpha-\tilde{\theta}^\alpha$
\begin{align}\label{eq10:bias}
	\rm{MSE}_{\alpha} &:= {\rm{E}}\left[\big(\hat{\theta}^\alpha-\tilde{\theta}^\alpha\big)^2\right] = \big(\Delta \theta^\alpha_{\rm bias}\big)^2+\sigma_\alpha^2\,,
\end{align} 
where ${\rm{E}}\left[\cdot\right]$ means the ensemble average over the noise, $\Delta \theta^\alpha_{\rm bias}:=
{\rm{E}}[\hat{\theta}]-\tilde{\theta}^\alpha=\Sigma^{\alpha\beta}\big(\tilde{h}\uf_{,\beta}\,,\tilde{h}\ud\big)$ is called bias. Note this derivation has assumed stationary and Gaussian noise, and used the relation ${\rm{E}}\left[(u\,,n)(n\,,v)\right] = (u\,,v)$~\cite{Finn:1992wt}.

Equation~(\ref{eq10:bias}) deserves a detailed discussion. Mathematically,
the MSE can
be decomposed into the sum of the square of the bias, and the variance of the
estimator itself. 
The other signal, $\tilde{h}\ud$, only contributes to the bias
part in the MSE via $\big(\tilde{h}\uf_{,\alpha}\,,\tilde{h}\ud\big)$, 
while the variance part has no relationship with
$\tilde{h}\ud$ and depends only on the sensitivity of the waveform to the
parameters (via $\tilde{h}\uf_{,\alpha}$) and the average strength of the noise
$S_n$.  

The MLE results are biased as long as
$\Sigma^{\alpha\beta}\big(\tilde{h}\uf_{,\beta}\,,\tilde{h}\ud\big) \neq 0$,
however, this is not the condition for biased OSs.  In practice, if the bias $\Delta
\theta^\alpha_{\rm bias}$ caused by $\tilde{h}\ud$ is much smaller than the
statistical uncertainty $\Delta \theta^\alpha_{\rm stat}$ caused by $n$, we can
not conclude the bias in the results at all, but instead will attribute it to random errors
caused by the noise or the sampling algorithm. Therefore, it is more
meaningful to define the (dimensionless) reduced bias
\begin{equation}\label{eq11:NB}
	B_\alpha:=\frac{\Delta \theta^\alpha_{\rm bias}}{\Delta \theta^\alpha_{\rm stat}} = \frac{\Sigma^{\alpha\beta}\big(h\uf_{,\beta}\,,h\ud\big)}{\sqrt{\Sigma^{\alpha\alpha}}}\,.
\end{equation}
Note that in Eq.~(\ref{eq11:NB}) only $\beta$ is summed. Also, we omit the tilde
sign since there is no essential difference whether the variable is $\bm \theta$
or $\tilde{\bm \theta}$. Once $h\uf$ and $h\ud$ are given, the reduced bias $\bm
B$ is determined by Eq.~(\ref{eq11:NB}).  Substituting $\bm B$ into the
expression of MSE, we obtain ${\rm{MSE}}_\alpha =
\sigma^2\big(B_\alpha^2+1\big)$.  Now, it is easy to give a natural and simple
criterion for biased OS: {\it for a certain OS, if there exists a component of $\bm B$
larger than $1$ or smaller than $-1$, the OS is a biased OS;  otherwise, the OS is an unbiased OS}.


\section{Bias in the Overlapping of Two Individual Signals}
\label{sec3:bias}

\subsection{Parameter selection and detector configuration}
\label{sec3.1}

Now we use the method developed in Sec.~\ref{sec2:method} to study the behaviors
of $\bm B$.  Although the FM method quickly calculates the bias for given
${\bt}\uf$ and ${\bt}\ud$, exploring a high-dimensional parameter space is still
not an easy task. As mentioned earlier, the focus of this work is not to
simulate several OS PEs as realistically as possible, but to discuss and explain
the origin of SPE biases within a simple parameter space. 

When generating the OSs, we get the nonspinning stellar-mass BBH signals using
the {\sc IMRPhenomD} waveform template for $h\uf$ and
$h\ud$~\cite{Husa:2015iqa,Khan:2015jqa}.  The variable parameters are the chirp
mass $\cal M$, the symmetry mass ratio $\eta$, the luminosity distance $d_L$,
and the merger time $t_c$. 
In addition, we use the angle-averaged waveforms over the sky localization
and inclination, as well as the polarization angle of GWs.  This corresponds to
a most conservative scenario, where two GW signals arrive at the detector from
the same direction with the same inclination and polarization.  Taking into
account these angle parameters may reduce the biases~\cite{Relton:2021cax}, but
this effect originates more from geometry than the intrinsic properties of PE.
As for the merger phase $\phi_c$, some studies pointed out that the merging
phase difference can affect the appearance of biases
significantly~\cite{Pizzati:2021apa,Relton:2021cax}, but no further analysis has
been done. Here we take the merger phase between two signals $\Delta \phi_c = 0$
temporarily.  After establishing the analysis process of biases, the $\Delta
\phi_c \neq 0$ case can be easily incorporated into our analysis, and how it
affects the biases can also be clearly explained.  Finally, all parameters
(especially the two mass parameters, ${\cal M}$ and $\eta$) are defined in the
detector frame, since the overlapping of GWs occurs in the detector rather than
at the GW source location.

As for the detector, we choose the designed sensitivity of
AdvLIGO\footnote{\url{https://dcc.ligo.org/LIGO-T1800044/public.}} for the
calculation of the inner product. The reason for this selection is essentially
the same as \citet{Pizzati:2021apa}.  On the one hand, although AdvLIGO is
unlikely to detect OSs  at present~\cite{Regimbau:2009rk,Relton:2021cax}, this
is due to the low total event rate in the detection horizon of AdvLIGO. If two
signals with close merging times are injected, a biased OS may occur. Also, the
effective duration of a signal in AdvLIGO is much shorter than that in XG
detectors, so the bias calculation is faster.  On the other hand, the shape of
the AdvLIGO sensitivity curve is similar to those of XG ground-based GW
detectors (such as CE and ET), except that the sensitivity of AdvLIGO is about
one order of magnitude lower.  This means that the behaviors of an OS in AdvLIGO
and XG detectors are similar, only differing roughly by a constant factor.
Therefore, we expect the properties of biases with AdvLIGO sensitivity also
apply to XG detectors. Moreover, as will be demonstrated in the following sections, the behaviors of bias in AdvLIGO is consistent with previous work, which considered OS in ET~\cite{Himemoto:2021ukb,Antonelli:2021vwg}. This further increases the confidence in our conclusions based on OS simulations in AdvLIGO. 

As a short summary, when generating an OS, we have eight free parameters, $\bt =
\{ {\cal M}\uf, \eta\uf, d_L\uf, t_c\uf, {\cal M}\ud, \eta\ud, d_L\ud,
t_c\ud\}$.  The reduced bias $\bm B = \{B_{\cal M},B_{\eta},B_{d_L},B_{t_c} \}$
is a $4$-dimensional function defined on this $8$-dimensional space. For
convenience, we use indices $\alpha = \{1,2,3,4\}$ for  $\{{\cal M},\eta,
d_L,t_c\}$ respectively.  In the following, we will analyze the properties of
$\bm B$ for each parameter.

\subsection{Luminosity distance, signal-to-noise ratio and merger time}\label{sec3.2}

In the {\sc IMRPhenomD} waveform, the amplitude is inversely proportional to the
luminosity distance $d_L$, while the phase is independent of $d_L$.  This is
relatively simple, so we analyze the dependence of $\bm B$ on $d_L\uf$ and
$d_L\ud$ first.  For $d_L\ud$, since $h\ud \propto 1/d_L\ud$, according to
Eq.~\eqref{eq11:NB} we have ${\bm B}\propto 1/d_L\ud$.  As for $d_L\uf$, since
for $\beta \neq 3$,
\begin{equation}
	h_{,\beta}\uf \propto 1/d_L\uf \,, \quad 
	h_{,3}\uf \propto 1/\big(d_L\uf\big)^2 \,, 
\end{equation}
we have, for $\alpha, \, \beta\neq 3$,
\begin{equation}
	\Sigma^{\alpha\beta}\propto \big(d_L\uf\big)^2 \,, \quad
	\Sigma^{\alpha3}\propto \big(d_L\uf\big)^3 \,, \quad
	\Sigma^{33}\propto \big(d_L\uf\big)^4 \,,
\end{equation}
from the definition of FM and its inverse [Eqs.~\eqref{eq5:FM} and
\eqref{eq6:Sigmas}].  Substituting them in Eq.~\eqref{eq11:NB}, we find that ${\bm B}\propto
\big(d_L\uf\big)^0\big(d_L\ud\big)^{-1}$. 

The luminosity distance directly affects the amplitude of GWs, which in turn
affects the signal-to-noise ratio (SNR). For a GW signal $h$, its SNR is $\rho =
\sqrt{(h\,,h)}$. Since SNR is inversely proportional to $d_L$, we have ${\bm
B}\propto \big(\rho \uf\big)^0\big(\rho \ud\big)^{1}$. At first glance, this conclusion may seem strange, since the bias should be smaller for louder $h\uf$. Note that the PE uncertainty is also smaller for louder signals, so the reduced bias $\bm B$ does not depend on the strength of signal $h\uf$. Similar results were mentioned by \citet{Antonelli:2021vwg}.  However, strictly speaking, this conclusion holds only
when $d_L$ varies or the sensitivity has been consistently increased or decreased by an overall factor. Signals with different internal parameters can have the same SNR, but lead to different biases.

Intuitively, for the merger time, $\bm B$ should be related only to the merger
time difference $\dt =t_c\ud-t_c\uf$ of the two signals. 
First,
we decompose the frequency domain waveform into amplitude and phase,
$h(f)=A(f)e^{i\phi(f)}$. It is easy to know that only $\phi(f)$ depends on
$t_c$, and can be further written as $\phi(f) = \phi_0(f)-2\pi ft_c$, where
$\phi_0$ is the phase when $t_c=0$.  On the other hand, $\Sigma$ is only related
to ${\cal M}\uf$, $\eta\uf$, and $d_L\uf$, so $\bm B$ depends on $t_c\uf$ and
$t_c\ud$ only through the inner product $\big(h\uf_{,\alpha}\,,h\ud\big)$, which
is written explicitly as 
\begin{align}\label{eq12:}
	\big(h\uf_{,\alpha}\,,h\ud\big) &=4\Re \int_{f_{\rm min}}^{{f_{\rm max}}} \frac{A\uf_{, \alpha } A\ud-i \phi\uf_{,\alpha } A\uf A\ud}{S_{n}(f)} e^{i\left(\phi_0\ud-\phi_{0}\uf-2 \pi f\big(t_c\ud-t_c\uf\big)\right)} {\rm d} f \notag\\
	&=\int_{f_{\rm min}}^{{f_{\rm max}}}  {\cal A}^{c}_\alpha  \cos \left(\Delta \phi_0-2 \pi \Delta t_{c} f\right) {\rm d} f+
	\int_{f_{\rm min}}^{{f_{\rm max}}}  {\cal A}^{s}_\alpha  \sin \left(\Delta \phi_0-2 \pi \Delta t_{c} f\right) {\rm d}  f \,,
\end{align}

where ${\cal A}^c_\alpha\equiv 4A\uf_{,\alpha}A\ud/S_n$, ${\cal
A}^s_\alpha\equiv 4\phi\uf_{,\alpha}A\uf A\ud/S_n$, and $\Delta \phi_0 \equiv
\phi_0\ud-\phi_{0}\uf$. The integration interval $[f_{\rm min}\,,f_{\rm max}]$
is determined by the sensitivity curve of AdvLIGO, independent of the OS
parameters. 

Mathematically, Eq.~\eqref{eq12:} is a modulated function integral, ${\cal
A}^c_\alpha$ and ${\cal A}^s_\alpha$ are the modulation amplitudes, and the
modulation phase $\Delta \phi_0$ is the phase difference between the two signals
at $t_c\uf = t_c\ud = 0$ in the frequency domain.  For the {\sc IMRPhenomD}
template, $d_L$ appears only in the amplitude, so we have ${\cal A}^c_3\neq 0$
and ${\cal A}^s_3=0$; $t_c$ appears only in the phase, so ${\cal A}^c_4= 0$ and
${\cal A}^s_4\neq0$.  The intrinsic parameters ${\cal M}$ and $\eta$ contribute
to both amplitude and phase, but numerical calculations show that ${\cal
A}^s_{1/2}$ is about two orders of magnitude larger than ${\cal A}^c_{1/2}$,
namely ${\cal A}^s_{1/2}\gg {\cal A}^c_{1/2}>0$ (see~\ref{appA:}). 
This is because that the current matched filtering technique is more sensitive
to the phase evolution of GWs. The phase $\phi_0$ is only related to the
internal parameters ${\cal M}$ and $\eta$, and is generally a nonlinear function
of $f$.  As for $t_c\uf$ and $t_c\ud$, they only contribute to the arguments of
trigonometric functions in the form of merger time difference $\Delta t_c$,
which agrees with our expectation. Moreover, the contribution to the phase is a
linear term of $f$. 

According to the above discussion, we only need to explore a $5$-dimensional
parameter space, and the corresponding parameters are $\{{\cal
M}\uf,\eta\uf,{\cal M}\ud,\eta\ud,t_c\ud-t_c\uf\}$. Therefore, unless specified,
afterwards we fix $d_L\uf = 500\,{\rm{Mpc}}$, $d_L\ud = 3000\,{\rm{Mpc}}$, and
$t_c\uf = 0$.


\subsection{Global dependence of biases on intrinsic parameters}
\label{sec3.3}

The dependence of $\bm B$ on $d_L$ is only a simple rescale factor and does not touch on the real
origin of biased OSs. 
To further
shrink the parameter space, this subsection will take several sets of $\{{\cal
M}\uf, \eta\uf, {\cal M}\ud, \eta\ud\}$ and study the generality of the
dependence of $\bm B$ on the internal parameters.

For $\{{\cal M}\uf,\eta\uf\}$, we will take $5$ sets of parameters, expressed in
terms of binary masses, with the Solar mass ${\rm M}_\odot$ as unit.  The first
set of parameters is intentionally taken as integers $\big(m_1\uf,m_2\uf\big) =
(30, 20)$. We call this configuration the ``{\sc Test Event}.''  For the other $4$
sets of parameters, we select the real GW events GW150914,
GW190602\underline{~}175927, GW190814, and GW190924\underline{~}021846. GW150914
is the first GW event.  GW190602\underline{~}175927, GW190814, and
GW190924\underline{~}021846 are the events with the second largest chirp mass,
the smallest mass ratio, and the lightest chirp mass, respectively, in GWTC-1
and GWTC-2. Finally, for each
set of the above parameters, we calculate $\bm B$ for $\Delta t_c \in [t_{{\rm
min}}\,,t_{{\rm max}}]=[-0.1\,,0.1]\, {\rm s}$. Most of the biased OSs occur in this
range of $\dt$~\cite{Pizzati:2021apa,Relton:2021cax,Himemoto:2021ukb}, which is
also verified by the calculations below.

\begin{figure}[htp]
	\centering
\subfigure{\includegraphics[width=7.5cm]{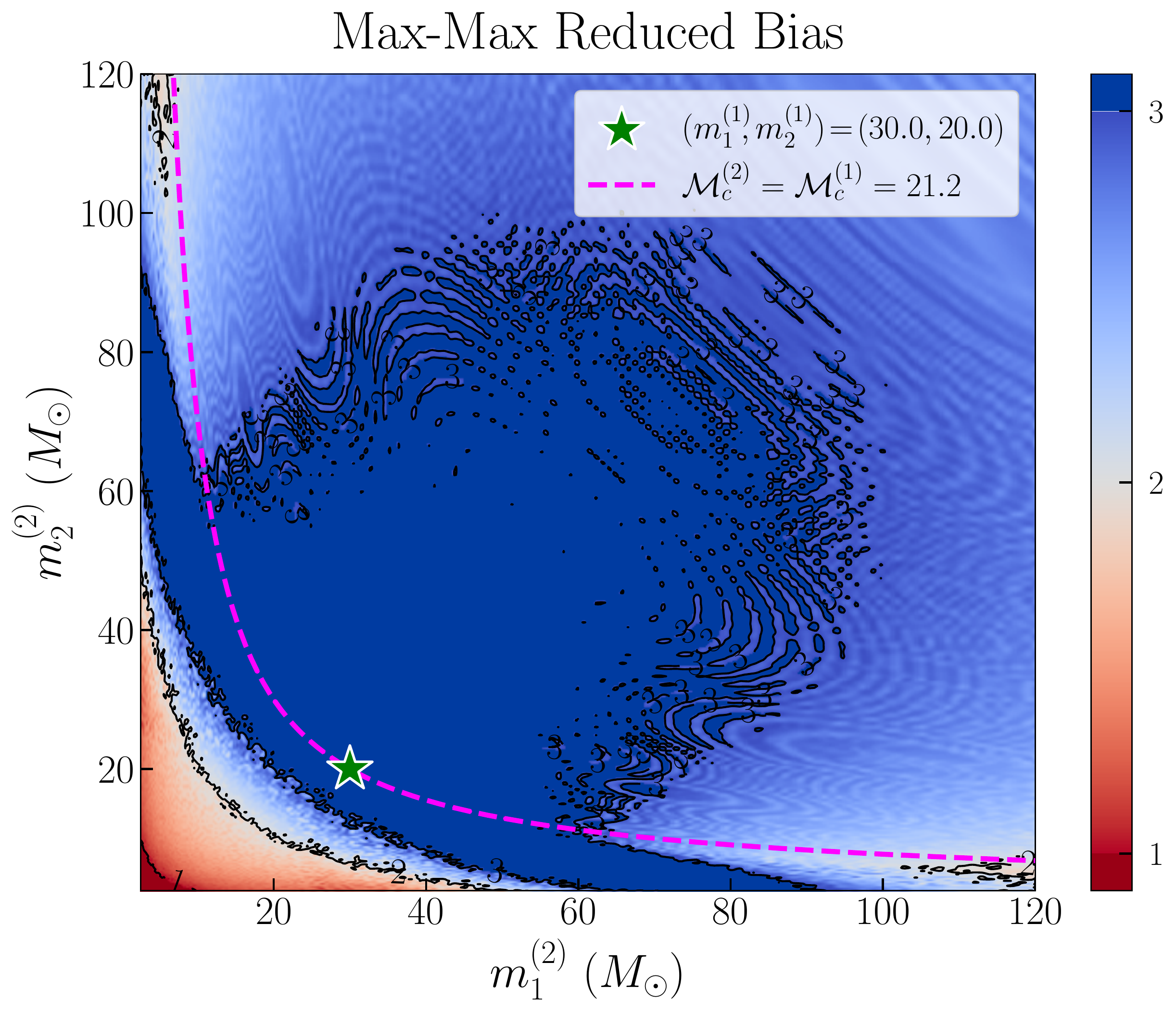}} 
	\centering
\subfigure{\includegraphics[width=7.8cm]{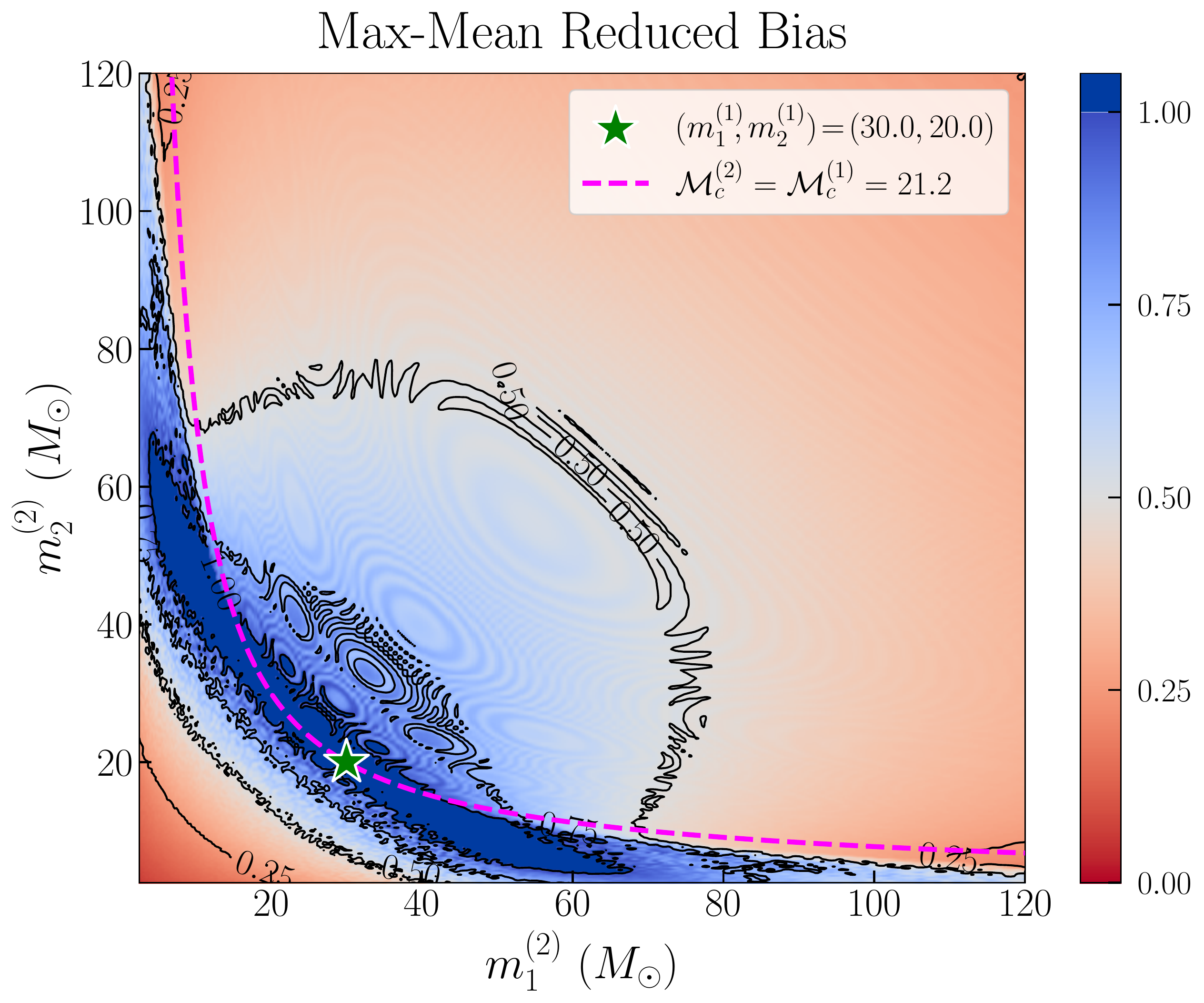}} 
\vspace{-0.5cm}
\\ 
	\centering
\subfigure{\includegraphics[width=8.0cm]{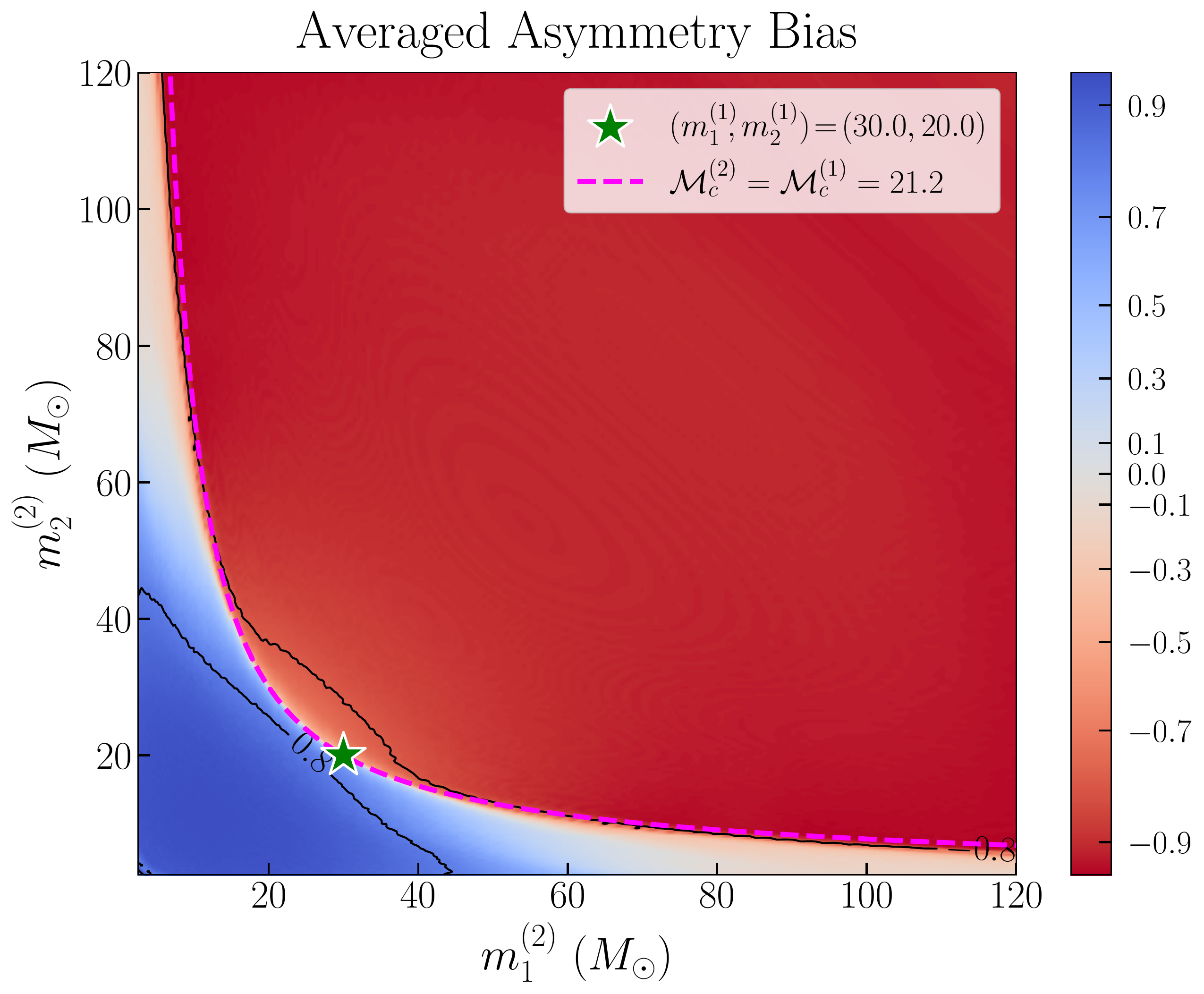}} 
\vspace{-0.5cm}
\caption{Dependence of the max-max value, max-mean value and average asymmetry
of ${\bm B}$ in the parameter space of $m_1\ud$ and $m_2\ud$ for the {\sc Test
Event}.  The location of $\big(m_1\uf,m_2\uf \big)$ is marked with green stars.
The magenta dashed lines indicate the values of $\big(m_1\ud,m_2\ud\big)$ for
which ${\cal M}\ud= {\cal M}\uf$.} 
\label{fig1}
\end{figure}

This subsection focuses on the dependence of $\bm B$ on internal parameters, which
means we need to choose some characteristic quantities to reflect the
distribution ${\bm B}$ in the range $\Delta t_c \in [-0.1\,,0.1]\, {\rm s}$. 
In
Fig.~\ref{fig1}, we show the dependence of ${\bm B}$'s ``max-max'' and
``max-mean'' values on $m_1\ud$ and $m_2\ud$ when $\big(m_1\uf,m_2\uf\big) =
(30,20)$. To obtain the max-max (max-mean) values, we first calculate the
maximum (mean) of $|B_\alpha|$ over $[t_{{\rm min}}\,,t_{{\rm max}}]$, then find
the maximum value in the $4$ (dimensionless) components of $\bm B$.  We take the
absolute value of $B_\alpha$ to identify biased OSs. In addition, to eliminate the
effects of signal strength on biases, we adjust $d_L\ud$ so that $\rho\ud = 8$,
which is the detection threshold of the ground-based detectors. 
The results of the other four sets of $m_1\uf$ and
$m_2\uf$ are shown in~\ref{appB:}. 

We find that for a wide range of $m_1\ud$ and $m_2\ud$, the max-max value of
$\bm B$ is greater than $1$. This means that for almost any two GW signals,
there exists a $\Delta t_c$ making the corresponding OS biased. 
For the max-mean value, it is greater
than $1$ only in the region around ${\cal M}\ud= {\cal M}\uf$. Therefore, in the
average sense, there is a higher probability of biased OS only when the chirp masses
of the two signals are close.  It should be noted that since $\bm B$ is larger
at $|\Delta t_c|\sim 0$ and decays to zero as $|\Delta t_c|$ increases, if the
length of the $\Delta t_c$ interval is shortened, e.g., to $|\Delta t_c| \in
[-0.05,0.05]\,{\rm s}$, the max-mean value will increase.  However, the global
behaviors of $\bm B$ remain unchanged. Also, the max-max and max-mean values involve the absolute size of the bias, and we fix the SNR $\rho\ud=8$. 

In addition to the absolute magnitude of the biases, we are also interested in
their distribution with respect to $\Delta t_c$.  One of the most important
questions is whether there is a significant difference in the biases when
$\Delta t_c>0$ and $\Delta t_c<0$, i.e., the dependence on the merger order. 
First, we define ${\rm Asy}_\alpha$ as the difference between the area enclosed
by the horizontal axis $|B_\alpha|$ on $[-0.1,0]\,{\rm s}$ and the area on
$[0,0.1]\,{\rm s}$,
\begin{equation}
	{\rm Asy}_\alpha\coloneqq \frac{\int_0^{t_{\rm max}}|B_\alpha|\,{\rm d} \Delta t_c-\int^0_{t_{\rm min}}|B_\alpha|\,{\rm d} \Delta t_c }{\int_0^{t_{\rm max}}|B_\alpha|\,{\rm d} \Delta t_c+\int^0_{t_{\rm min}}|B_\alpha|\,{\rm d} \Delta t_c }\,.\notag
\end{equation}
Then, we average ${\rm Asy}_\alpha$ over the parameters of $\bt\uf$ to obtain
the average asymmetry $\overline{\rm Asy}$.  $\overline{\rm Asy}$ is a function
of internal parameters $\{{\cal M}\uf,\eta\uf,{\cal M}\ud,\eta\ud\}$. When
$\overline{\rm Asy}>0$, it means that the OS is more likely to be biased when
$h\ud$ merges after $h\uf$.  
In addition, the closer
$|\overline{\rm Asy}|$ is to $1$, the stronger the dependence of the bias on the
merger order.  For example, $\overline{\rm Asy}=0.8 = (9-1)/(9+1)$ means that
the average bias on cases where $h\uf$ merges first is $9$ times of that where
$h\ud$ merges first. We calculate the average asymmetries of the above five sets
of GW events and put the results of the {\sc Test Event} and the other four real
GW events in  Fig.~\ref{fig1} and~\ref{appB:} respectively.  There is a
clear boundary between $\overline{\rm Asy}>0$ and $\overline{\rm Asy}<0$. The
boundary coincides with the ${\cal M}\ud= {\cal M}\uf$ curve. Roughly speaking,
it is easier to generate a biased OS when the signal with heavier masses merges first. 
In addition, we found that the regions of small $|\overline{\rm Asy}|$ (the
white regions in the average asymmetry plot) are consistent with the regions of
large bias (the dark blue regions in the max-max, max-mean plots).  
\begin{table*}[htp]
\def\arraystretch{1.3}
\tabcolsep=0.07cm
\centering
\caption{Parameter configurations of the six cases of $h\ud $, where $M=m_1+m_2$
is the total mass of binary. For $h\uf $, we have $m_1=30.00\,{\rm M}_\odot$,
$m_2=20.00\,{\rm M}_\odot$ and $d_L=500 \,{\rm Mpc}$. \label{tab1:} }
\begin{tabular}{cccccccc }
\hline\hline
Name & $m_1\,({\rm M}_\odot)$&$m_2\,({\rm M}_\odot)$&$M\,({\rm M}_\odot)$&${\cal M}\,({\rm M}_\odot)$
 &$\eta$&$d_L\,({\rm Mpc})$&$\rho$\\
 \hline
{\sc Equal}&30.00&20.00&50.00&21.24&0.240&3000&5.5\\
{\sc Asymmetric}&27.00&18.00&45.00&19.11&0.240&3000&5.1\\
{\sc Asymmetric2}&33.00&22.00&55.00&23.36&0.240&3000&5.9\\
{\sc Random}&40.00&34.00&74.00&32.08&0.248&3000&7.7\\
{\sc Symmetric and close}&33.00&18.20&51.20&21.15&0.229&3000&5.4\\
{\sc Symmetric and not close}&60.00&9.11&69.11&18.82&0.114&3000&3.8\\
\hline
\end{tabular}
\end{table*}

\subsection{Origin of biases}
\label{sec3.4}

In Sec.~\ref{sec3.3}, we vary $\big\{m_1\uf,m_2\uf,m_1\ud,m_2\ud \big\}$ and
summarize the generality of the behaviors of $\bm B$ with different internal
parameters. Now we present the mathematical characteristics and physical
interpretation of $B_\alpha\big(\dt\big)$ and draw the most important conclusion
in our work: {\it the SPE biases originate from the overlapping of frequency
evolution processes of the two signals}.  To obtain this, we fix the internal
parameters and investigate the relationship between $\bm B$ and $\Delta t_c$. 
For $h\uf$, we take ${\bm \theta}\uf$ as the parameter in the {\sc Test Event}.
For $h\ud$, we take six representative sets of parameters, called {\sc Equal},
{\sc Asymmetric}, {\sc Asymmetric2}, {\sc Random}, {\sc Symmetric and close},
{\sc Symmetric and not close}, according to the relationship between ${\bm
\theta}\ud$ and ${\bm \theta}\uf$ and the behavior of ${\bm B}\big(\Delta
t_c\big)$. The parameters are listed in Table~\ref{tab1:}. In the main text, we only discuss the first two configurations to draw the main conclusion. The other four additional configurations provide a more comprehensive study in the parameter space, and we discuss them in the ~\ref{appC:}.

\begin{figure*}[tp]
	\centering
\subfigure{\includegraphics[width=11cm]{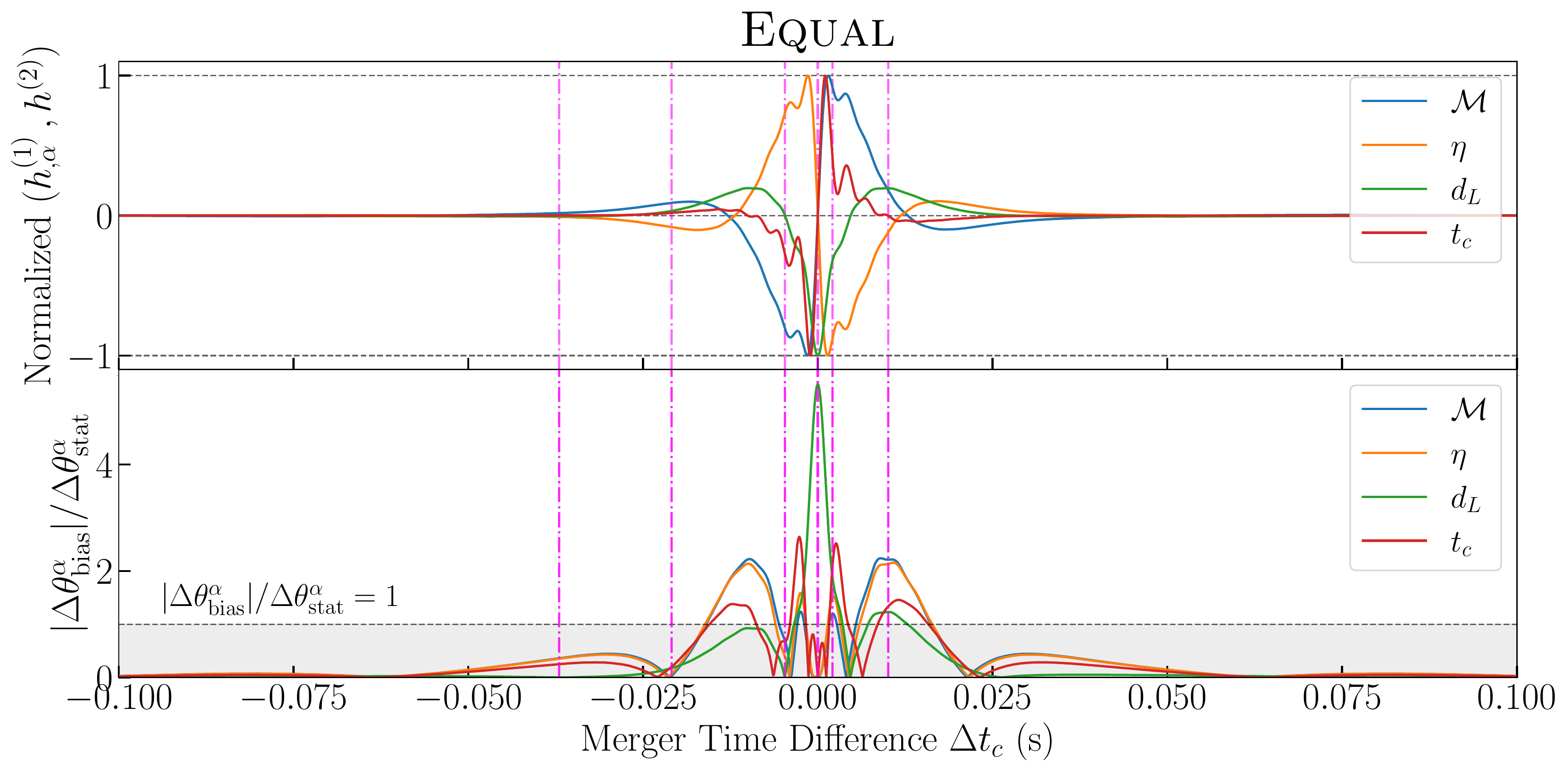}}
\\
\vspace{-0.4cm}
	\centering
\subfigure{\includegraphics[width=11cm]{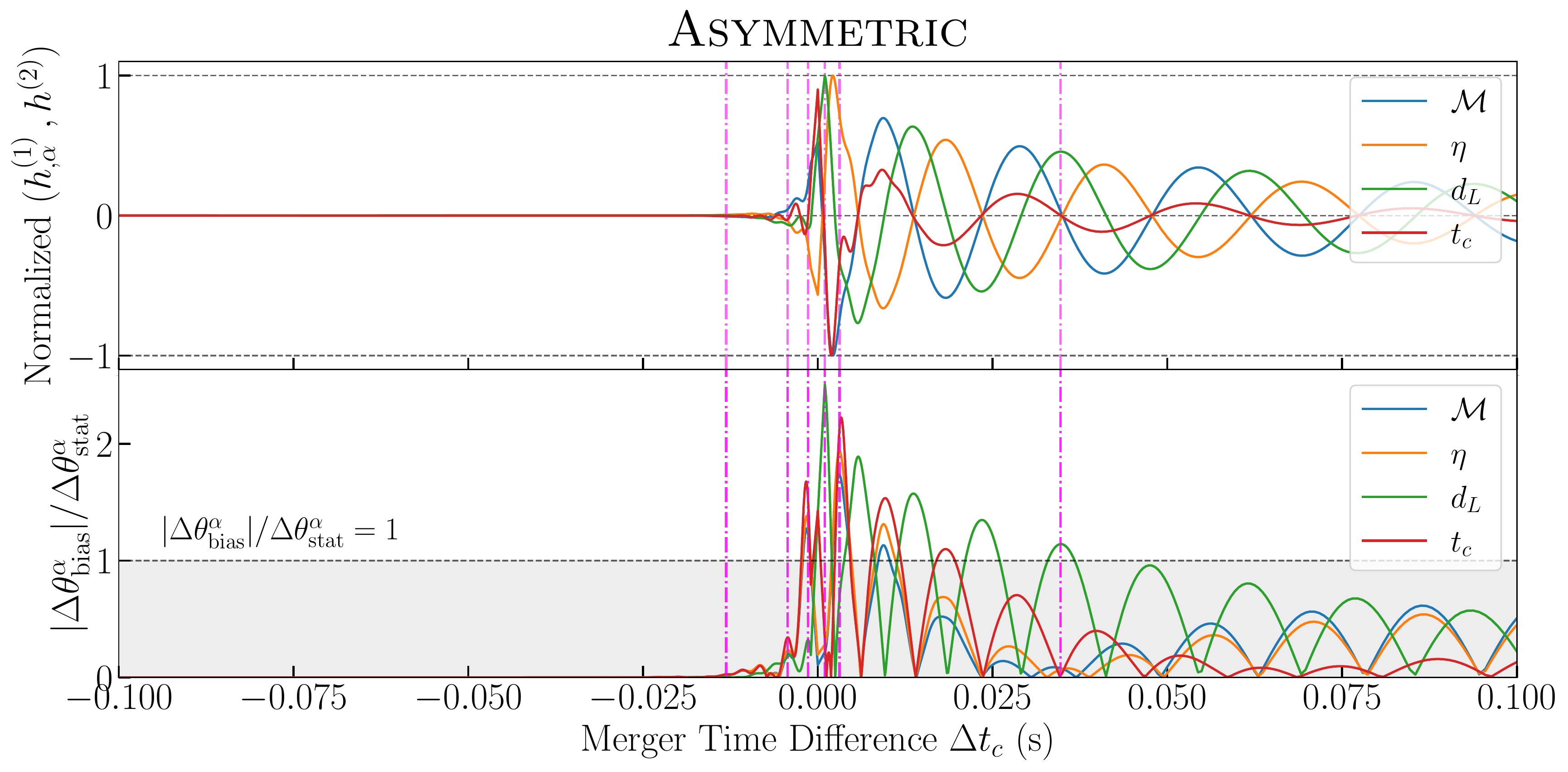}} 
\caption{The dependence of $\big(h\uf_{,\alpha}\,,h\ud\big)$ and $|B_\alpha| =|\Delta \theta^\alpha_{\rm bias}|/\Delta \theta^\alpha_{\rm stat} $ on $\Delta t_c$ for {\sc Equal} and
{\sc Asymmetric} configurations. In the shaded area, $|B_\alpha|<1$, which corresponds to unbiased OSs; the other region corresponds to biased OSs. The magenta dashed lines mark some representative $\Delta t_c$ for reference.} \label{fig2}
\end{figure*}

According to Eqs.~\eqref{eq11:NB} and \eqref{eq12:}, the reduced bias $B_\alpha$
is just a linear combination of $\big(h\uf_{,\alpha}\,, h\ud\big) $, and the
combination coefficients only depend on ${\bm \theta}\uf$.  When $h\uf$ is
fixed, a large bias is roughly equivalent to a large
$\big(h\uf_{,\alpha}\,,h\ud\big)$, and the latter can be expressed as the sum of
the integrals of two modulated trigonometric functions, which is easier to
analyze.  
Thus, for each parameter configuration of $h\ud$, we vary $\Delta t_c$
within $[-0.1,0.1]\,{\rm s}$, calculate $\big(h\uf_{,\alpha}\,,h\ud\big)$ and
the corresponding ${B}_\alpha$, and display them in Fig.~\ref{fig2}.


For the convenience of presentation, we define
\begin{align*}
	D_\alpha(f) \coloneqq &{\cal A}^{c}_\alpha  \cos \left(\Delta \phi_0-2 \pi \Delta t_{c} f\right) 
	 + {\cal A}^{s}_\alpha  \sin \left(\Delta \phi_0-2 \pi \Delta t_{c} f\right) \,,\\
	R_\alpha(f) \coloneqq &\Sigma^{\alpha\beta}D_\alpha/\sqrt{\Sigma^{\alpha\alpha}}\,,
\end{align*}
which are the integrand functions of $\big(h\uf_{,\alpha},\,h\ud\big)$ and ${B}_\alpha$, respectively. According to the dependence of $\cac$ and $\cas$ on the parameters, we have 
\begin{align}
	D_1 &\approx\cas_1\sin\Delta \phi  \,,\qquad
	D_2 \approx\cas_2\sin\Delta \phi  \,, \\
	D_3 &\approx\cac_3\cos\Delta \phi  \,,\qquad
	D_4 \approx\cas_4\sin\Delta \phi  \,,
\end{align}
where $\Delta \phi = \Delta \phi_0-2\pi \Delta t_c f$.

When $m_1\uf$ and $m_2\uf$ are fixed, the most special OS consists of two GWs
with the same internal parameters, i.e., $\big(m_1\uf,m_2\uf\big) =
\big(m_1\ud,m_2\ud\big)$. We call this configuration {\sc Equal}. In this case,
$\Delta\phi_0 =0$, and Eq.~\eqref{eq12:} degenerates to
 \begin{align*}
 	\big(h\uf_{,\alpha}\,,h\ud\big)=& \int_{f_{\rm min}}^{{f_{\rm max}}}  {\cal A}^{c}_\alpha  \cos \left(2 \pi \Delta t_{c} f\right) {\rm d}  f
	 - 
	\int_{f_{\rm min}}^{{f_{\rm max}}}  {\cal A}^{s}_\alpha  \sin \left(2 \pi \Delta t_{c} f\right) {\rm d}  f \,.
\end{align*}
Thus, the modulation phase becomes a linear function of $f$, whose slope is
proportional to $\Delta t_c$.

In Fig.~\ref{fig2}, $\big(h\uf_{,3}\,,h\ud\big)$ is an even function of $\dt$,
while the other three components behave like odd functions. The $d_L$ only
contributes to the amplitude part of the waveform, then $\cas_3 = 0$. 
Therefore, $D_3(f;-\dt) = D_3(f;\dt)$ is an even function of $\dt$, and remains
an even function after integration over $f$.  Noting that $t_c$ only contributes
to the phase, we have $\cac_4 = 0$, so $\big(h\uf_{,4}\,,h\ud\big)$ is an odd
function.  For ${\cal M}$ and $\eta$, we have $\cas\gg\cac$. So $D_\alpha$ is
dominated by $\cas$, and both $\big(h\uf_{,1}\,,h\ud\big)$ and
$\big(h\uf_{,2}\,,h\ud\big)$ are almost odd functions. As for
$\left|B_\alpha\right|$, they  are all basically even functions of $\dt$. The
slight asymmetry originates from the linear combination of
$\big(h\uf_{,\alpha}\,,h\ud\big)$. Physically, the intrinsic parameters of the two signals are the same, and only
the amplitudes are different.  Therefore, the relationship between biases and
merger time difference will inherit the symmetry. 

\begin{figure*}[htp]
	\vspace{-0.4cm}
	\centering
\subfigure{\includegraphics[width=12cm]{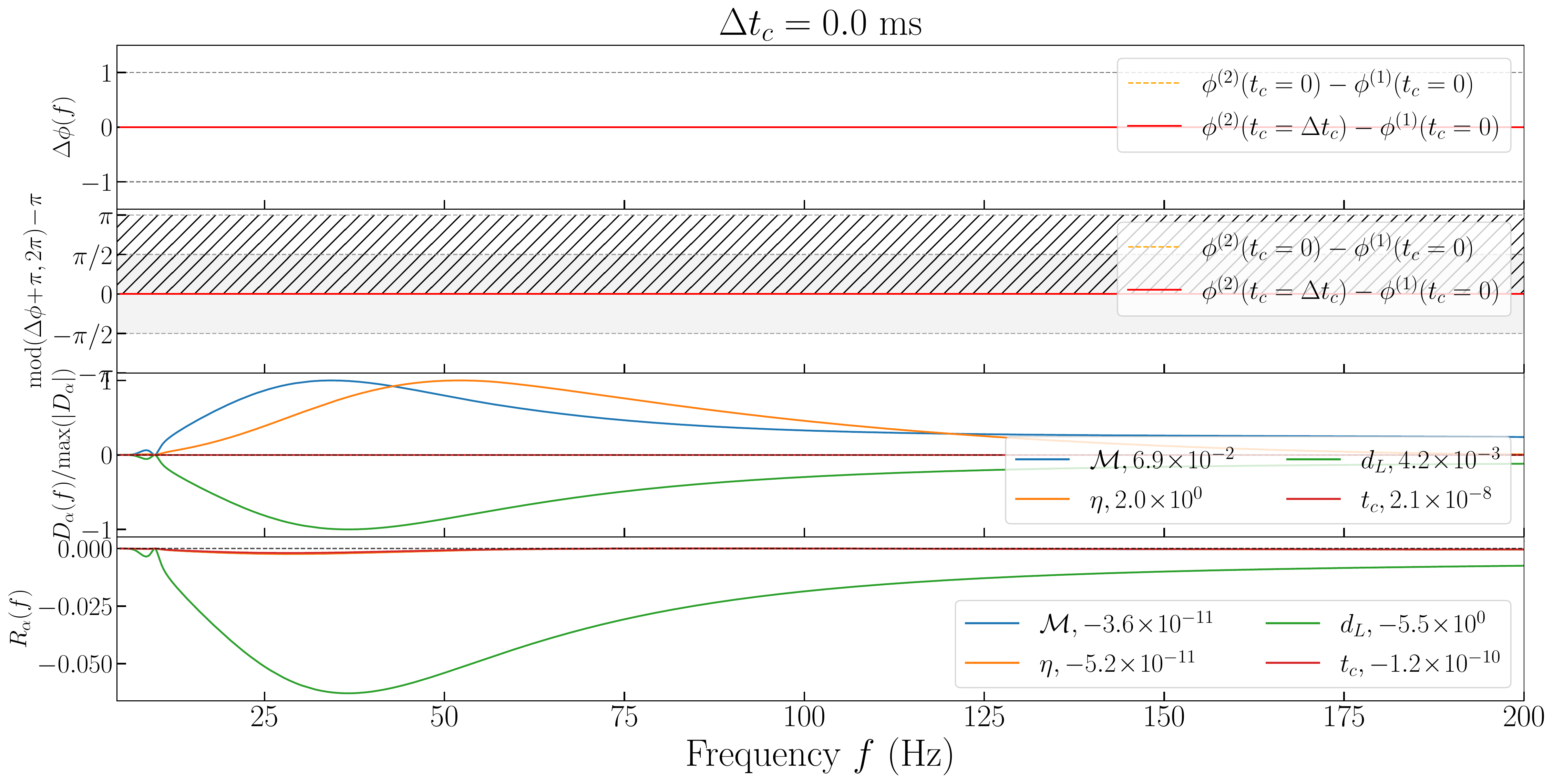}}
\\
\vspace{-0.4cm}
	\centering
\subfigure{\includegraphics[width=12cm]{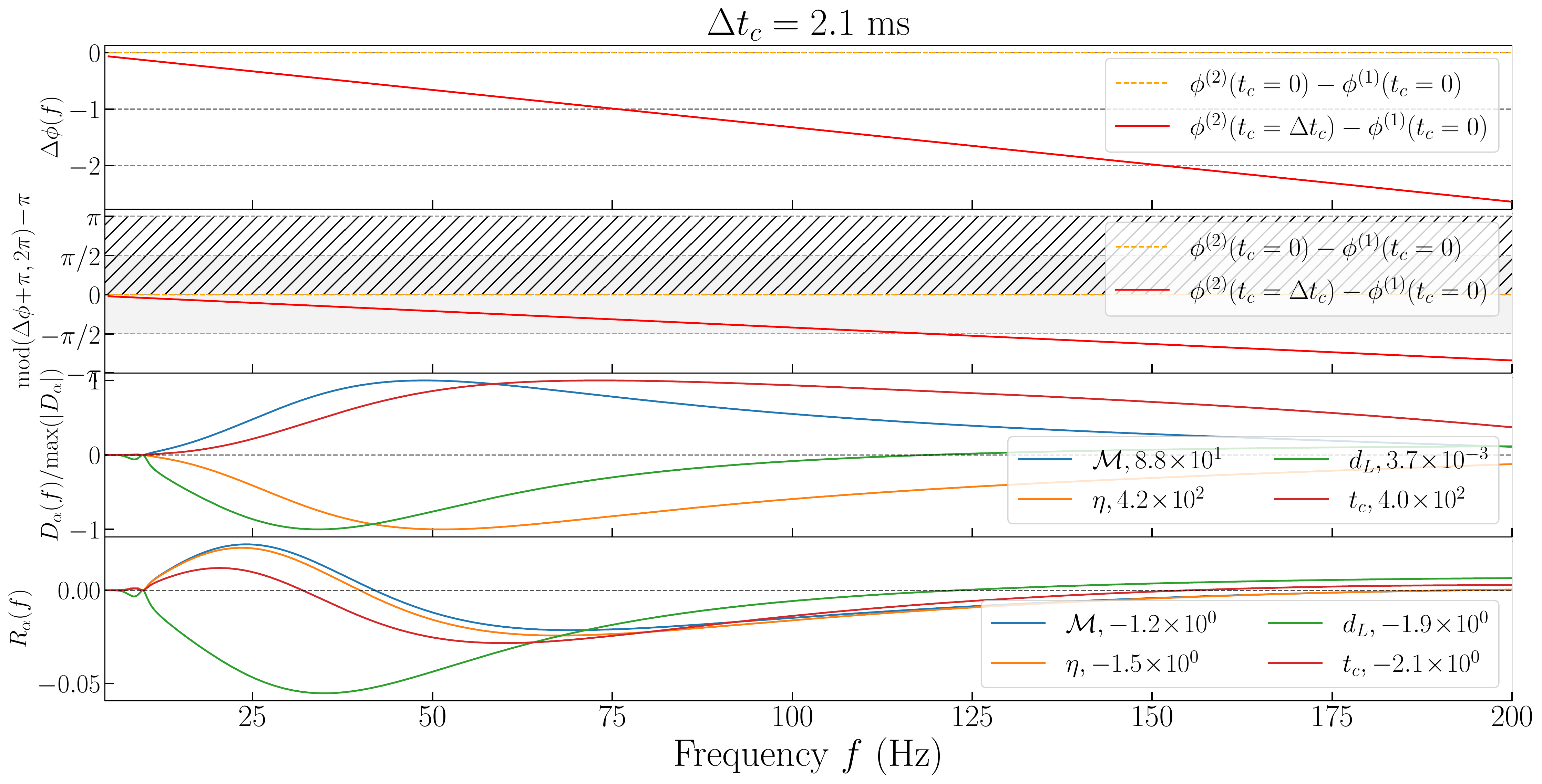}} 
\\ 
\vspace{-0.4cm}\hspace{0.07cm}
	\centering
\subfigure{\includegraphics[width=12cm]{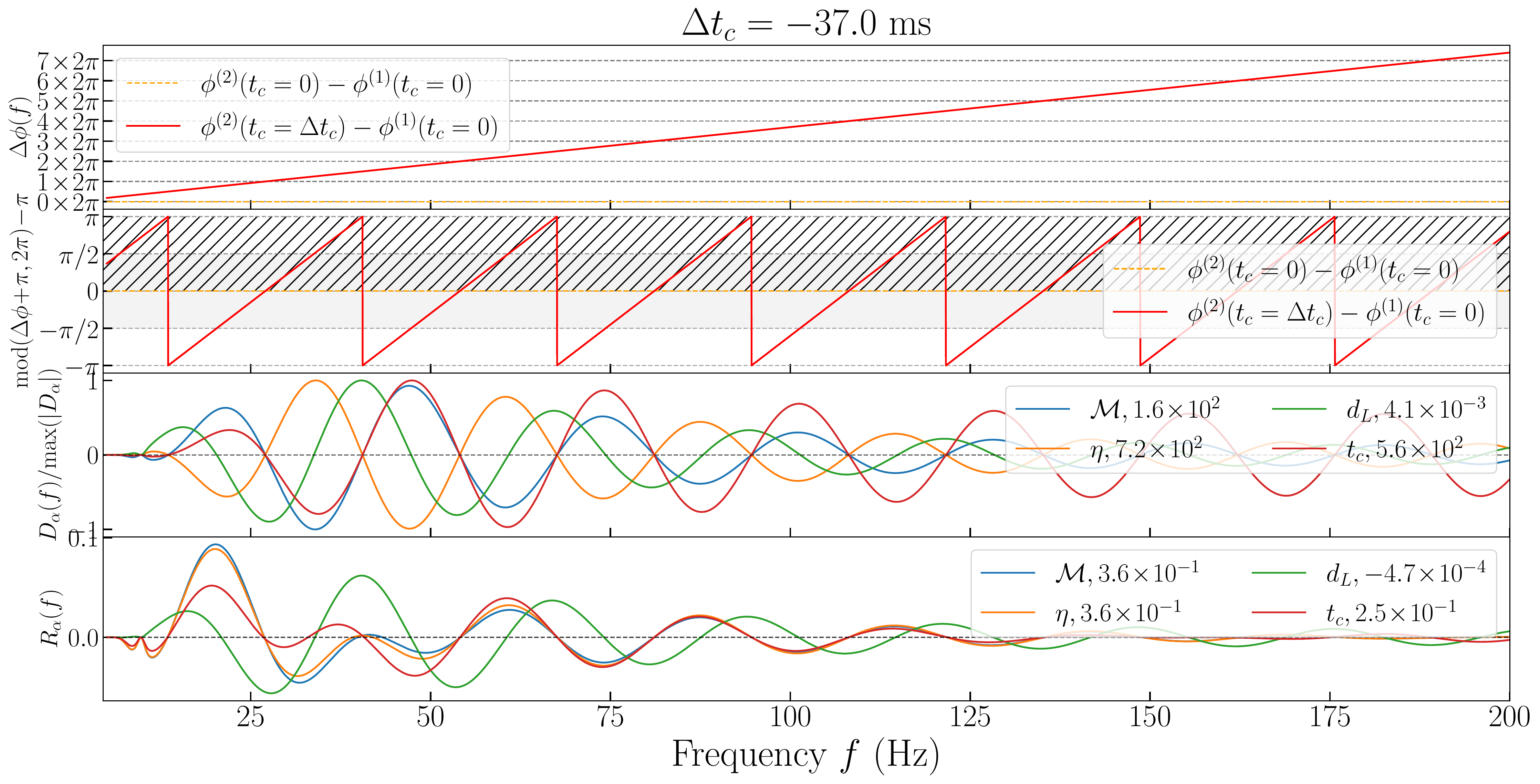}}
\vspace{-0.6cm}
\caption{The dependence of $\Delta \phi(f)$, $D_\alpha$, and $R_\alpha$ on the
frequency $f$ in the {\sc Equal} configuration. The merger time differences are
also marked. Note that the vertical variable ranges, such as $\Delta \phi$, in each subfigure are different. In the second row of each subfigure, the shaded area represents $-\pi/2<\Delta \phi(f)<\pi/2$, while the hatched area means $0<\Delta \phi(f)<\pi$. In the third row, the number after each parameter in the legend is
the maximal value of $|D_\alpha|$. In the forth row, the number means the
integration of $R_\alpha$ over $f$, i.e. $B_\alpha$.} \label{fig3}
\end{figure*}

Another thing is that the biases are large at $\dt  \approx 0 $, but small when $|\dt|$ is far
away from $0$. Obviously, the larger the difference in the merger times is, the
smaller the bias will be. To explain this point quantitatively, we take some $\dt$ and draw the
corresponding $\Delta \phi(f)$, $D_\alpha$, and $R_\alpha$ in Fig.~\ref{fig3}. 
When $\dt = 0$, $|B_{d_L}|>5$, while the other three parameters are almost
unbiased. In this case, $D_\alpha =  \cac_\alpha $, and only $\cac_3 $
significantly deviates from $0$. In the {\sc Equal} configuration and $\dt =
0$, the two signals only have the difference in amplitude, $h\uf+h\ud  \propto
h\uf $. Therefore, the existence of $h\ud $ is equivalent to slightly increasing
the amplitude of $h\uf$, leading to a large impact on the PE of $d_L$. The difference between $\dt=2.1 \mms $ and
$\dt=-37 \mms$ is only reflected in the slope of $ \Delta \phi$ with respect to
$f $. For the case of small $\dt$, $\Delta \phi$ changes slowly by $\pi$ in the range of $0\sim 200\,{\rm Hz}$.  For all
parameters, $D_\alpha$ is significantly nonzero, which eventually leads to large
biases. However, in the $\dt = -37 \mms$ case, the rapid increase of $\Delta
\phi$ leads to violent oscillations of $D_\alpha$, which can not generate large
biases after integration.  This quantitatively explains why the biases are small
for large $\dt$. 
At last, it is worth noting that $\Delta \phi$
always contains the $2\pi\dt f$ term that grows linearly with frequency, so this
behavior can be applied to all mass configurations.

\begin{figure*}[htp]
\vspace{-0.4cm}
	\centering
\subfigure{\includegraphics[width=12cm]{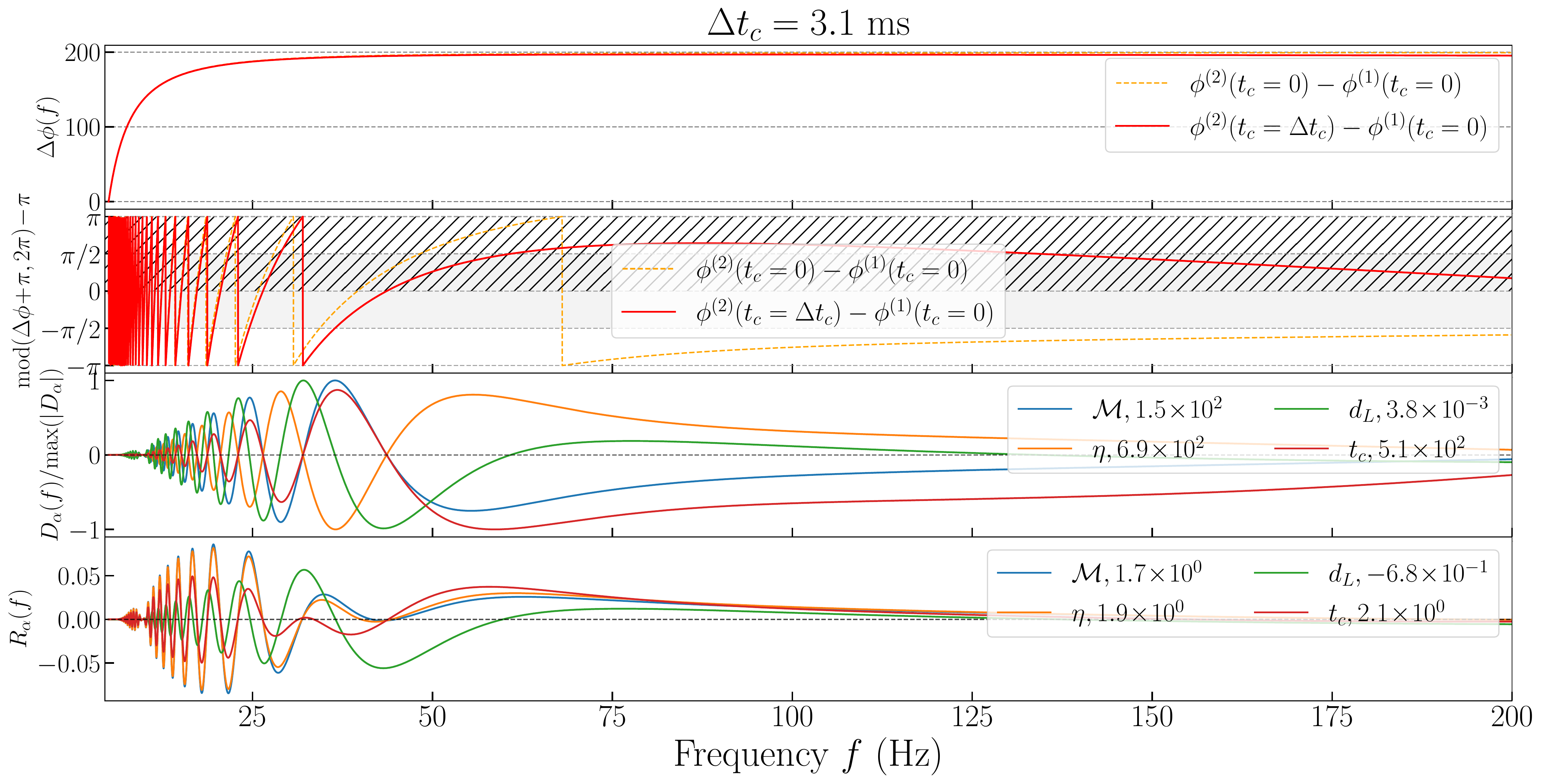}} 
\\ 
\vspace{-0.4cm}\hspace{0.07cm}
	\centering
\subfigure{\includegraphics[width=12cm]{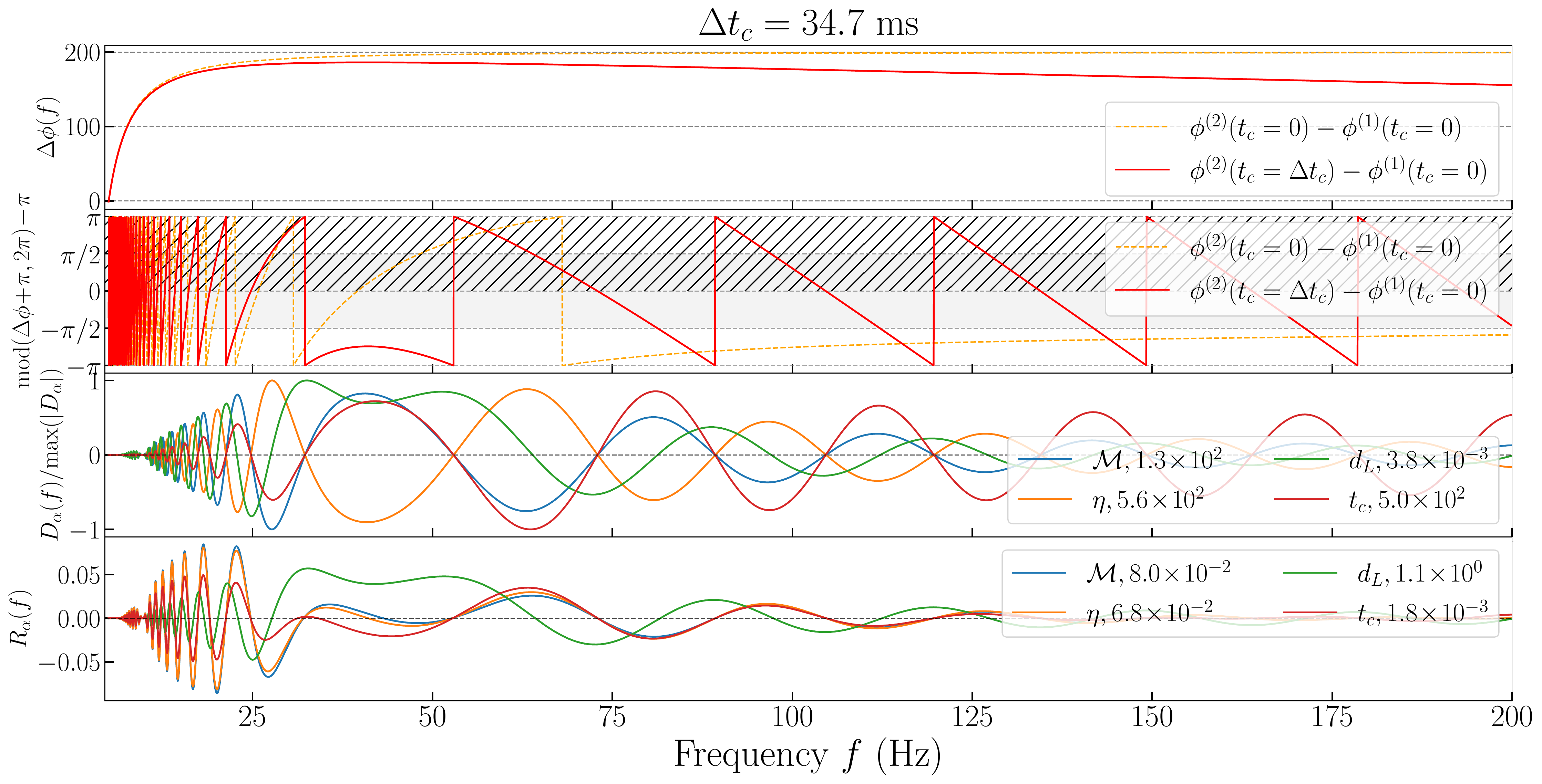}}
\\
\vspace{-0.4cm}
	\centering
\subfigure{\includegraphics[width=12cm]{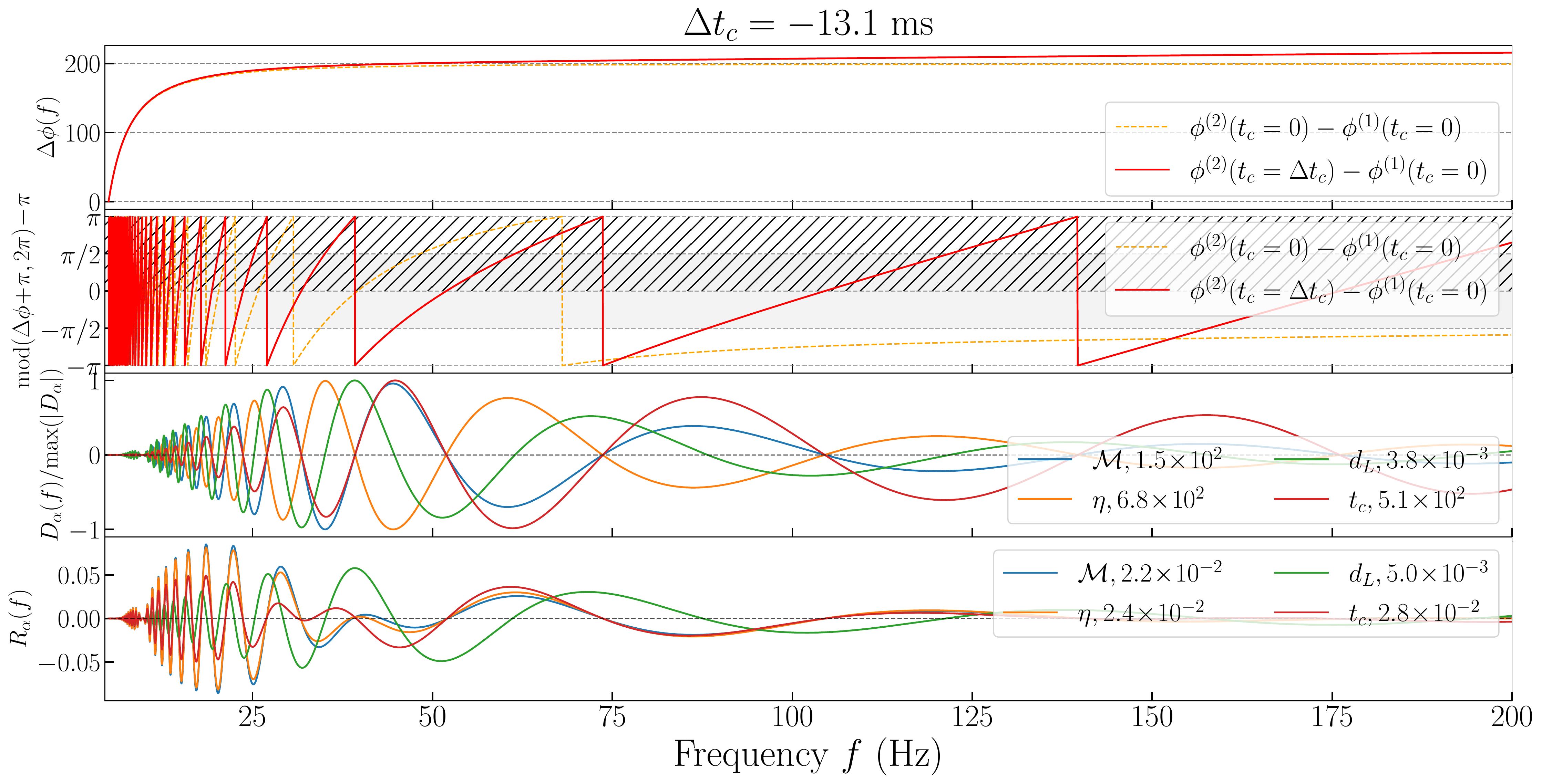}} 
\vspace{-0.3cm}
\caption{Same as Fig.~\ref{fig3}, but for the {\sc Asymmetric} configuration.}
\label{fig4}
\end{figure*}

In the {\sc Equal} case, $\Delta \phi$ is completely determined by $\dt$, but
for general mass configurations, we have to consider the contribution of a
non-zero $\Delta \phi_0$.  In the {\sc Asymmetric} ({\sc Asymmetric2})
configuration, we decrease (increase) $m_1\ud$ and $m_2\ud$ simultaneously,
while keeping the mass ratio unchanged. Taking {\sc Asymmetric} as an example,
we immediately find the asymmetry of $\bm B$ on $\dt$, i.e., the significant
dependence of the merging order.  When $h\ud$ merges after $h\uf$, the biases
are generally large. Otherwise, the biases can be ignored. Same as the {\sc
Equal} case, we also take some $\dt$ as examples and show them in
Fig.~\ref{fig4}.  
Intuitively, we do expect the largest
bias when two signals are ``closest.'' However, from the $B_\alpha\big(\dt\big)$
plot, we find that  when $\dt = 0$ the bias  is not maximum. In the following,
we discuss how these larger biases arise, which correspond to those OSs with
$\dt > 0$.

Mathematically, $D_\alpha$ is a modulated trigonometric function. The modulation
amplitude is $\cac$ ($\cas$), and the phase argument is $\Delta \phi = \Delta
\phi_0-2\pi \dt f$. In order to generate a significant non-zero integral, one
needs a large modulation amplitude and a stable $\Delta \phi(f)$. Since $\cac$
($\cas$) is large in the low-frequency band and slowly decays to $0$ in the
high-frequency band, combined with the behavior of $\Delta \phi_0$, we can
divide the large bias cases into two categories according to the frequency band
as follows.
\begin{enumerate}[(i)]
  \item Low-frequency band. In this case, $\Delta \phi_0$ changes drastically,
  but adding a larger linear term can make $\Delta \phi_0$ stable in a small
  range. Because $\cac$ ($\cas$) is relatively large in the low-frequency band,
  a large bias can also be integrated in a short frequency range. In order to
  offset the drastic changes in $\Delta \phi_0$, the corresponding $|\dt|$ is
  large when biased in this way. 
  \item High-frequency band. When $f$ is high, $\Delta \phi_0$ changes
  approximately linearly, so it is easy to keep $\Delta \phi(f)$ to be a
  constant in a large frequency range by adding a linear term. Although $\cac$
  ($\cas$) is small at this time, a large bias can still be accumulated when
  integrating in a wide frequency band where $\Delta \phi(f)$ is stable. Since
  the change of $\Delta \phi_0$ in the high-frequency band is slow, the
  corresponding $|\dt|$ is also small.
\end{enumerate}

In Fig.~\ref{fig4}, we give examples of the two cases. When $\dt = 3.1\mms$, due
to the existence of the linear term $-2\pi \dt f$,
 $\Delta \phi$ first increases
and then decreases in the $50\sim 150 \,{\rm Hz}$ band.  The change of
monotonicity implies that $\Delta \phi$ basically stays around $\pi/2$ in the
corresponding frequency band, leading to a large bias after integration. As for
$ \dt = 34.7 \mms $, the monotonicity of $\Delta \phi$ changes in the $30\sim 60
\,{\rm Hz}$ band, producing a short stable stage for $D_\alpha$ and $R_\alpha$. 
In this frequency band, both $\cac$ and $\cas$ also reach their maxima, so large
biases can occur. 

To sum up, as long as there is a relatively stable stage in the phase argument
$\Delta \phi$, biased OSs are likely to arise. The location and even the presence of the
stable stage are determined by $\dt$. With this in mind, we can explain why
$B_\alpha$ oscillates and tends to $0$ with the increase of $|\dt|$. Similar phenomenon has been noticed before, but not explored  in detail~\cite{Himemoto:2021ukb}. Taking
$d_L$ as an example, the growth rate of $\Delta \phi_0$ about $f$ gradually
decreases, while the linear term $-2\pi \dt f$ contributes a constant negative
growth rate. Therefore, increasing $\dt$ will cause the stable stage to shift
left.  On the other hand, since $\Delta \phi_0$ is changing between $-\pi$ and
$\pi$ (or equivalently, between $0$ and $2\pi$), the change of the horizontal
position of the stable stage will be accompanied by the change of its vertical
position.  This means that, the stable value of $\Delta \phi$, which is denoted
as $\phi_{\rm sta}$, will oscillate between $-\pi$ and $\pi$. For $d_L$, in the
stable stage, $D_3 = \cac_3 \cos \Delta \phi \approx \cac_3 \cos\phi_{\rm sta}$.
If $\phi_{\rm sta}\approx 0$ or $\pi$, we have $D_3\approx 0$ and
$\big(h\uf_{,4}\,,h\ud\big)\approx 0$ after integration.  As for $B_3$, it is a
linear combination of $\big(h\uf_{,\alpha} , h\ud\big) $, and there still exists
$\phi_ {\rm sta}$ making $B_3\approx 0$. Therefore, when $\dt$ increases from
$0$, the stable stage moves to the left, and $\phi_{\rm sta} $ oscillates
between $-\pi$ and $\pi$, so $B_3$ also oscillates.  When $\dt$ further
increases, the stable stage is very short, and the $\cac$ ($\cas $) in this
frequency band is small, then the cumulative bias is small in the stable stage.
Outside the stable stage, $\Delta \phi $ changes rapidly with the frequency, and
it is impossible to generate a large bias.  Finally, we have ${\bm B} \to 0 $. 

As for the other three parameters, their corresponding $D_\alpha$ is dominated
by $\cas  \sin \Delta  \phi$. When $D_3=0$, the other $D_\alpha$
corresponding to $\alpha = 1,2,4$ will reach the extreme value. Therefore, when the bias of $d_L$
is very small, the bias of the other three parameters is close to the maximum
and vice versa. This is consistent with the results of numerical calculation, as
shown in Fig.~\ref{fig2}.  Further, it is easy to explain why the biases are
generally small when $\dt<0$. Taking $\dt=- 13.1\mms$ as an example (see
Fig.~\ref{fig4}), both $- 2  \pi  \dt f$ and $\Delta \phi_0$ are monotonically
increasing with respect to $f$, and the phase argument changes more rapidly.  So
the oscillation of $D_\alpha$ is more violent than that when $\dt>0$, and there
is no stable stage and large bias. This explains the strong dependence of
$B_\alpha$ on the merge order of the two signals. 

Using the fact that biases are mainly generated at the stable stage, it is easy
to analyze the dependence of biases on $\phi_c$. The numerical results of
\citet{Pizzati:2021apa} indicate that when $|\dt|$ is small the bias is very
sensitive to the difference of merger phase  $\Delta \phi_c = \phi_c\ud-\phi_c
\uf$.\footnote{Note that the merger phase is originally defined in the time
domain.} One can even turn a BOS into an UOS by adjusting $\Delta \phi_c$. Here we give an explanation. Adding the parameter $\Delta
\phi_c  \neq 0 $, the phase argument becomes $ \Delta  \phi= \Delta  \phi_
0+\Delta \phi_ c-2\pi \dt f$, and the stable value becomes
$\phi_{\rm sta}+\Delta \phi_c$.  If $\phi_{\rm sta}+\Delta \phi_c$ is close to $0$ or $\pi$,
then the $D_\alpha$ of ${\cal M}$, $\eta$, and $t_c$ are also very close to $0$.
Even if there is a long stable stage, it is difficult to generate a large bias.
On the contrary, the bias of $d_L$ will be close to the extremum.  Similar
analysis can be applied to $\phi_{\rm sta}+\Delta \phi_c \approx \pm \pi/2$,
where the bias of $d_L$ is small while biases of ${\cal M}$, $\eta$, and $t_c$
reach their extrema.

\begin{figure*}[t]
    \hspace{-0.4cm}
        \centering
    \subfigure{\includegraphics[width=13.2cm]{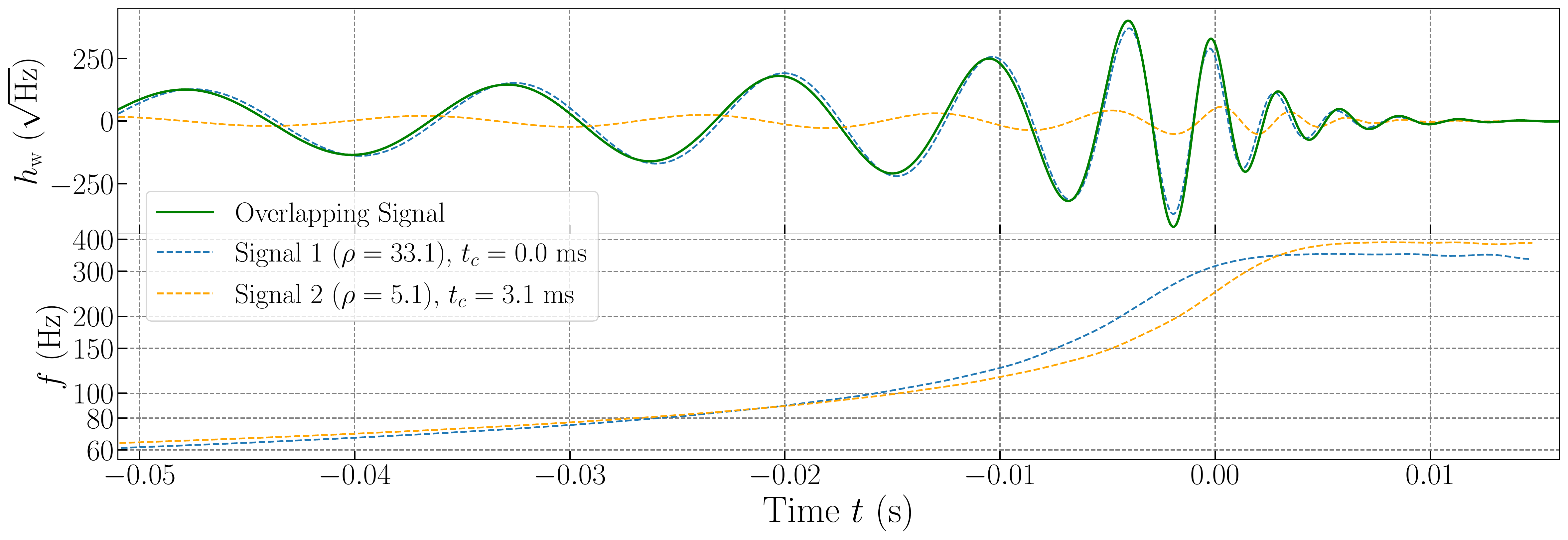}}
    \\
        \centering
    \subfigure{\includegraphics[width=13.5cm]{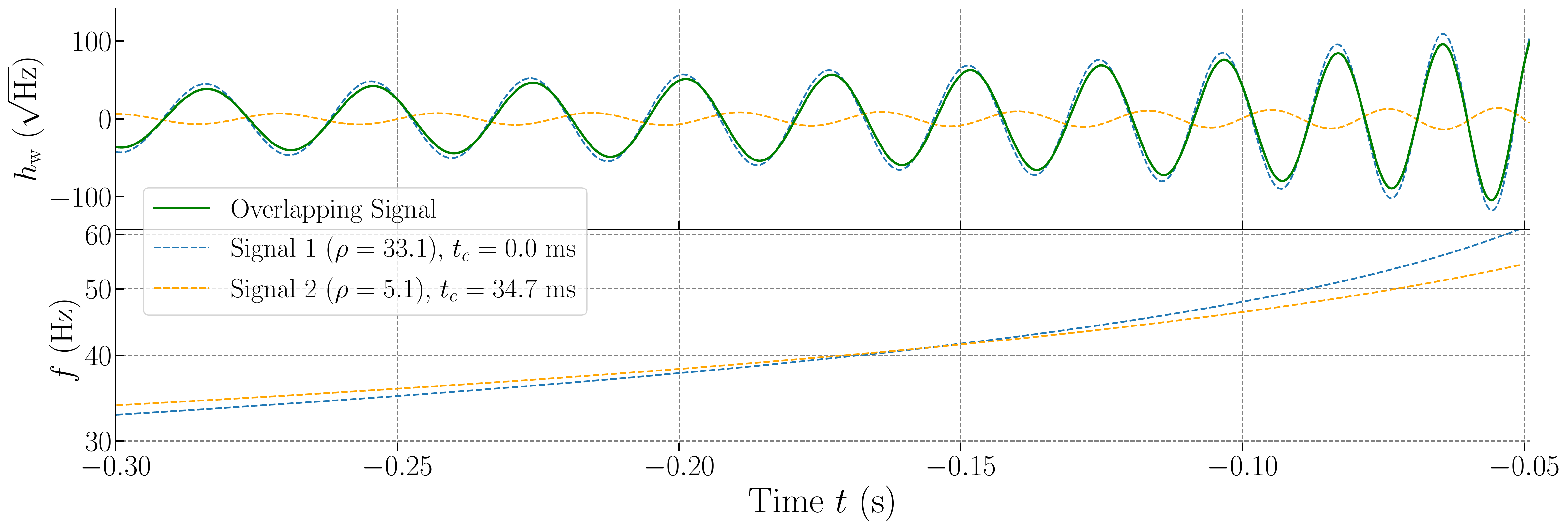}} 
    \caption{The whitened waveforms and frequency evolution diagrams of two signals
    in the time domain for the {\sc Asymmetric} configuration when $\dt$ takes
    $3.1\,\mms$ and $34.7\,\mms$.} \label{fig5}
    \end{figure*}

Through the above analysis, we have understood the origin of biases
mathematically. Physically, the biases come from the overlapping of the
frequency evolution processes of the two signals, that is, in the same time
period, the frequencies of the two signals are relatively close to each other.  As an example, in the {\sc Asymmetry}
configuration, we find that it is easier to generate BOS when $h\ud$ merges
before $h\uf$. The chirp frequency of the lighter binary is higher, so
the frequency evolution processes are easier to ``overlap'' when it merges after
the heavier one. Further, in Fig.~\ref{fig5}, we show the whitened waveform,
$h_{\rm w}$,  and the frequency evolution, $f (t) $, of two signals in the time
domain for some representative $\dt$.  For the case of $ \dt=3.1  \,\rm{ms} $ ($34.7 \,\rm{ms} $), we note that the frequency band $50 \sim 150 \, { \rm Hz} $ ($30  \sim
60 \, { \rm Hz} $), where the chirp frequency of the two signals are relatively
close, is also the frequency band where the phase difference in frequency domain
$ \Delta  \phi$ is stable.  The physical and
mathematical explanations of the origin of biases are (and they must be)
consistent. Generally speaking, the frequency of GW changes rapidly in the
merger and ringdown stages, and only in the inspiral stage can there be a long
overlapping period of the frequency evolution processes.  On the other hand, in
the inspiral phase, the frequency domain phase and the frequency evolution have
the relationship~\cite{Buonanno:2009zt}
\begin{equation}
	\frac{{\rm d} \phi}{{\rm d}  f} = -2\pi t(f)\,. 
\end{equation}
Applying to $h\uf$ and $h\ud$, we have
\begin{equation}
		\frac{{\rm d} \Delta \phi}{{\rm d} f} \approx 0
		\Leftrightarrow \frac{{\rm d}\phi\uf}{{\rm d} f} \approx \frac{{\rm d}\phi\ud}{{\rm d} f}
		\Leftrightarrow t\uf(f)\approx t\ud(f)\,,
\end{equation}
which shows that {\it the phase difference in the frequency domain keeps
constant in a certain frequency band} and that {\it the frequency evolution
processes overlap in the time domain} are equivalent. They just explain the
origin of bias from different aspects. Both explanations are consistent with our intuitive
understanding of the matched filtering technology. If the frequency evolution
over time of the two signals is completely the same, it is impossible to
separate the OS into individual ones. The strong degeneracy will influence the
SPE, leading to large biases.


\section{Other methods of analyzing biases}\label{sec4:other}

In this section, we verify the results and corresponding conclusions of biases
in Sec.~\ref{sec3:bias}, using the correlation coefficients and the full
Bayesian inference. We also point out the feasibility of identifying biased OSs and
unbiased OSs with FM in our analysis. 

\subsection{The correlation between two signals}\label{sec4.1:}

In the previous section, we have used the SPE biases to characterize the effect
of signal overlapping. When large biases occur, it indicates that the two
signals have strong influences on the PEs of each other, 
i.e., the correlation between the two signals is
large.  \citet{Pizzati:2021apa} briefly studied the relationship between the
correlation coefficient of the parameters of the two signals, as well as $\dt$
under some specific configurations. However, they did not discuss the
relationship between the correlation coefficients and the biases. 
\citet{Antonelli:2021vwg} calculated some correlation coefficients and biases,
but the result is limited to a rough qualitative analysis due to the high
dimensionality of the parameter space. Here we calculate the correlation
coefficients between the parameters of the two signals by the FM method, and
focus on the relationship between the correlation
coefficients and the
biases.  Here we only consider the correlation coefficients between the
corresponding parameters of the two signals (for example, ${\cal M}\uf$ and
${\cal M}\ud$), which represent the main correlation between the two signals.
The systematic analysis of the correlation coefficient between any two
parameters (e.g., between ${\cal M}\uf$ and $\eta\ud$) is left to future work.

When calculating the correlation coefficients of the parameters between the two
signals, the FM approximation should be applied to the JPE. The parameters are
$\{\bt\uf,\bt\ud\}$, while the model takes $h\big(\bt\uf,\bt\ud\big) =
h\uf\big(\bt\uf\big)+h\ud\big(\bt\ud\big)$. According to Eq.~\eqref{eq5:FM}, the
FM is
\begin{equation}
	F = \begin{bmatrix}
  F\uf & M\\
  M^\intercal &F\ud
\end{bmatrix}\,,
\end{equation}
where $F\uf$ and $F\ud$ are the FM calculated when the parameters are taken as
$\{\bt\uf\}$ and $ \{\bt\ud\}$, respectively.  The inverses of them are noted as
$ \Sigma \uf$ and $  \Sigma \ud $. For the off-diagonal term, we have $M_{\alpha
\beta} = \big(\partial\uf_\alpha h\uf \,,\partial\ud_\beta h\ud\big) =
\big(h\uf_{,\alpha} \,,h\ud_{,\beta}\big)$, and $M^\intercal$ is the
transposition of $M$.  When $|\dt|$ is large, the correlation between the two
signals is small, so each element of $M$ will be much smaller than those of
$F\uf$ and $F\ud$. In this case, the inverse of $F$ can be written as
\begin{equation}\label{eq16:}
	\Sigma = \begin{bmatrix}
  \Sigma \uf & -\Sigma\uf M \Sigma\ud \\
  -\Sigma\ud M^\intercal \Sigma\uf &\Sigma \ud
\end{bmatrix}+{\cal O}\big(M^2\big)\,.
\end{equation}
Ignoring the ${ \cal O} \big(M^2\big)$ terms, the correlation coefficients
between the corresponding parameters of the two signals are
\begin{equation}\label{eq17:}
	c_{\theta^{(1)\alpha}\theta^{(2)\alpha}} = c_{\alpha,\alpha+4} = \frac{\Sigma^{(1)\alpha\beta}\big(\partial\uf_\beta h\uf \,,\partial\ud_\gamma h\ud \big)\Sigma^{(2)\gamma \alpha}}{\sqrt{\Sigma^{(1)\alpha\alpha}\Sigma^{(2)\alpha\alpha}}}\,.
\end{equation}
Note that $\alpha$ is not summed in the above equation, and both $\{\bt\uf\}$
and $\{\bt\ud\}$ have $4$ independent parameters. Correlation coefficients
$c_{\alpha,  \alpha+4}$ are linear combinations of the inner products
$\big(h\uf_ {,\beta} \,, h\ud_{,\gamma}\big)$, and the combination coefficients
are independent of $\dt$.

The expression of the correlation coefficients in Eq.~\eqref{eq17:} is very
similar to the expression of the biases in Eq.~\eqref{eq11:NB}, except for the
element in the inner product. Therefore, we can use the analysis method in
Sec.~\ref{sec3:bias} to deal with them.  According to Fig.~\ref{fig2}, we find
$B_\alpha$ and $\big(h\uf_{,\alpha}\,,h\ud\big)$ have similar behaviors with
respect to $\dt $, so are the positions of zeros and extrema. Mathematically,
this means that using $\Sigma^{(1)\alpha\beta}$ to linearly combine
$\big(h\uf_{,\alpha} \,, h\ud\big)$ is similar to using  a diagonal matrix.
Applying this conclusion to Eq.~\eqref{eq17:}, we only need to analyze the
dependence of $\big(h\uf_{,\beta} \,,h\ud_{,\gamma}\big)$ on $\dt$. 

Like in Eq.~\eqref{eq12:}, we explicitly write the expression of
$\big(h\uf_{,\beta} \,,h\ud_{,\gamma}\big)$
\begin{equation}\label{eq18:}
\begin{aligned}
	\big(h\uf_{,\beta} \,,h\ud_{,\gamma}\big) =& 4 \int_{f_{\rm min} }^{f_{\rm max} } \frac{A_{,\beta}^{(1)} A_{, \gamma}^{(2)}+A^{(1)} A^{(2)} \phi_{, \beta}^{(1)} \phi_{, \gamma}^{(2)}}{S_{n}} \cos \left (\Delta \phi_0-2\pi \Delta t_c f \right) {\rm d}  f\\
	&- 4 \int_{f_{\rm min} }^{f_{\rm max} } \frac{A_{,\beta}^{(1)} A^{(2)}\phi_{, \gamma}^{(2)} +A^{(1)} A^{(2)}_{, \gamma} \phi_{, \beta}^{(1)} }{S_{n}} \sin \left(\Delta \phi_0-2\pi \Delta t_c f \right) {\rm d}  f\,,
	\end{aligned}
\end{equation}
where $\Delta \phi_0$ is the same as in Eq.~\eqref{eq12:}. Since the linear
combinatorial effects of $\Sigma\uf$ and $\Sigma\ud$ are equivalent to diagonal
matrices, we only need to consider the contribution of the diagonal elements. 
Take $\beta = \gamma = \alpha$, substitute them in Eq.~\eqref{eq12:} we have
$c_{\alpha,\alpha+4} \propto \big(h\uf_{,\alpha} \,,h\ud_{,\alpha}\big)$. For
the luminosity distance, it does not contribute to the phase, i.e.\ $\phi_{,3} =
0$. For the merger time, $A_{,4} = 0$. As for ${\cal M}$ and $\eta $, we have
$A_{,\alpha}\ll A\phi_ {,\alpha}$. Therefore, in the special case of $ \beta=
\gamma= \alpha $, the integral of $\sin  \Delta  \phi$ in Eq.~\eqref {eq18:} can
always be insignificant, that is
\begin{equation*}
	\big(h\uf_{,\beta} \,,h\ud_{,\gamma}\big) \approx 4 \int_{f_{\rm min} }^{f_{\rm max} } \frac{A_{,\beta}^{(1)} A_{, \gamma}^{(2)}+A^{(1)} A^{(2)} \phi_{, \beta}^{(1)} \phi_{, \gamma}^{(2)}}{S_{n}} \cos \left (\Delta \phi_0-2\pi \Delta t_c f \right) {\rm d}  f\,,
\end{equation*}
which is the modulation integral of $\cos  \Delta  \phi $.
Therefore, we can expect the dependence of $c_ { \alpha,
\alpha+4}$ and $B_3$ (the bias of luminosity distance) on $\dt$ to be almost the
same.

\begin{figure*}[htp]
	\vspace{-0.4cm}
	\centering
\subfigure{\includegraphics[width=12cm]{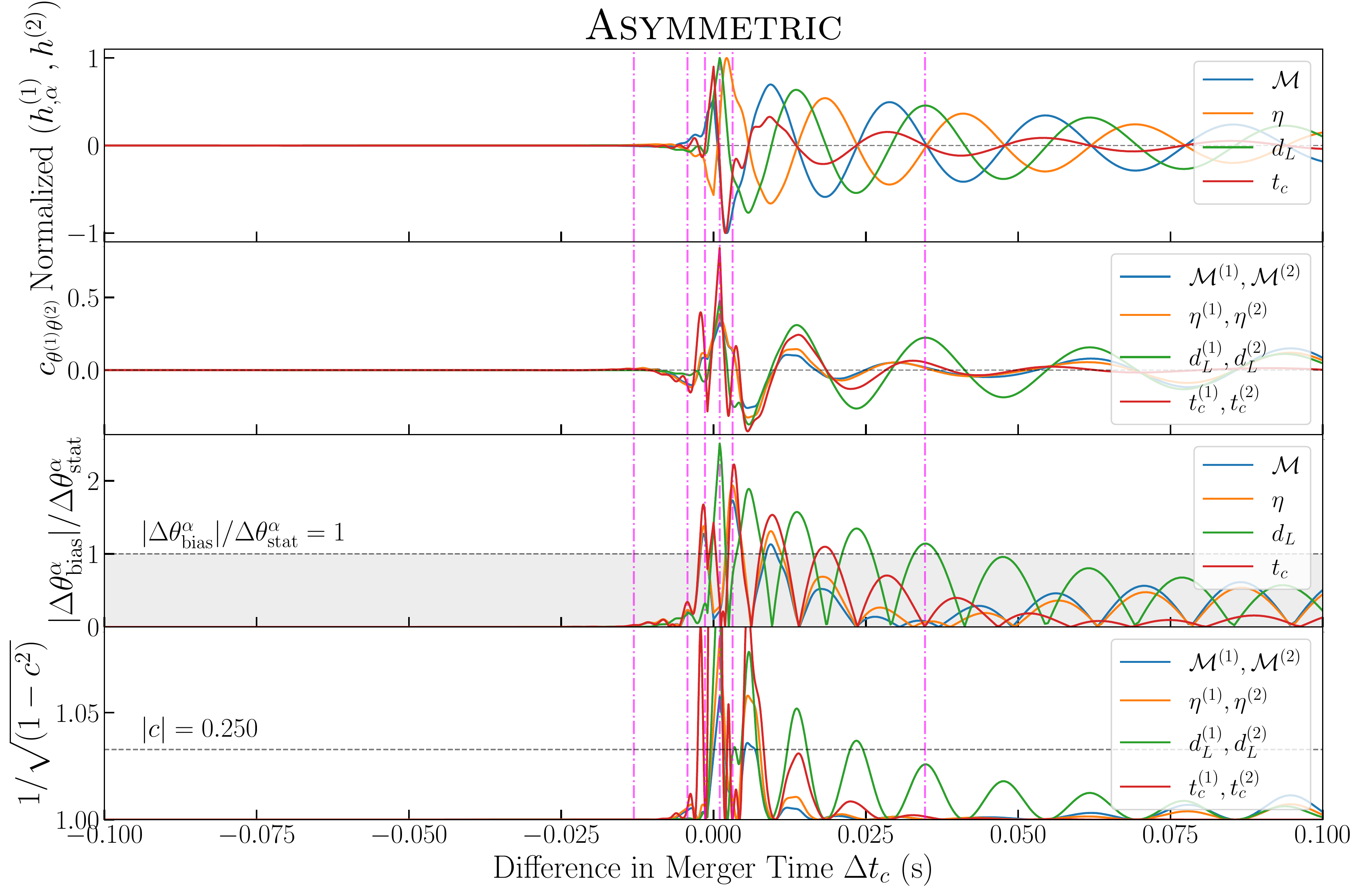}}
\\
\vspace{-0.4cm}
	\centering
\subfigure{\includegraphics[width=12cm]{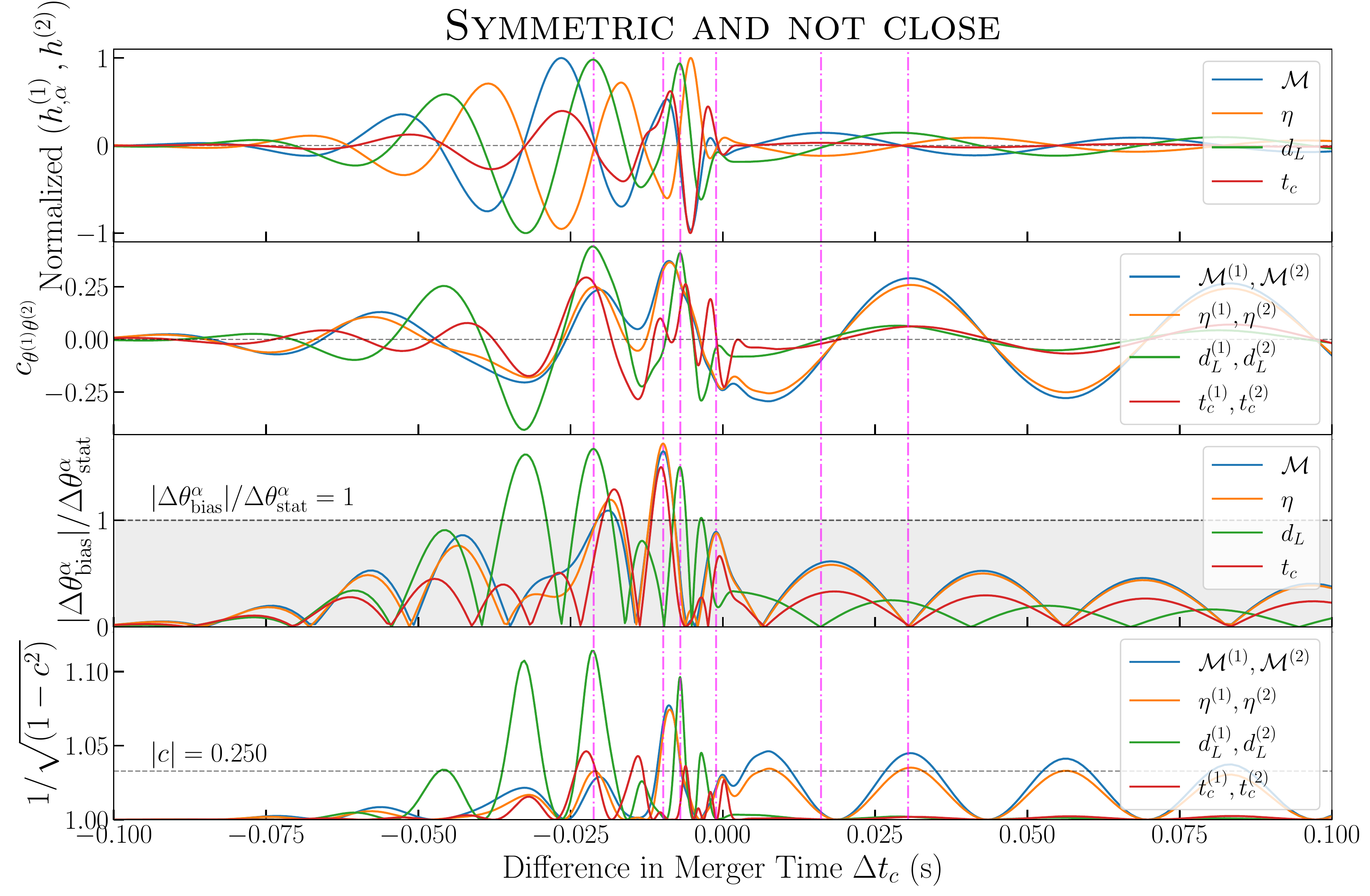}} 
\vspace{-0.3cm}
\caption{The dependence of biases and correlation coefficients on $\dt$ in the
{\sc Symmetric and not close} configuration. The behaviors of
$\big(h\uf_{,\alpha} \,,h\ud\big)$ and $1/\sqrt{1-c^2}$ are also shown for
comparison.} \label{fig8}
\end{figure*}

\refereeA{In Fig.~\ref{fig8}, we show the dependence of $\big(h\uf_{,\alpha}\,,h\ud\big)$, correlation coefficients $c_{\alpha,\alpha+4}$, and $B_\alpha$ on the merger time difference
$\dt$.}  For the sake of brevity, we only show the results of two configurations,
{\sc Asymmetric} and {\sc Symmetric and not close}, which can clearly reflect
the oscillation of $B_\alpha$.  When $| c | $ is small ($\lesssim 0.5 $), the
behaviors of $c_ {\alpha,  \alpha+4}$  versus $\dt$ are highly consistent with
$B_3$, and the positions of extrema and zeros are also similar. 
Correspondingly, for ${\cal M}$,  $\eta$, and $ t_c $, the zeros of their biases
basically coincide with the extrema of $c_{\alpha, \alpha+4}$.  When $|c|$ is
large (such as the region around $\dt\sim 0$ in the {\sc Asymmetric}
configuration), the oscillations of correlation coefficients and biases are more
severe, but their basic trends are still similar, namely, $c$ increases
(decreases) when $\bm B$ increases (decreases).

In
this work, we only discuss the dependence of the relative size of
correlation coefficients and biases on $\dt$. In fact, it is incorrect to expect
that there can not be any biased OS when $|c|$ is small.  For example, in the {\sc
Symmetric and not close} configuration, there is $| c |\lesssim 0.4 $, but the
maximum of $B_\alpha$ is about $1.5$. For the {\sc Asymmetric} configuration,
the maximum value of $|c|$ is close to $1$, but the maximum of $B_\alpha$ is
only slightly greater than $2$.  After rescaling the biases by the SNRs of the
two configurations, their maxima of biases have only small difference. On the
other hand, when $|c|$ is large, the higher-order term of $M$ in
Eq.~\eqref{eq16:} cannot be ignored, and the relationship between correlation
coefficients and biases will become more complicated. 

The correlation between any two parameters can also be
analyzed using Eq.~\eqref{eq18:}.  For example, if one wants to study the
correlation between $d_L \uf$ and $t_c\ud $, taking $ \beta=3$ and $\gamma=4 $,
only the $A_ {,\beta}^{(1)} A^{(2)}\phi_ {,  \gamma} ^ {(2)} $ term is nonzero. 
Thus, Eq.~\eqref{eq18:} becomes the modulation integral of $\sin \Delta \phi$,
so the correlation coefficient will behave similarly to $B_4$. In addation,  through simple calculation we have
$c \propto \big(d_L\uf\big)^0 \big(d_L\ud\big)^0$. This means that the
correlation coefficients do not depend on the strength of the two signals, which
is consistent with our intuitive understanding.  

\subsection{Full bayesian inference}\label{sec4.2}

\begin{figure*}[tp]
        \centering
    \subfigure{\includegraphics[width=15cm]{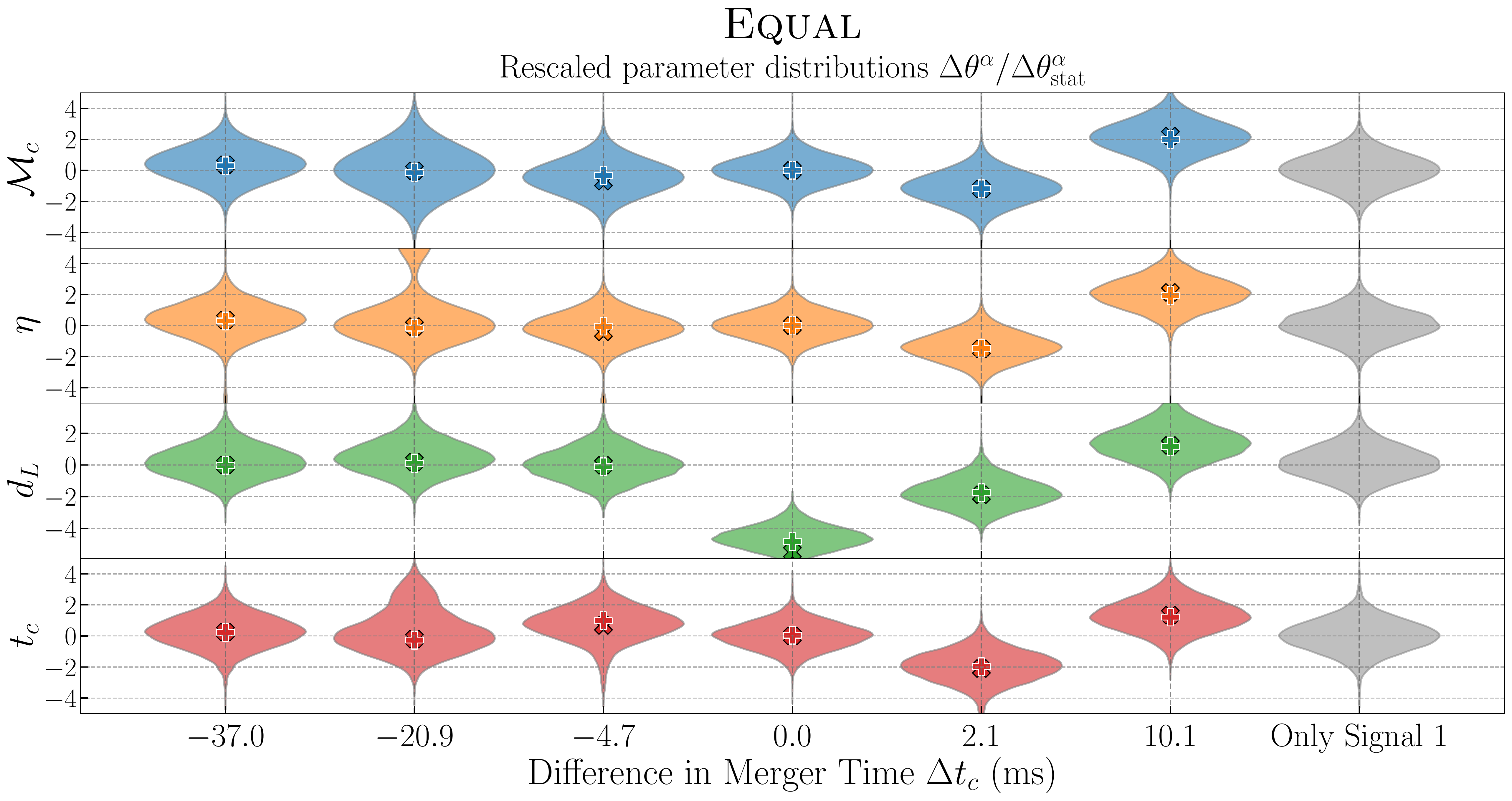}}
    \vspace{-0.4cm}
    \caption{The SPE results in the {\sc Equal} configuration, rescaled by the
    statistical uncertainty $\Delta \theta_{\rm stat}$ from FM. The maximum
    likelihood points in the full Bayesian analysis and biases forecasted by the FM
    are labeled ``$+$'' and ``$\times$'' respectively. For reference, the last
    column shows the SPE results when the data stream contains only the louder
    signal.} \label{fig9}
    \end{figure*}

So far, we have used the FM method to discuss the biases in SPE. This requires
that the SNR of the target signal $h\uf$ is large enough,\footnote{When
calculating the correlation coefficients with the FM method, the SNRs of both
$h\uf$ and $h\ud$ need to be large.} and the biases are small enough to make linearized-signal approximation valid~\cite{Finn:1992wt,Vallisneri:2007ev}.  Although the results obtained by FM
in the previous sections are consistent with our physical intuition, in
principle, it is always necessary to verify the applicability of the FM
approximation.  In this section, we select some representative points for the
$6$ group parameter configurations in Sec.~\ref{sec3.4}, such as the extrema and
zeros of $B_\alpha$. We use Eq.~\eqref{eq8:real likelihood} to perform full
Bayesian PEs and compare them with the results in Sec.~\ref{sec3:bias}. 

We use open source software package {\sc Parallel Bilby,
UltraNest}~\cite{Ashton:2018jfp,Smith:2019ucc,Romero-Shaw:2020owr,2021arXiv210109675B}
and the nested sampling Monte Carlo algorithm {\sc
MLFriends}~\cite{Johannes:2016,2019PASP..131j8005B} to conduct SPEs on $\bt =
\big\{{\cal M}\uf,\eta\uf,d_L\uf,t_c\uf \big\}$.  For ${\cal M}$, $\eta$, and
$t_c$, we choose uniform priors. As for $d_L$, we first generate a uniform prior
for the comoving volume $V_c$ in the source frame, and then convert it in the
parameter space of $d_L$.  Different prior choices may affect PE results, but
the difference between them can be ignored when $\rho\uf$ is sufficiently
large~\cite{Vallisneri:2007ev}. 

When injecting mock GW data, $g(t)$, different noise realizations also affect
the PE results. Here we take $n=0$, which is the most convenient case for
generating data.  More importantly, as can be seen from Eq.~\eqref{eq9:}, the
deviation of the MLE result from the true value in this case is precisely the
bias, $\Sigma^{\alpha\beta}\big(h\uf_\beta \,,h\ud\big)$. It is not only reasonable, but also necessary to use the SPE results when $n=0$ to verify the validity of FM~\cite{Nissanke:2009kt}.

\begin{figure*}[htp]
        \centering
    \subfigure{\includegraphics[width=13cm]{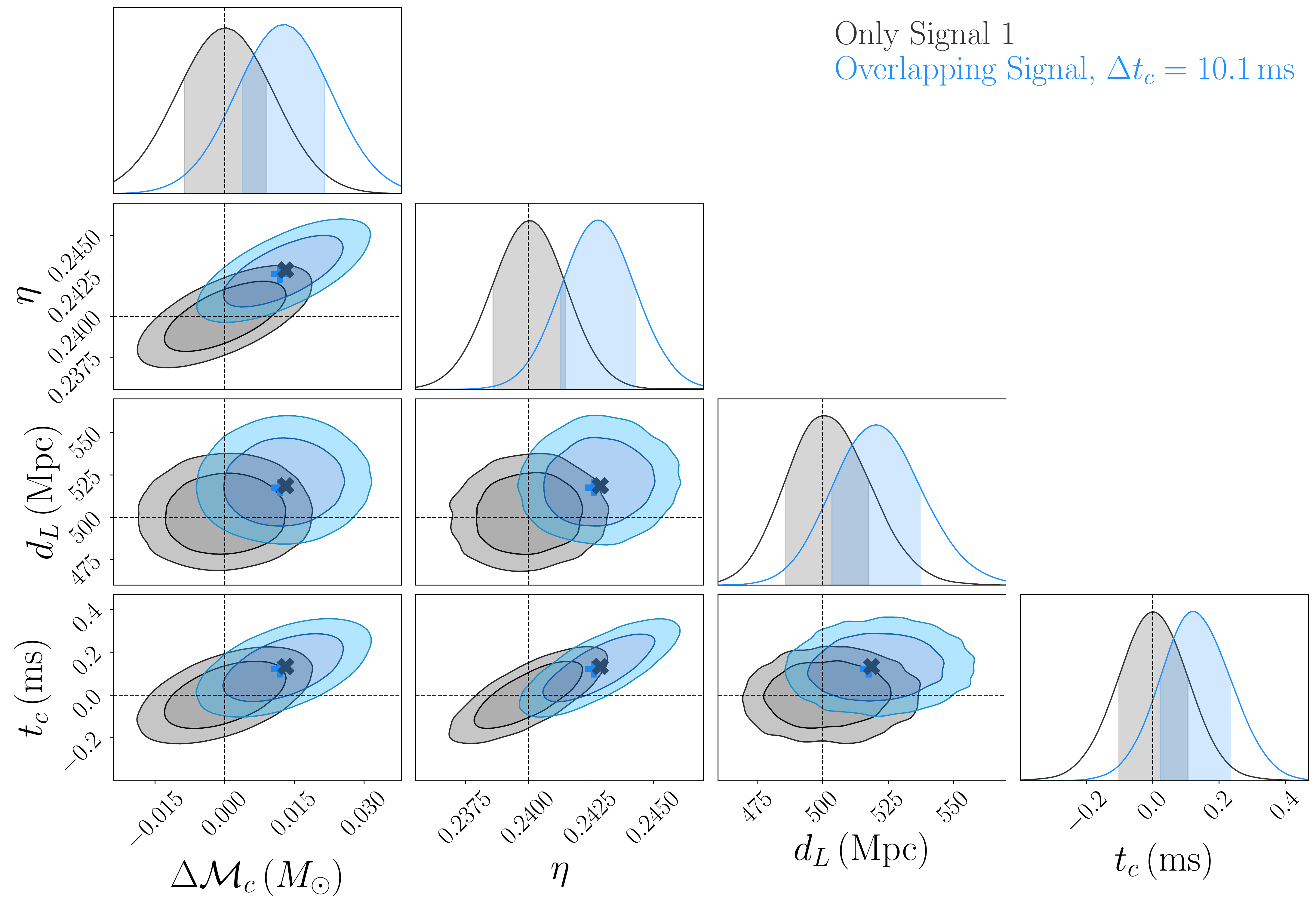}}
    \vspace{-0.3cm}
    \caption{The contour plot of the parameters when $\dt = 10.1\mms$ in the {\sc
    Equal} configuration. The maximum likelihood points in the full Bayesian
    analysis and biases forecasted by the FM are labeled ``$+$'' and ``$\times$''
    respectively.} \label{fig10}
    \end{figure*}
    
In Fig.~\ref{fig9}, we show the marginalized distribution of each parameter in
the {\sc Equal} configuration, with $\dt$ taken as those values corresponding to
the magenta reference lines in Fig.~\ref{fig2}.  The biases calculated using FM
in Sec.~\ref{sec3:bias} are marked in the figure. The results of the other $5$
groups of configurations are shown in~\ref{appD:}.  For PE results with
small biases, the difference between MLE and the truth is almost the same as the
biases forecasted by the FM approximation. Furthermore, the contour plot of the
parameters when $\dt = 10.1\mms$ in the {\sc Equal} configuration is shown in
Fig.~\ref{fig10}.  In this case, all parameters have a large bias
($|B_\alpha|>1$), but the forecasted biases of FM still highly coincide with the
MLE of SPE. 
\section{Conclusion and Summary}
\label{sec5:}

With the continuous upgrading of GW detectors, more and more GW events will be
detected. For next generation ground-based detectors, a large number of GW signals will
overlap with each other~\cite{Regimbau:2009rk, Samajdar:2021egv,
Pizzati:2021apa, Relton:2021cax, Himemoto:2021ukb}.  In order to extract
physical information from these OSs more efficiently and accurately, it is
necessary to study the PE technology of OSs. However, before performing complex
PE on the whole OS, it is important to understand how the PE of a specified
signal in the OS is affected by other signals.  This effect is reflected in the
bias between the SPE result and the true value of the specified signal. If the
overlapping effects on each SPE are small, then the signals in the OS can be
analyzed separately, and there is no need for the time-consuming JPE.  So far,
many works have discussed the SPE biases of OSs, covering various combinations
of detectors, waveforms, and GW
sources~\cite{Regimbau:2009rk,Samajdar:2021egv,Pizzati:2021apa,Relton:2021cax,Himemoto:2021ukb,Antonelli:2021vwg,Janquart:2022nyz,Smith:2021bqc,Hu:2022bji}.
However, calculating lots of SPEs will cost considerable computing resources,
and it is difficult to study the systematic dependence of bias on various
parameters, resulting in our limited understanding of their behaviors.

In this work, we use the FM method to investigate the origin of biases in SPE
and their dependence on GW parameters. Thanks to the fast computational speed of
the FM method~\cite{Finn:1992wt, Vallisneri:2007ev, Wang:2022kia}, we can
explore the parameter space in more detail, and gain a deeper understanding of
the biases.  In Refs.~\cite{Pizzati:2021apa, Antonelli:2021vwg}, the FM method
was considered, but not treated as the primary tool to analyze biases. Here,
with the help of FM, we establish an analytical expression of SPE biases in
Sec.~\ref{sec2:method}.  We mainly studied the dependence of bias on the mass,
luminosity distance, coalescence time, and phase, where OSs are generated by two
BBH systems in AdvLIGO.  Since the direct cause of OS is the overlapping of two
signals in the time domain, the relationship between $\dt$ and biases is
particularly discussed in detail.

In Sec.~\ref{sec3:bias}, we find that there are large biases in the PE when the
frequency evolutions of the two signals overlap to a large extent.  In the
frequency domain, this is equivalent to a constant phase difference between the
two signals in the corresponding frequency band. This conclusion can also be
intuitively obtained by considering the properties of matched filtering
technology, but in this work, we give a rigorous and quantitative demonstration.
For the whole parameter space, the existence of large biases is common, but
biases are more likely to occur when the masses of the two signals, especially
the chirp masses, are close.  For most OSs, the biases strongly depend on the
merger order of the two BBHs. When the binary with heavier component masses
merges first, its frequency evolution is more likely to overlap with the other
one, leading to large biases.  The biases always reach their global extrema near
$\dt \simeq 0 $, and then oscillate and tend to zero with the increase of $|\dt|
$.  For a specific parameter, the behavior depends on how this parameter
contributes to the waveform. The zero points of the bias of the parameter that
mainly contributes to the frequency-domain phase are often the extreme points of
the bias of the parameter that mainly contributes to the amplitude.

We also used other methods to analyze the bias behaviors. In
Sec.~\ref{sec4:other}, we calculate the correlation coefficient of the
corresponding parameters of the two signals, and compare their behaviors with
biases.  Using the explicit expressions in Eqs.~\eqref{eq12:} and \eqref{eq18:},
we find that the behavior of correlation coefficients is similar to the bias of
the parameter that mainly contributes to the amplitude.  The numerical results
verified this result (see Fig.~\ref{fig8}). We also conducted full Bayesian
inference on some representative parameter configurations to check the validity
of the FM approximation.  In conclusion, using FM to forecast the biases
beforehand provides a powerful guidance for a full JPE or SPEs.

The main purpose of this work is to explain the origin of biases. For the
conciseness and universality of our conclusion, some secondary effects are ignored.  In the future, the results of this work should be further
extended before practical application. For parameters, we only selected the most
critical ones including the binary masses, luminosity distance, merger time, and
merger phase. When considering more parameters, such as the spins and the sky location, the PE error will likely increase, and the absolute size of the bias will also be affected. \refereeB{For
detectors, we only considered OSs in AdvLIGO in this work and expect the bias behaviors to be similar to those in XG
ground-based detectors, which is implied by recent works \cite{Antonelli:2021vwg,Dang:2023xkj,Himemoto:2021ukb}.} For long
signals, one needs to consider the rotation of the Earth.  For space-based
detectors, the duration of each signal will be even longer, and biases in these
detectors may have some special properties. Besides, the utilization of detector
networks can also have a great impact on biases, depending on the merger times of two signals in different detectors~\cite{Relton:2021cax}. For
waveforms, the influence of waveform selection on bias can be studied. \refereeB{In
addition, PE biases can come from other effects, such as inaccurate waveform modeling~\cite{Hu:2022bji} and existence of glitches~\cite{Ghonge:2023ksb}.} For the correlation between two
signals, we pointed out the similarity between correlation coefficients and
biases more quantitatively, extending the studies by \citet{Pizzati:2021apa} and
\citet{Antonelli:2021vwg}. Whether there is a numerical relationship between
them remains to be discussed.  Finally, the conclusions of this work can be
probably extended to  cases where more GW signals overlap. In the FM
approximation, the bias caused by different GW signals can be linearly
superimposed, so one will need to consider the biases led by every other signal
one by one. 

It should be pointed out that, this work focuses on understanding the PE
behaviors from OSs, rather than seeking specific algorithms for analyzing OSs. 
In real GW detection, we do not know the true values of each signal, and even not know how many individual signals are present in the OS. The simple FM method
can provide some insights for PE, but cannot replace a focused analysis. 
On the contrary, we expect that, the understanding of the mechanism of biases in OS will help develop a faster and more efficient PE algorithm, and serve in analyzing real OS events.

\section*{Acknowledgements}
This work was supported by the National Natural Science Foundation of China
(11975027, 12147177, 11991053, 11721303), the China Postdoctoral Science Foundation
(2021TQ0018), the National SKA Program of China (2020SKA0120300), the Beijing
Municipal Natural Science Foundation (1242018), the Max Planck Partner Group
Program funded by the Max Planck Society, and the High-Performance Computing
Platform of Peking University.  J.\ Z. is supported by the ``LiYun''
Postdoctoral Fellowship of Beijing Normal University, and C.\ L. is supported by
the China Scholarship Council (CSC).

\appendix
\clearpage
\markboth{Anatomy of parameter-estimation biases in overlapping gravitational-wave
signals}{Anatomy of parameter-estimation biases in overlapping gravitational-wave
signals}


\section{Waveforms and derivatives of waveforms}
\label{appA:}
\begin{figure*}[t]
	\centering
\subfigure{\includegraphics[width=11cm]{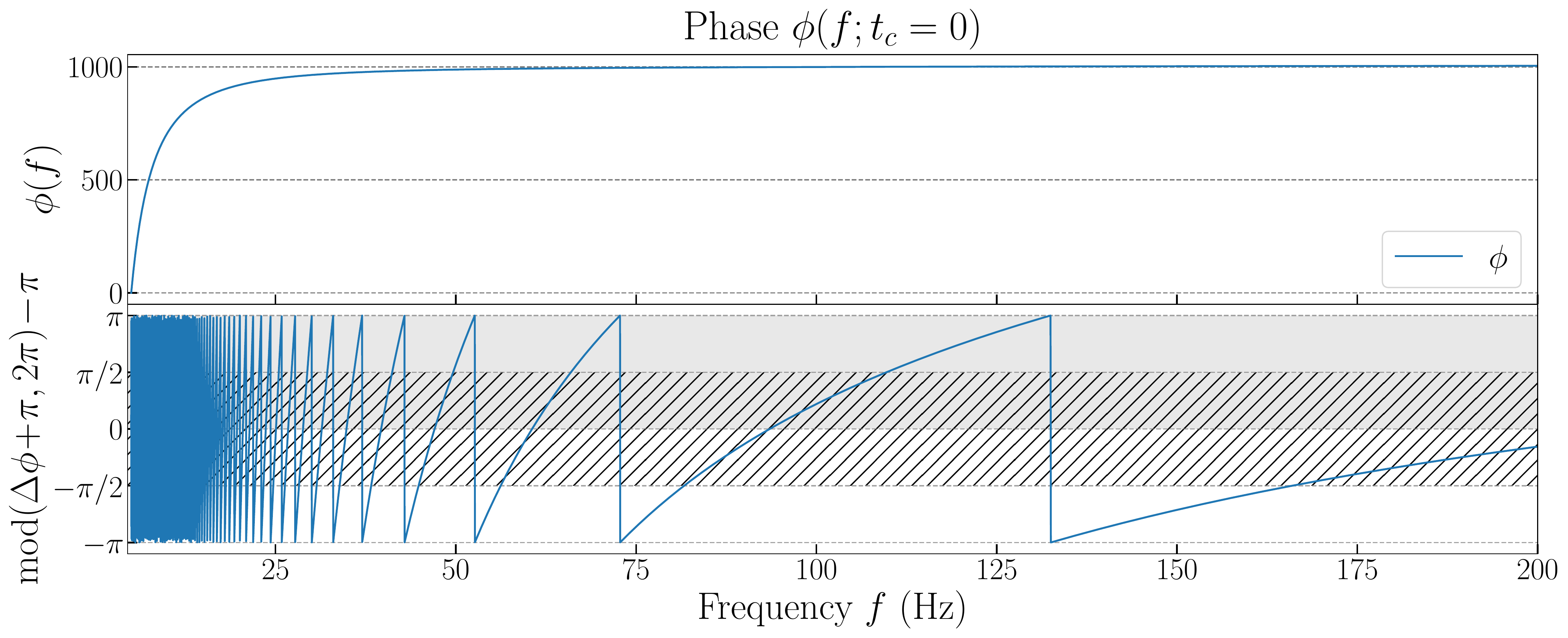}} 
\vspace{-0.2cm}
\\ 
	\centering
\subfigure{\includegraphics[width=11cm]{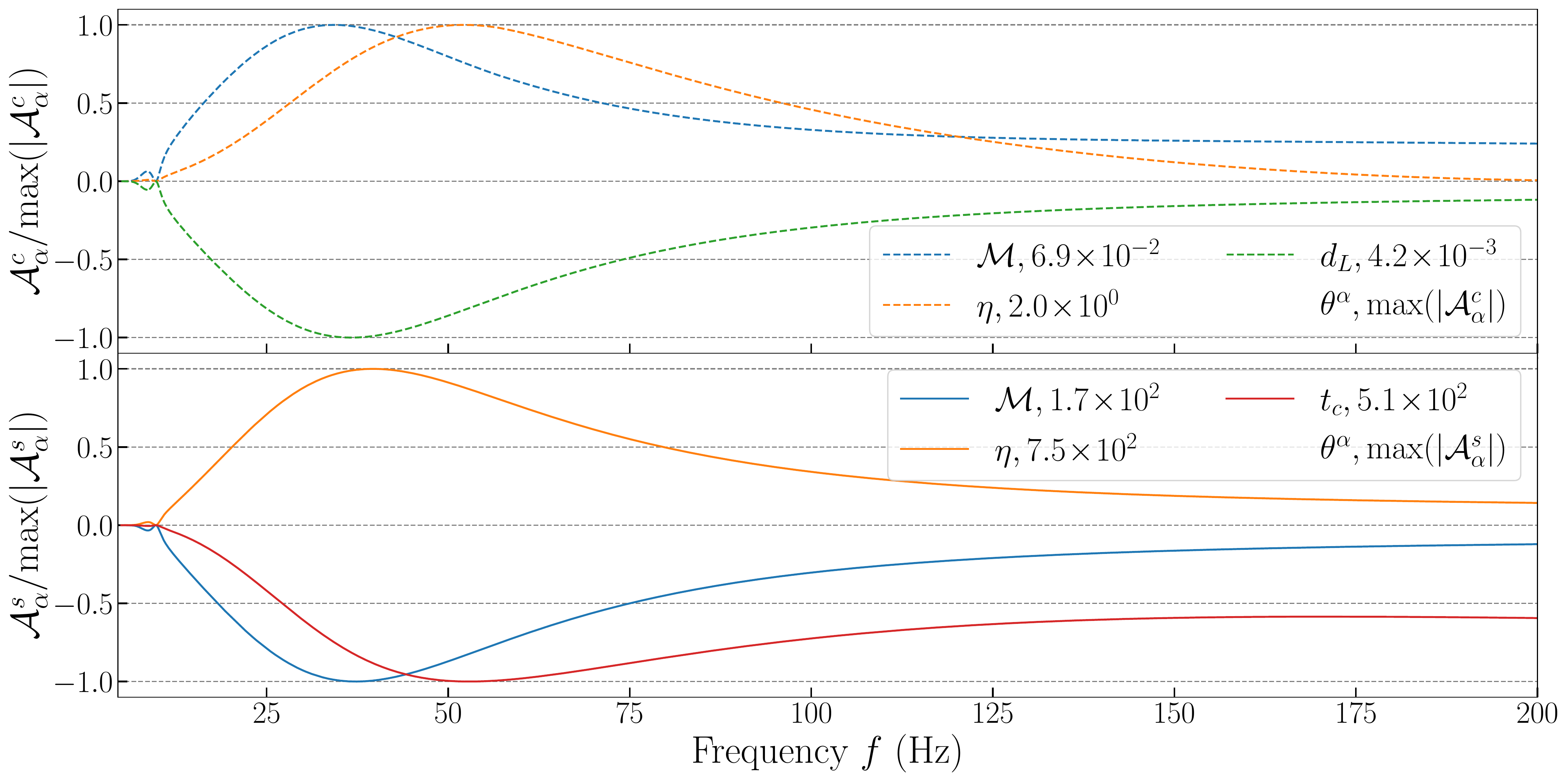}} \vspace{-0.2cm}
\\ 
	\centering
\subfigure
{\includegraphics[width=11cm]{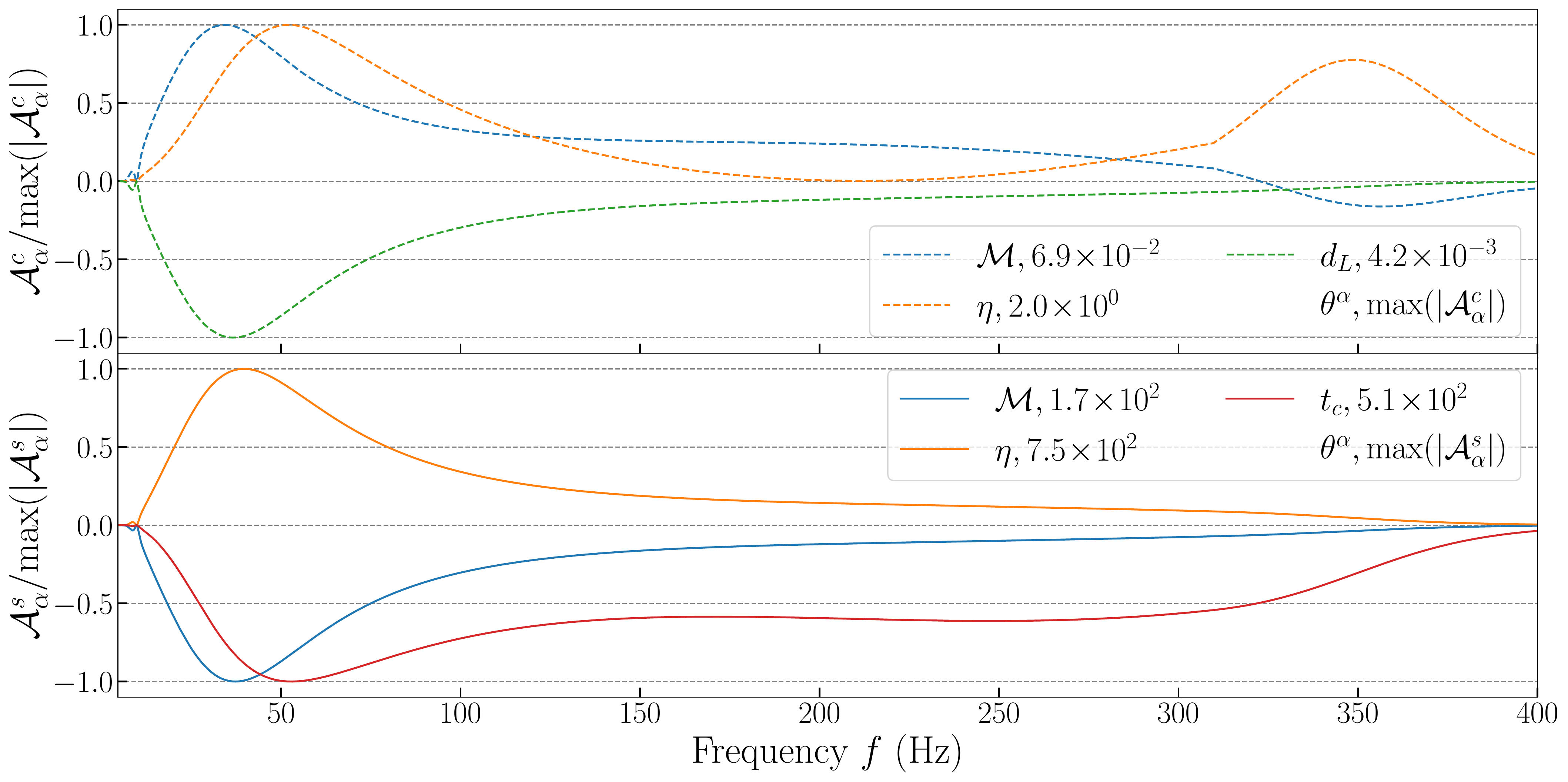}}  \vspace{-0.2cm}
\caption{The frequency-domain phase, $\phi(f)$, and the normalized modulation
amplitudes, ${\cal A}^c_\alpha$ and ${\cal A}^s_\alpha$, of the waveform. The
number after each parameter is the maximum value of the corresponding normalized
modulation amplitude.} \label{figA1}
\end{figure*}

Figure~\ref{figA1} shows the frequency-domain phase of the {\sc IMRPhenomD}
waveform template, as well as the derivatives of the waveform (expressed by
${\cal A}^c_\alpha$ and $ {\cal A}^s_\alpha$). Note that ${\cal A}^s_3 = 0$ and
${\cal A}^c_4=0$ are not drawn.  The injected parameters are $\big\{m_1 =
30\,{\rm M_\odot},m_2 = 20\,{\rm M_\odot}, d_L = 500\, {\rm Mpc},t_c = 0\big\}$.
The frequency-domain phase grows rapidly in the low frequency band ($\lesssim
25\,{\rm Hz}$), while in the high frequency band ($\gtrsim 100 \, {\rm Hz}$) it
grows slowly and approximately linearly.  The modulation amplitude typically
reaches the maximum value within $30$--$50\,{\rm Hz}$, consistent with the most
sensitive frequency band of AdvLIGO. Also, for ${\cal M}$ and $\eta$, ${\cal
A}^s_\alpha$ is at least two orders of magnitude larger than ${\cal
A}^c_\alpha$.

\section{Max-max and max-mean of reduced biases}\label{appB:}

\begin{figure*}[h]
	\centering
	\hspace{-1cm}
\subfigure{\includegraphics[width=8.0cm]{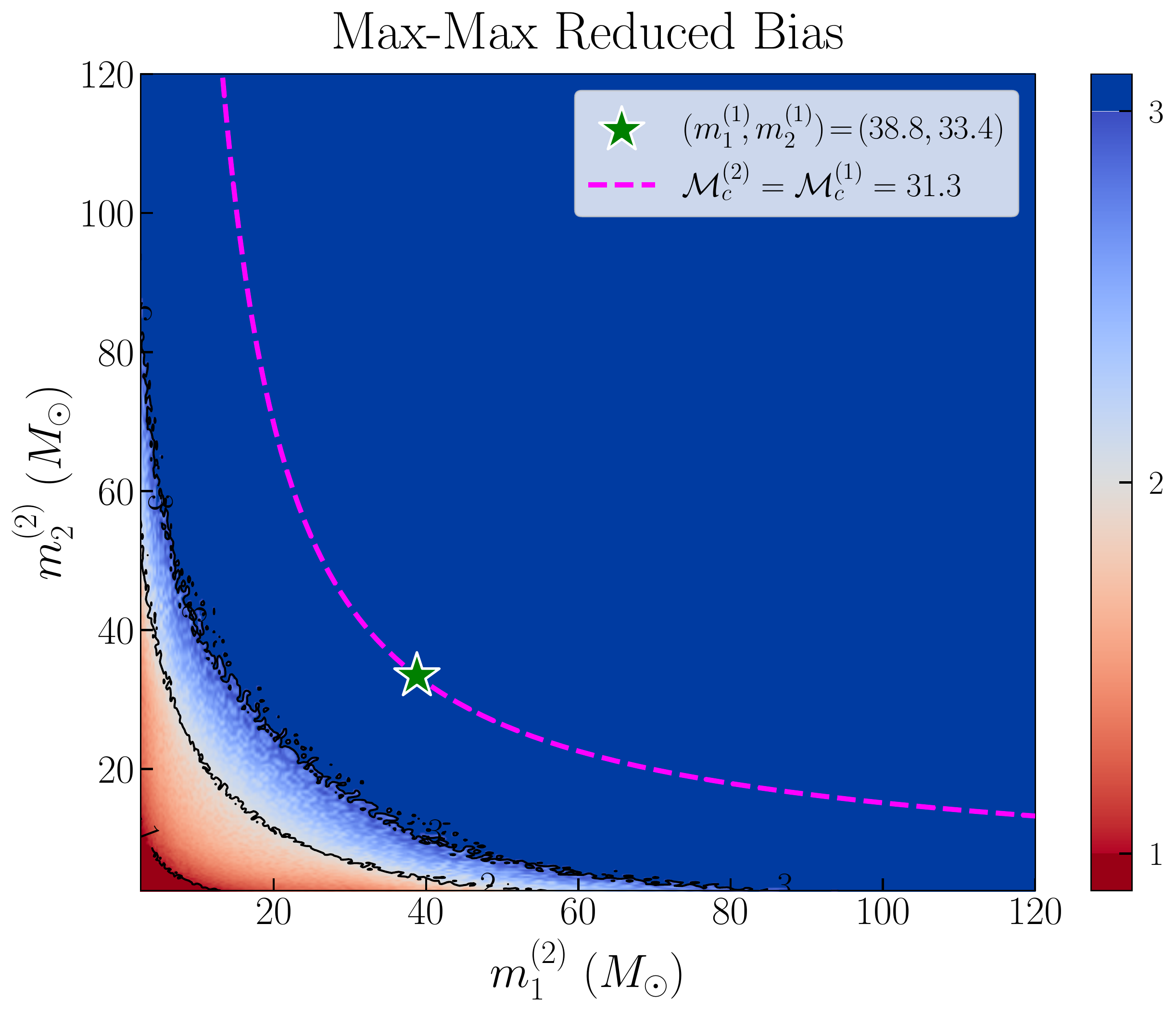}} 
	\centering
\subfigure{\includegraphics[width=8.0cm]{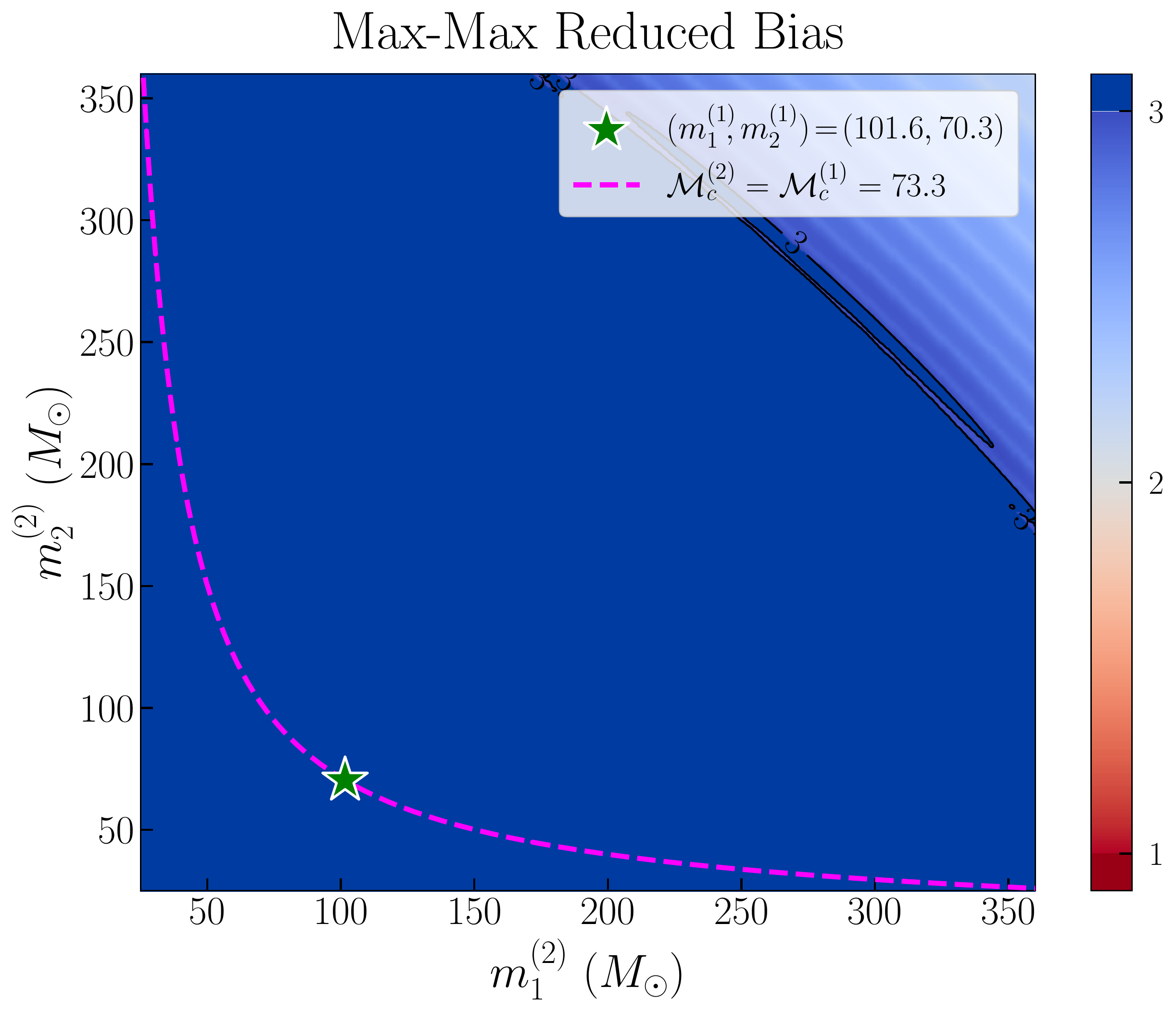}} 
\vspace{-0.1cm}
\\
\centering\hspace{-1cm}
\subfigure{\includegraphics[width=8.0cm]{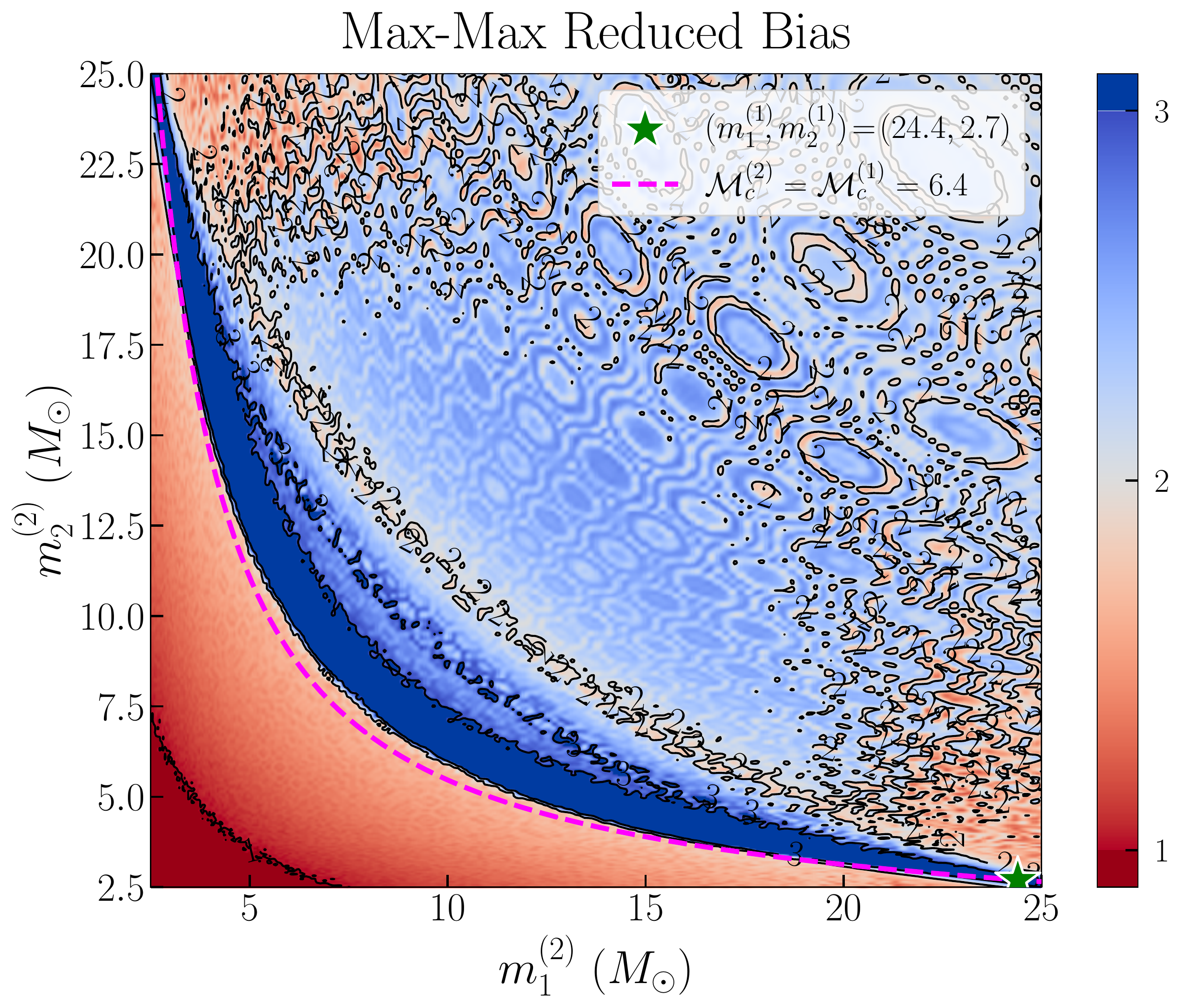}} 
	\centering
\subfigure{\includegraphics[width=8.0cm]{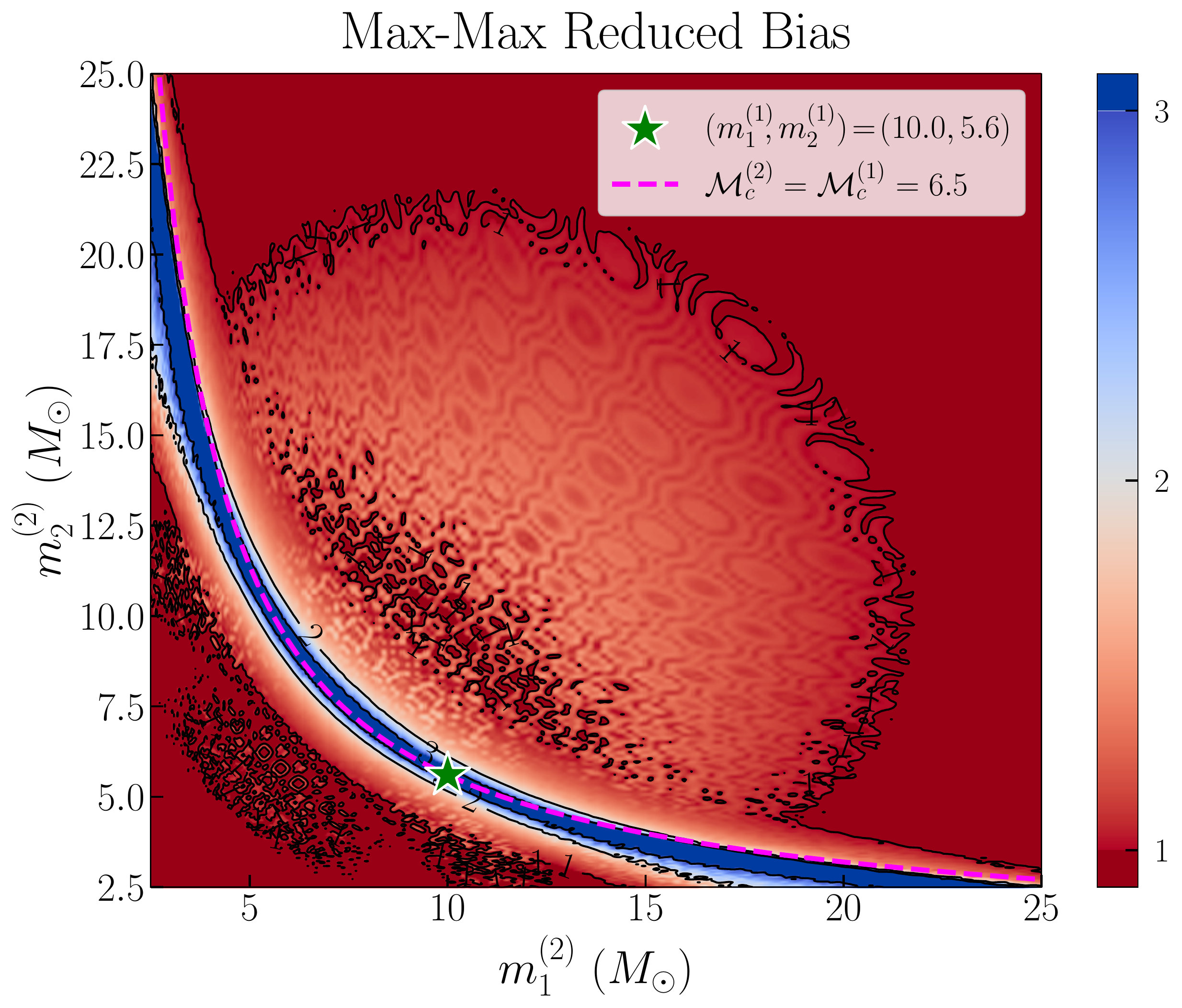}} 
\vspace{-0.1cm}
\\
\centering\hspace{-0.7cm}
\subfigure{\includegraphics[width=8.0cm]{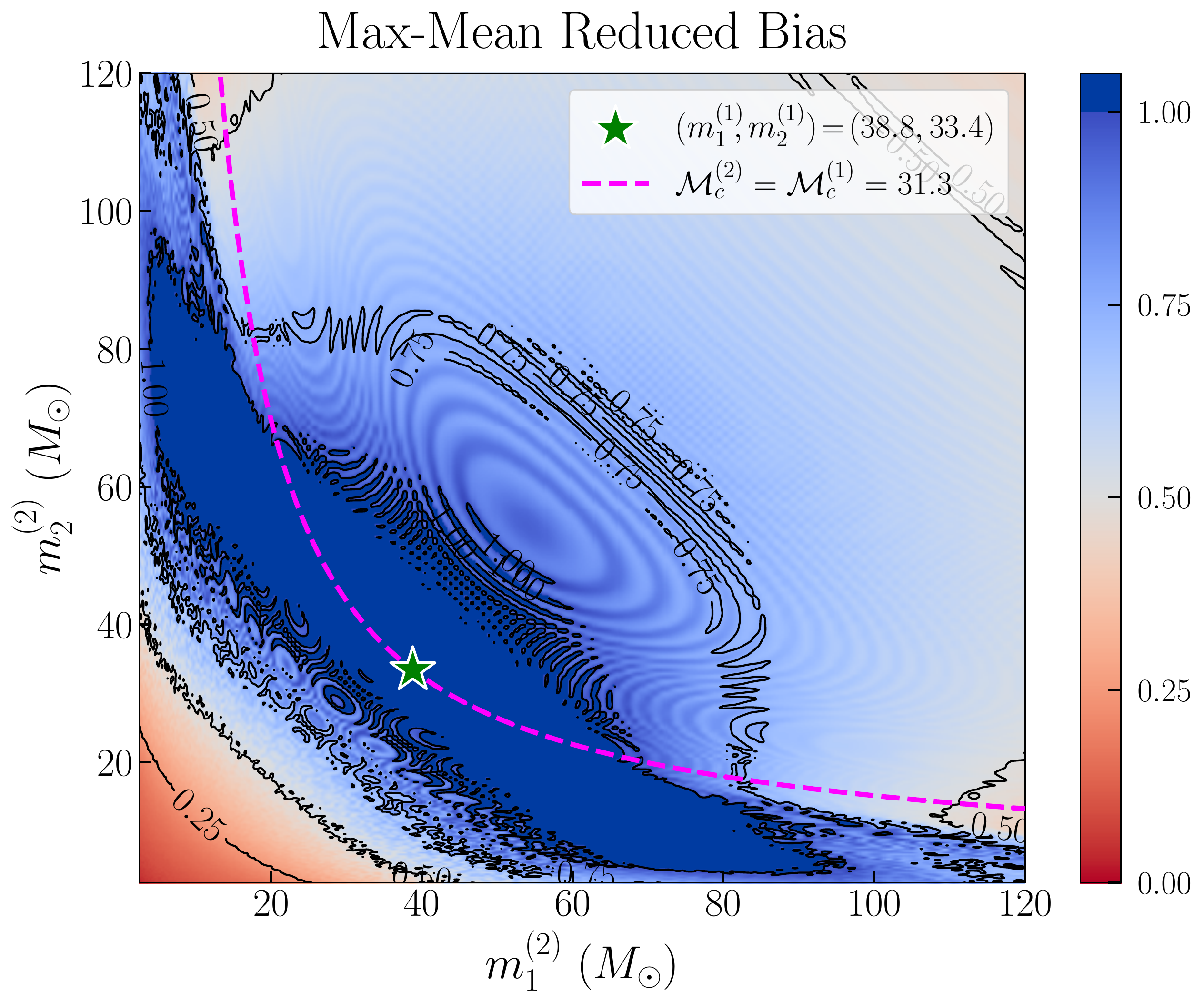}} 
	\centering
\subfigure{\includegraphics[width=8.0cm]{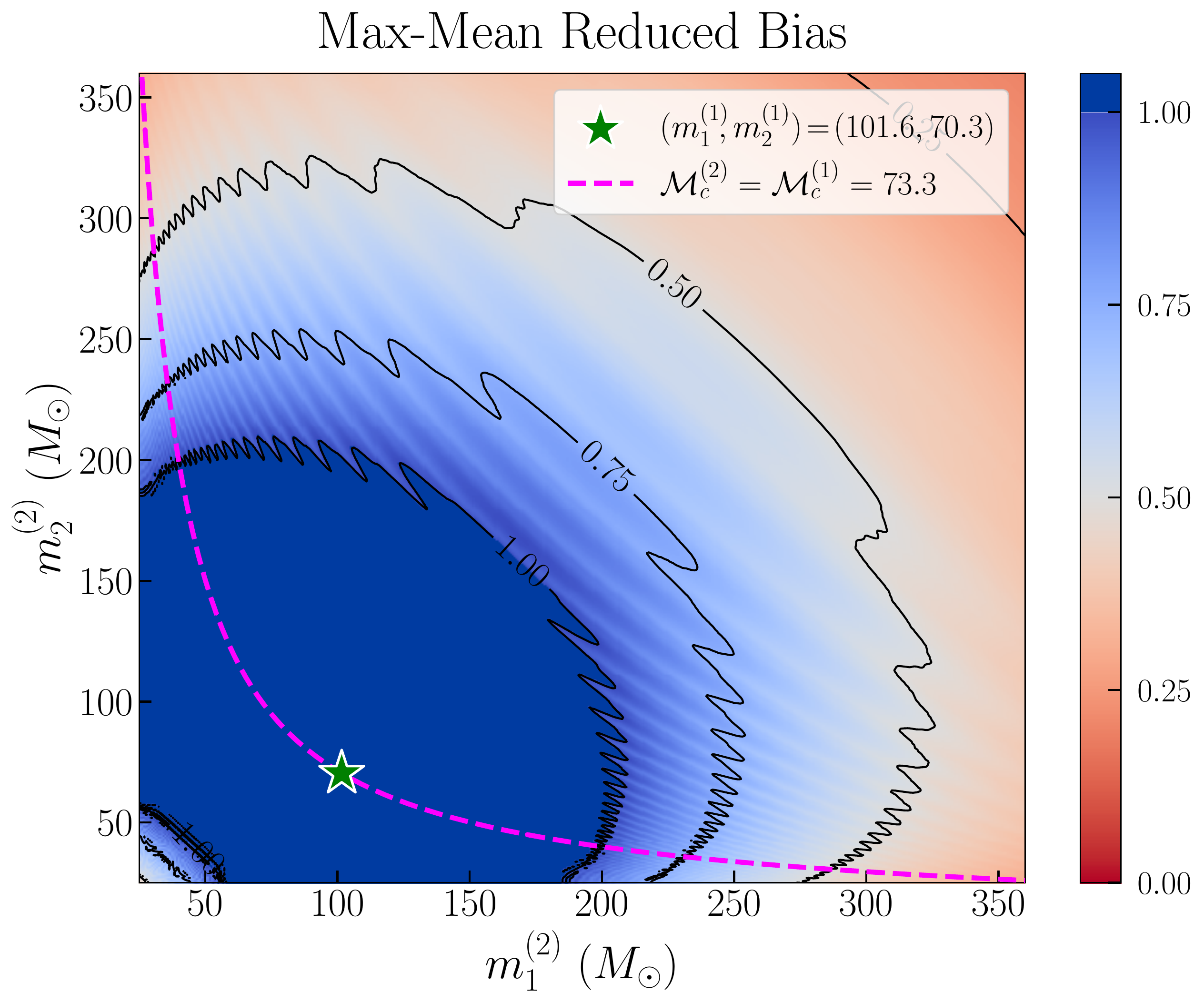}}
\vspace{-0.1cm}
\caption{Same as Fig.~\ref{fig1}, but the parameters are taken from real GW
events.} \label{figB1}
\end{figure*}

\begin{figure*}[h]
\ContinuedFloat
\centering\hspace{-1cm}
\subfigure{\includegraphics[width=8.0cm]{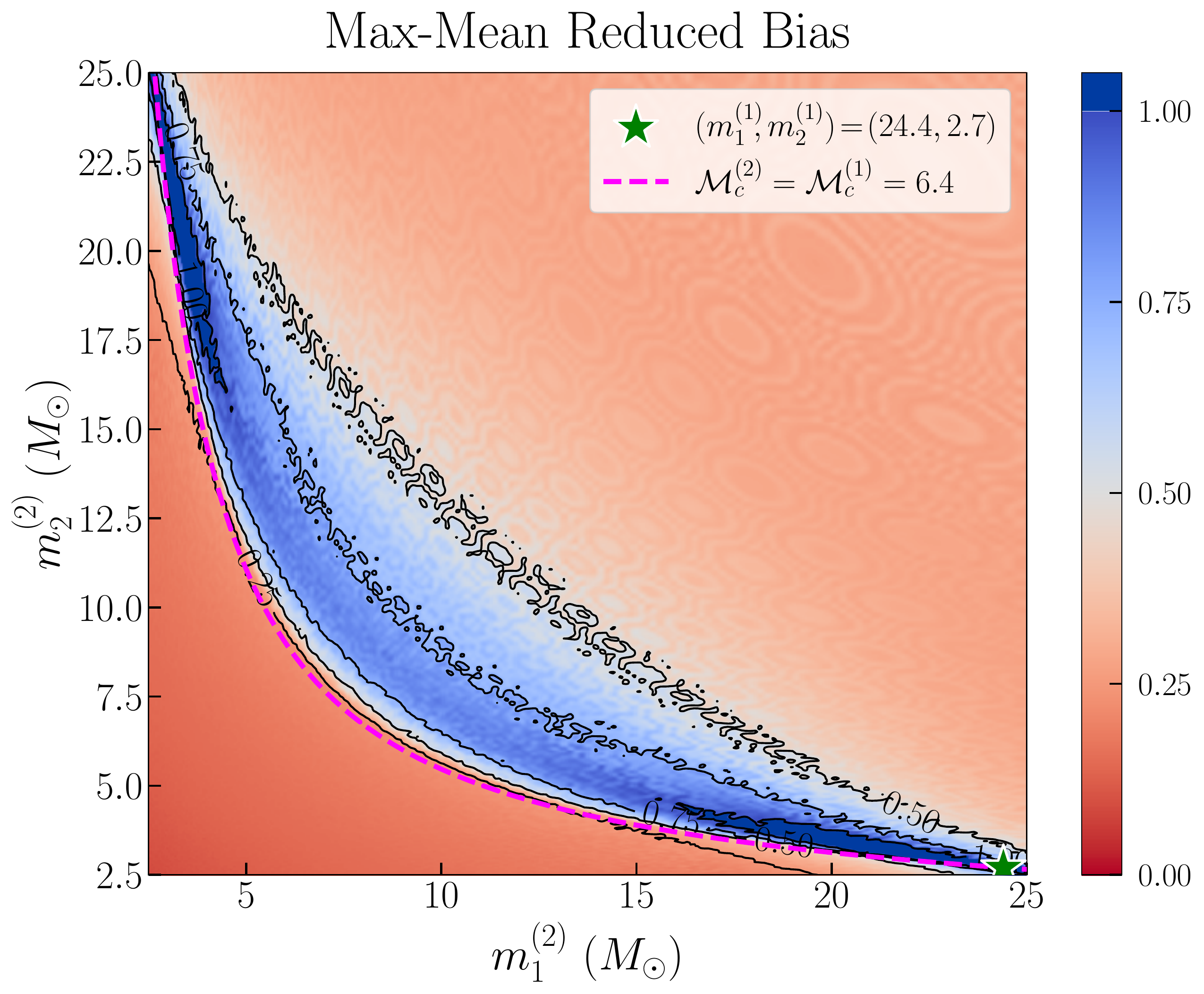}} 
	\centering
\subfigure{\includegraphics[width=8.0cm]{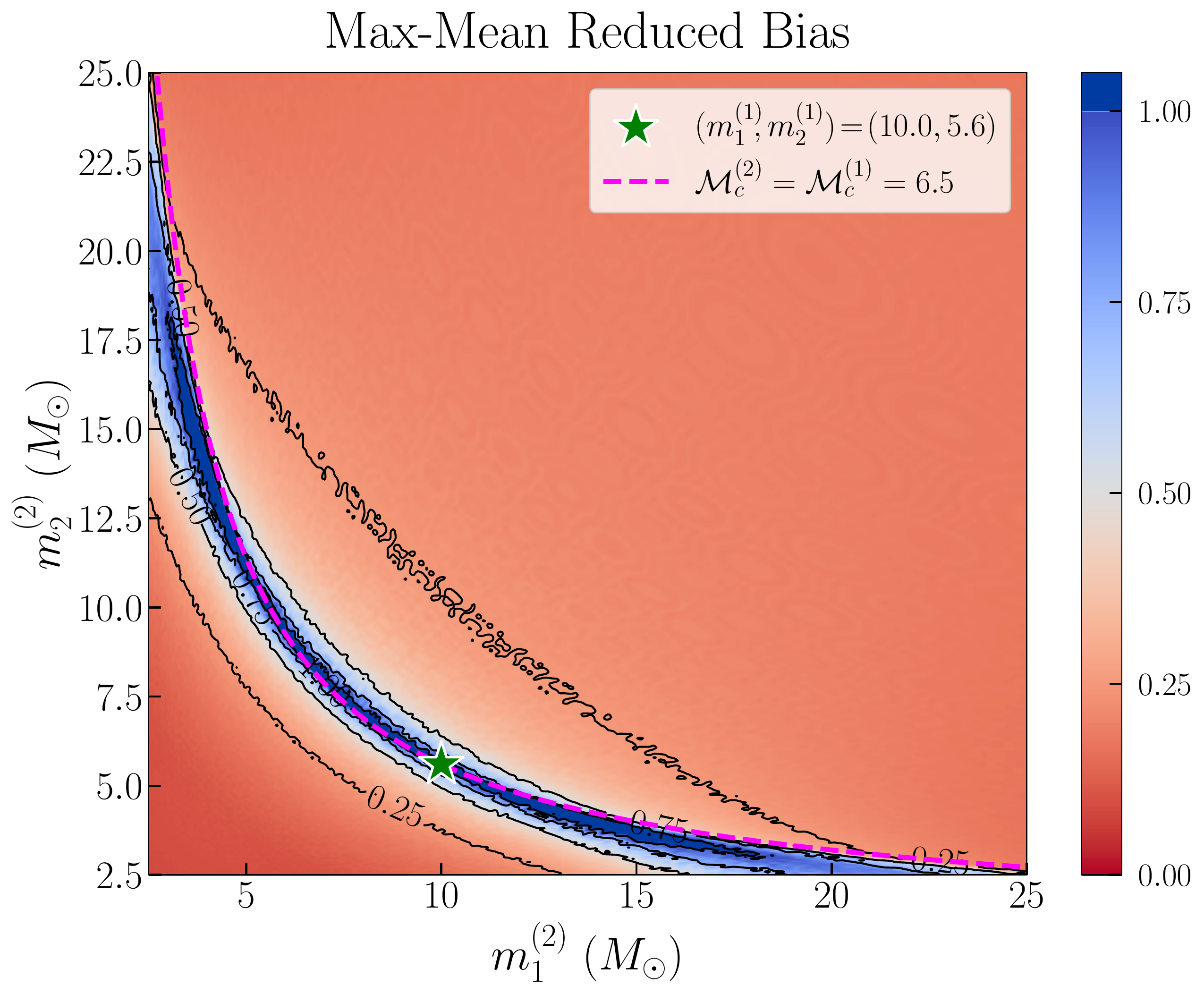}} 
\vspace{-0.1cm}\\
\centering\hspace{-1cm}
\subfigure{\includegraphics[width=8.0cm]{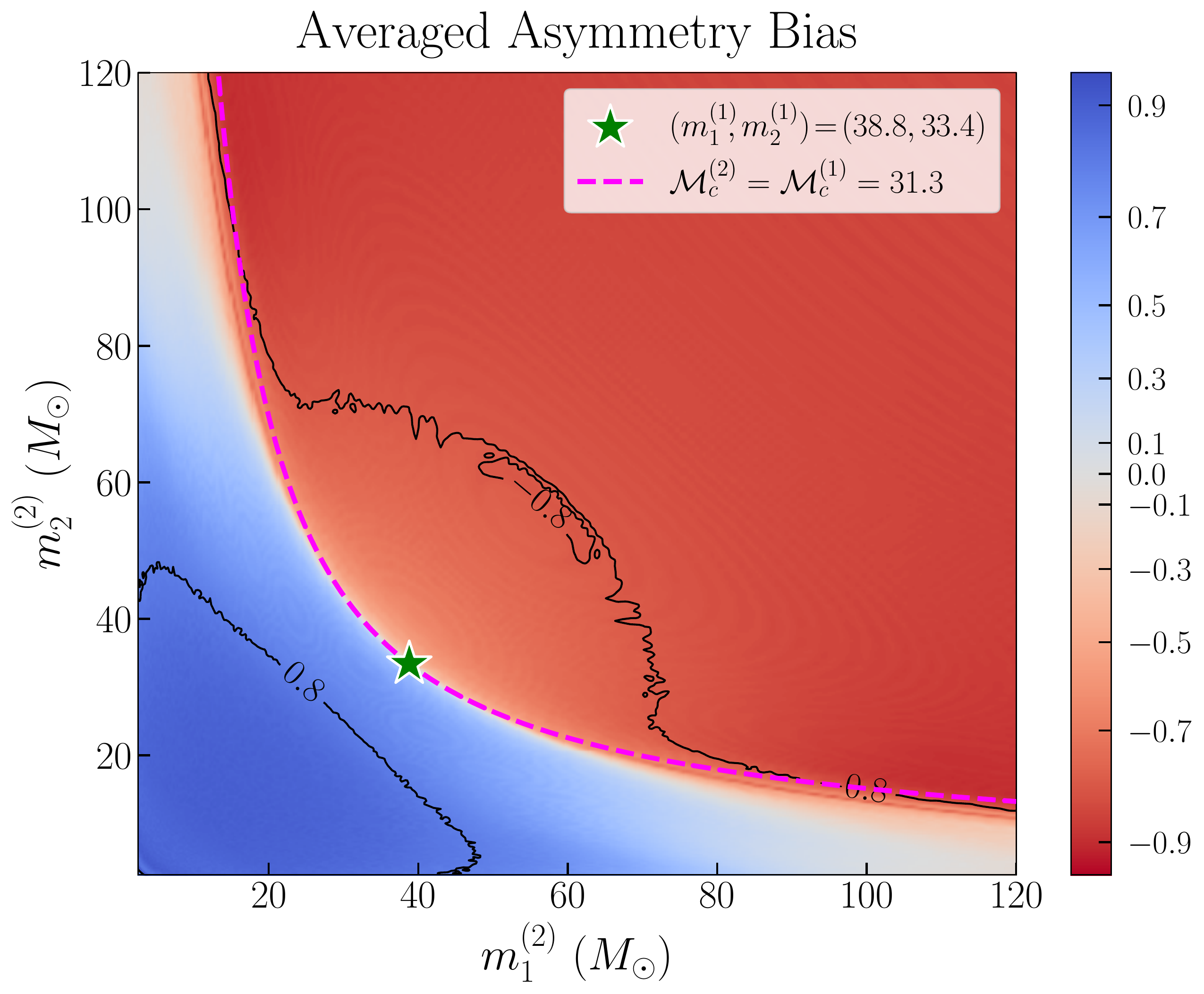}} 
	\centering
\subfigure{\includegraphics[width=8.0cm]{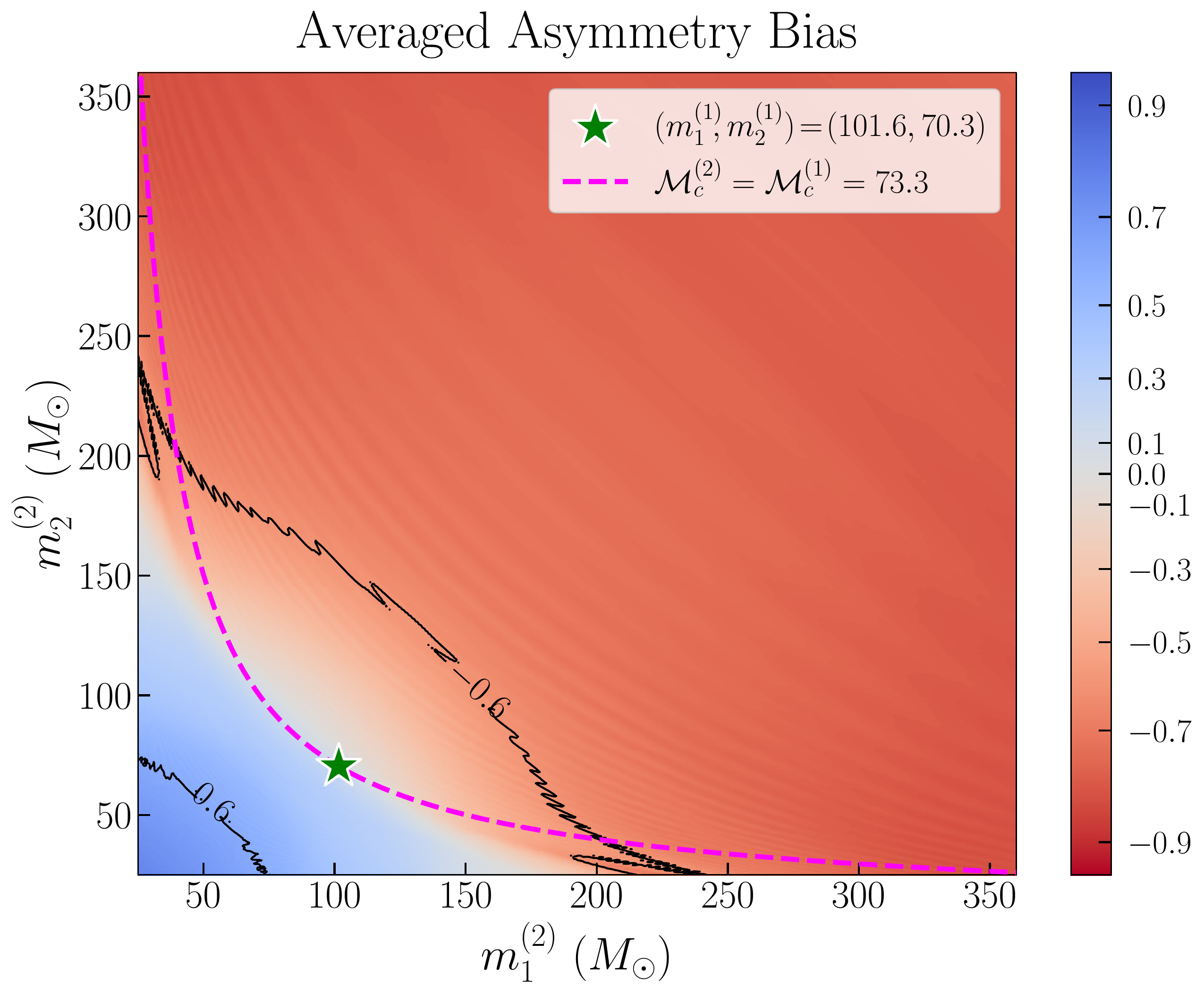}} 
\vspace{-0.1cm}\\
\centering\hspace{-1cm}
\subfigure{\includegraphics[width=8.0cm]{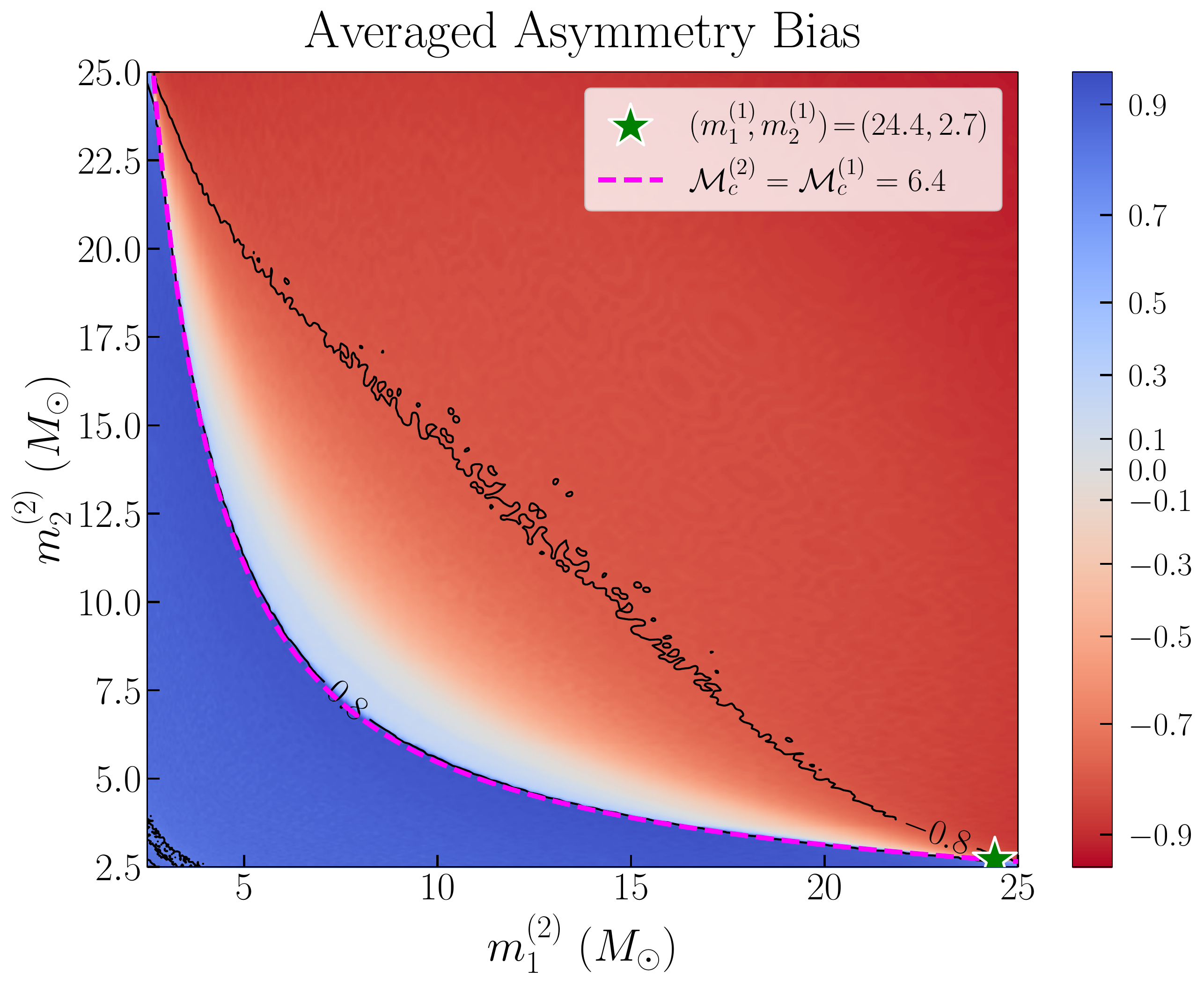}}  
	\centering
\subfigure{\includegraphics[width=8.0cm]{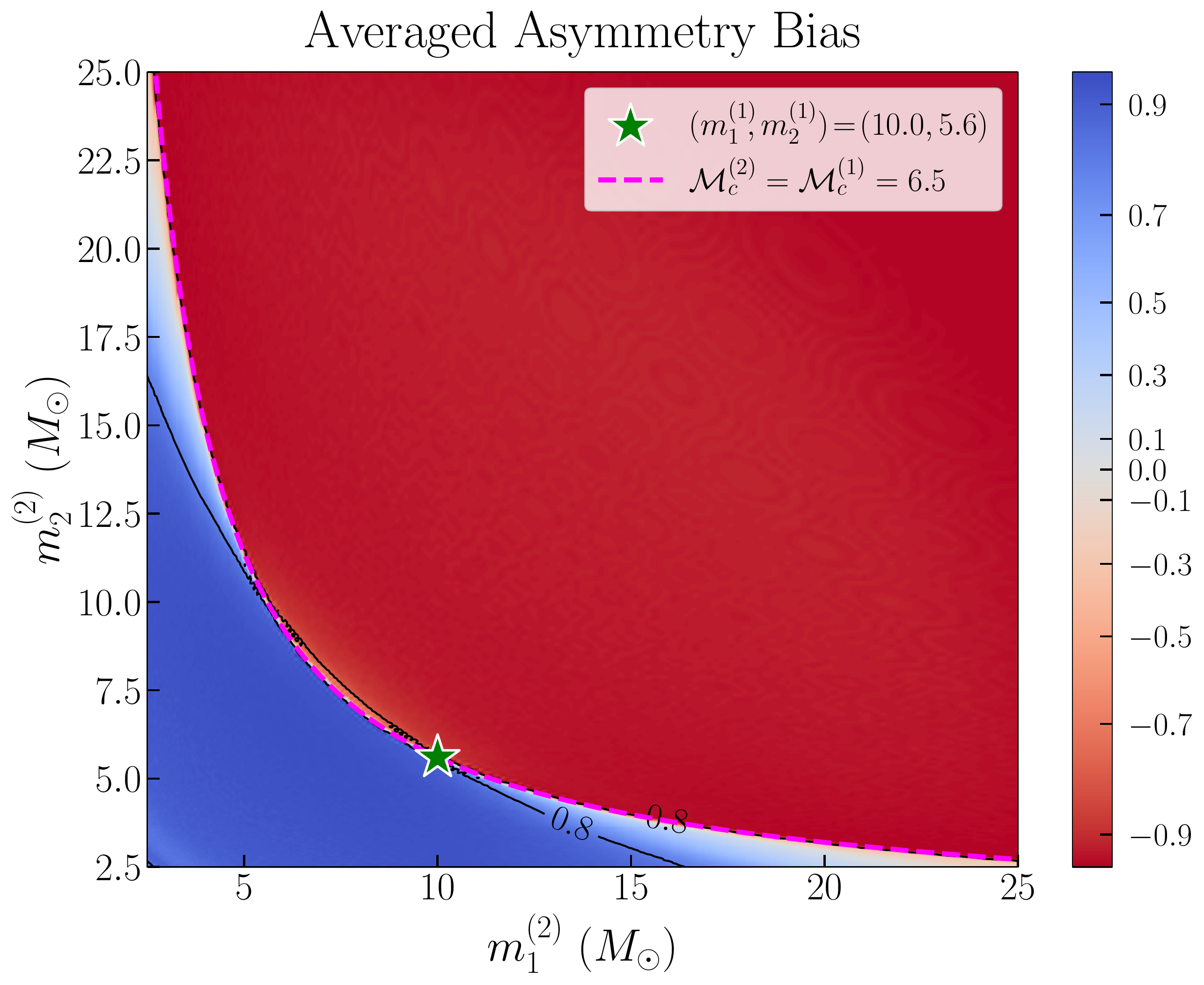}}  
\vspace{-0.1cm}\\
\caption*{FIG. B1 (continued).}
\end{figure*}

Figure~\ref{figB1} shows the dependence of the max-max value, max-mean average
value, and average asymmetry of $\bm B$ on $\big(m_1\ud,m_2\ud\big)$. The mass
components $\big(m_1\uf,m_2\uf\big)$ are taken as the masses in the detector
reference system of GW events GW150914, GW190602\underline{~}175927, GW190814,
and GW190924\underline{~}021846. Their behaviors are similar to those in
Fig.~\ref{fig1}. When ${\cal M}\ud\approx {\cal M}\uf$, large biases occur, and
the average asymmetry is close to $0$.

\section{SPE biases behaviours in other internal parameter configurations}
\label{appC:}

\begin{figure}[htp]
    \vspace{-0.4cm}
        \centering
    \subfigure{\includegraphics[width=11cm]{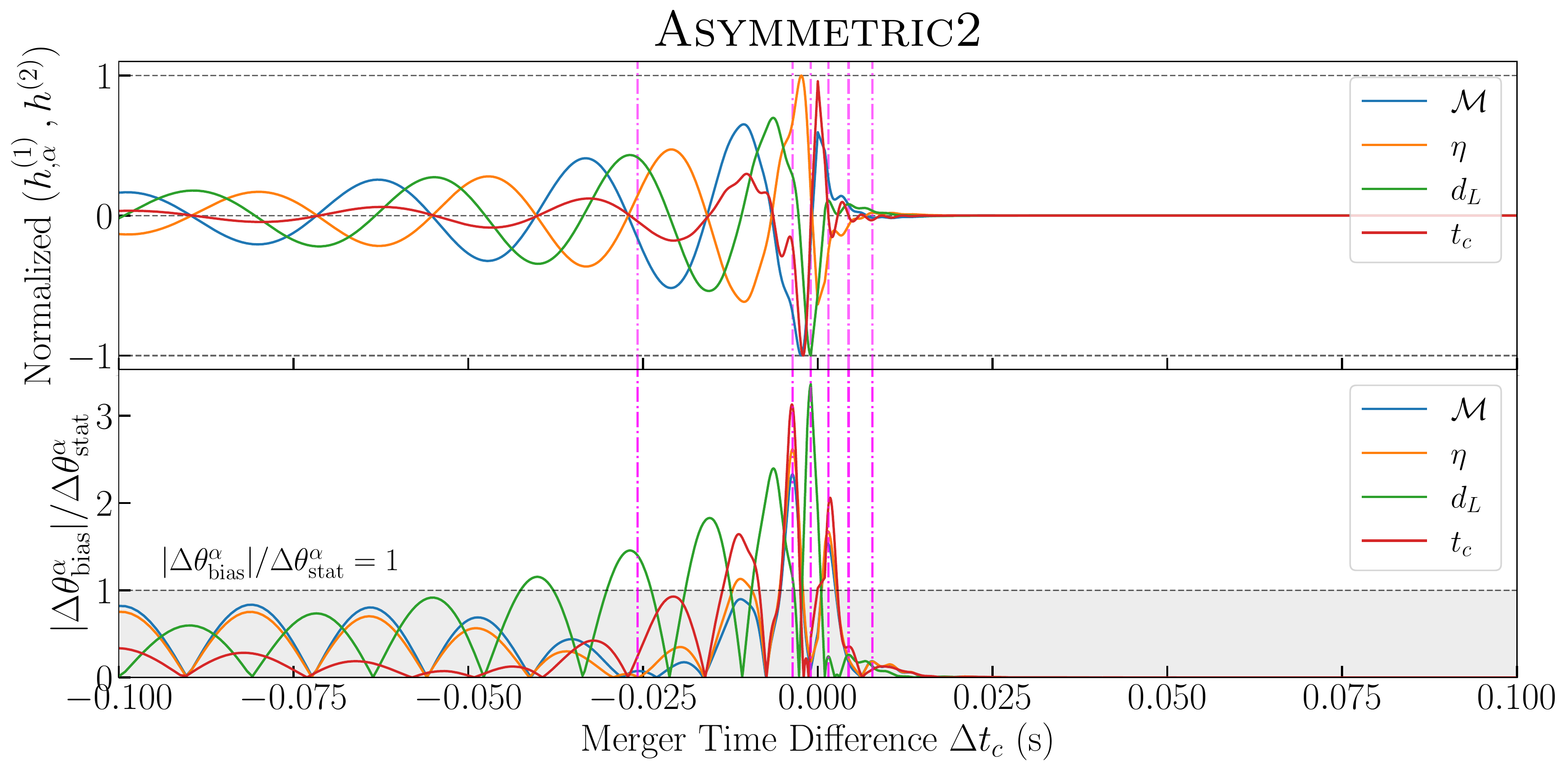}}
    \\
    \vspace{-0.4cm}
        \centering
    \subfigure{\includegraphics[width=11cm]{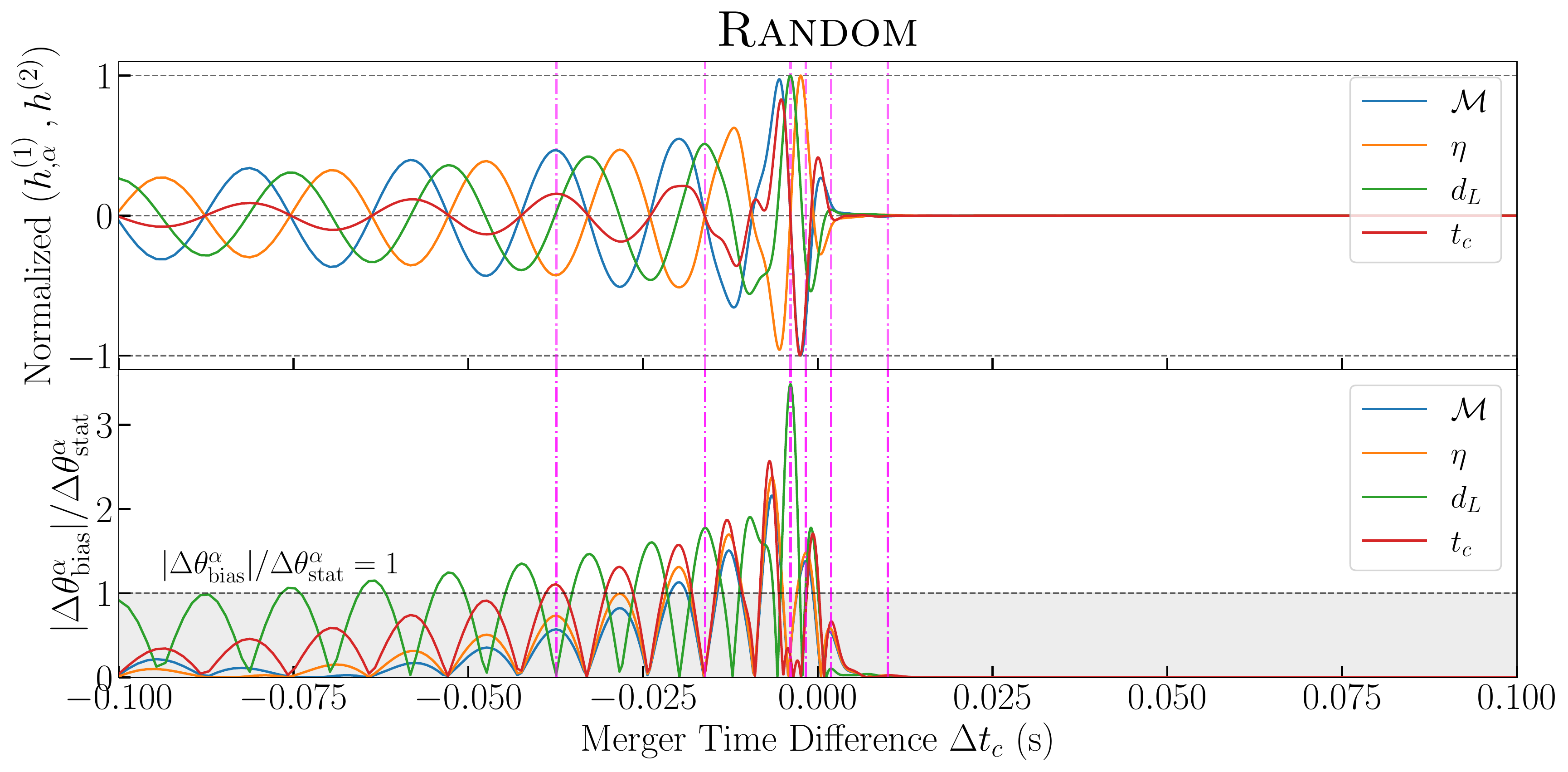}}
    \\
    \vspace{-0.4cm}
        \centering
    \subfigure{\includegraphics[width=11cm]{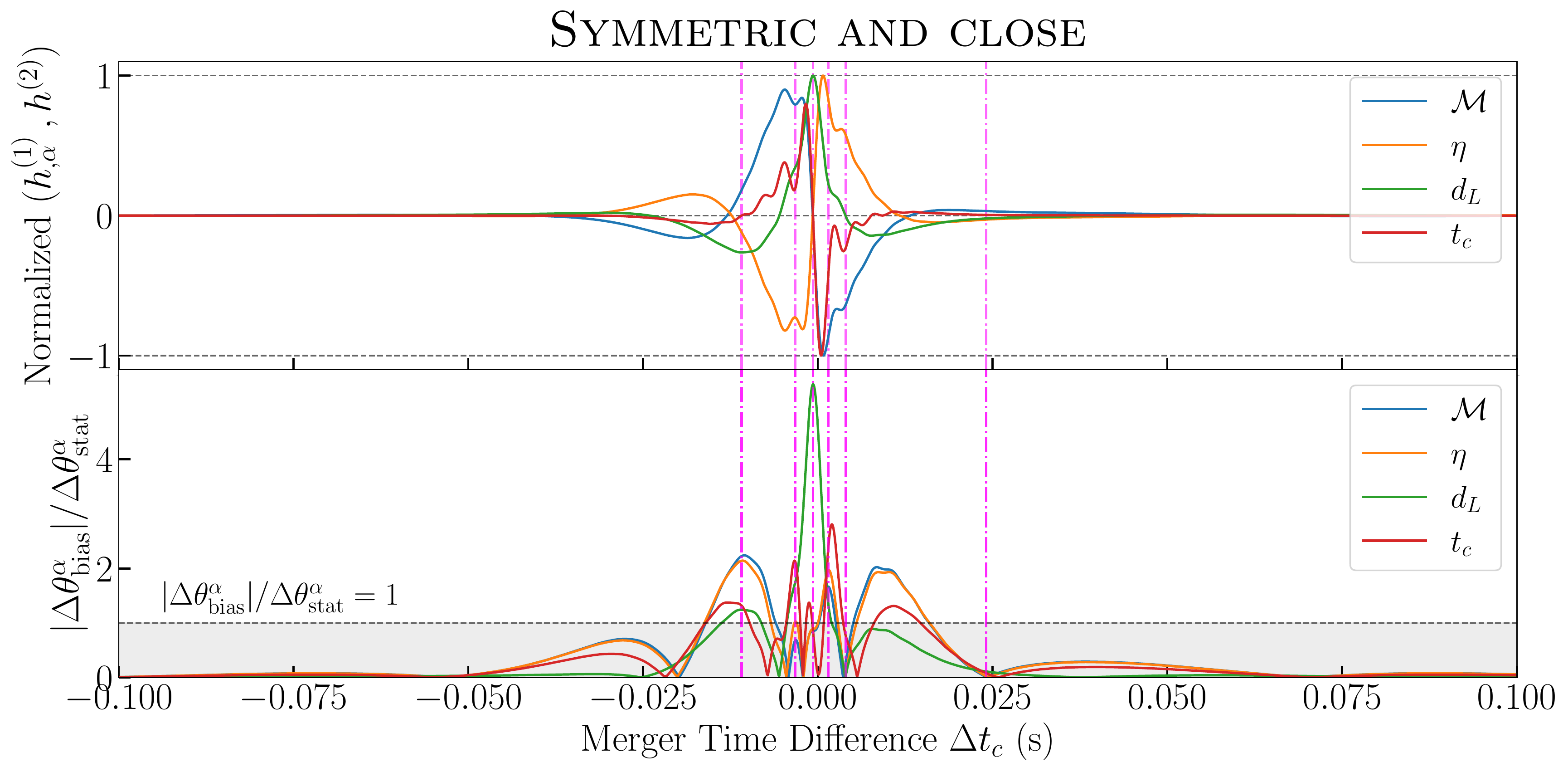}} 
    \\ 
    \vspace{-0.4cm}
        \centering
    \subfigure{\includegraphics[width=11cm]{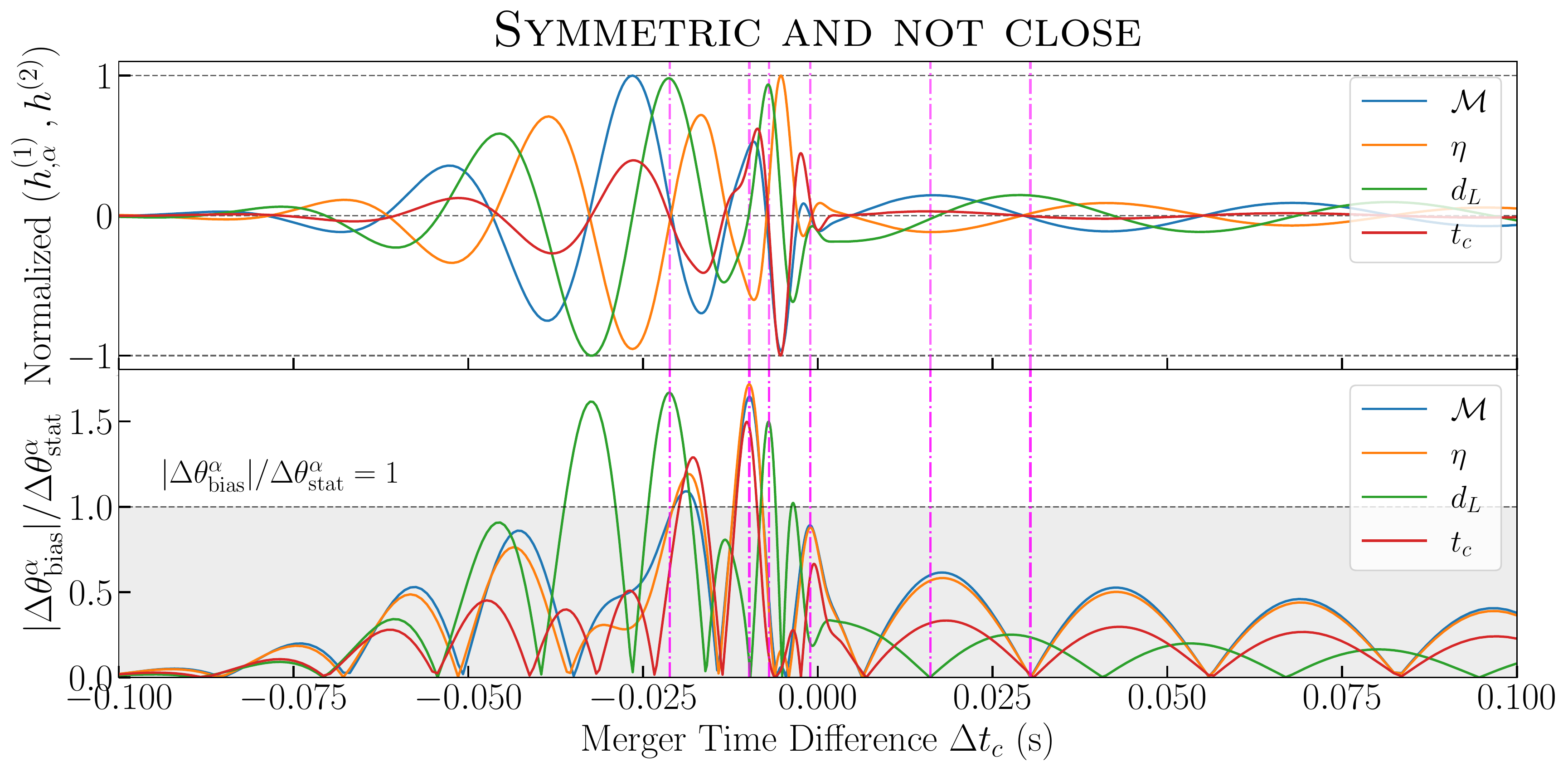}}
    \vspace{-0.5cm}
    \caption{Same as Fig.~\ref{fig2}, but for the {\sc Asymmetric2}, {\sc Random}, {\sc Symmetric and close}, and
    {\sc Symmetric and not close}
    configurations.}\label{figC1}
    \end{figure}

In the main text, we demonstrate that the biases come from the overlapping of frequency evolution. 
After understanding the origin of biases, it is easy to
understand the global dependence of biases on intrinsic parameters in Sec.~\ref
{sec3.3} as follows.
\begin{enumerate}[(i)]
	\item For a given detector, its sensitive frequency band is fixed. Due to
	the selection effect of the detector, the characteristic frequencies of the
	detected signals are similar ($\sim 100\,\rm{Hz}$ for AdvLIGO). As long as
	the merger time difference is appropriate, it is easy to make the
	corresponding frequency evolutions overlap in the time domain. This explains
	why biased OSs are inevitable.
	\item For two GW signals with similar mass components, the frequency
	evolutions are also similar. Therefore, they can easily overlap for a wide
	range of $\dt$, and the stable stage of $\Delta \phi $ is longer, which is
	more likely to produce large biases on average.
    \item The frequency evolution $f (t) $ is strongly dependent on the masses
    of the GW source. As long as the masses of two GW signals are not
    particularly close, their evolution behaviors will be significantly
    different. Obviously, when the signal with a lower frequency merges first,
    it is easier to overlap with the frequency evolution of the signal with a
    higher frequency, which produces a strong dependence of the biases on the
    merger order. On the contrary, if the masses of the two signals are close,
    it does not matter which one merges first, so $|\overline { \rm Asy} | $ is
    small. At the same time, according to (ii), the biases in these cases are
    large in the average sense. This explains the consistency between the
    regions where $|\overline {\rm Asy} |  \sim 0 $ and those where the bias is
    large.
\end{enumerate}

Without loss of generality, we selected three representative configurations in
the $\big(m_1\ud,m_2\ud\big)$ parameter space, namely {\sc Asymmetric2}, {\sc Random}, {\sc
Symmetric and close}, {\sc Symmetric and not close} to further validate these conclusions.

For
the {\sc Asymmetric2} configuration, it is equivalent to exchanging the
positions of $h\uf $and $h\ud $, and $\Delta  \phi$ is monotonically decreasing
with respect to $f$. Therefore, biased OSs will mainly occur when $\dt<0 $, which is
consistent with the results in Fig.~\ref{figC1}.
For the {\sc Random} configuration, the source masses of $h\ud$ are
significantly larger than that of $h\uf$, so we expect that large biases mainly
occur when $\dt<0$, and the dependence of the bias on $\dt$ should be similar to
the {\sc Asymmetric2} configuration.  The results in Fig.~\ref{fig2} verified
this. Further, for most $\big(m_1\ud,m_2\ud\big)$ in the parameter space, their
$|\overline{\rm Asy}|$ is close to $1$, showing a strong asymmetry of their bias
behaviors. Therefore, {\sc Asymmetric} and {\sc Asymmetric2} configurations can
represent the dependence of biases on $\dt$ for most OSs whose masses of two
signals are not similar. 

\begin{figure*}[t]
	\centering
\subfigure{\includegraphics[width=12cm]{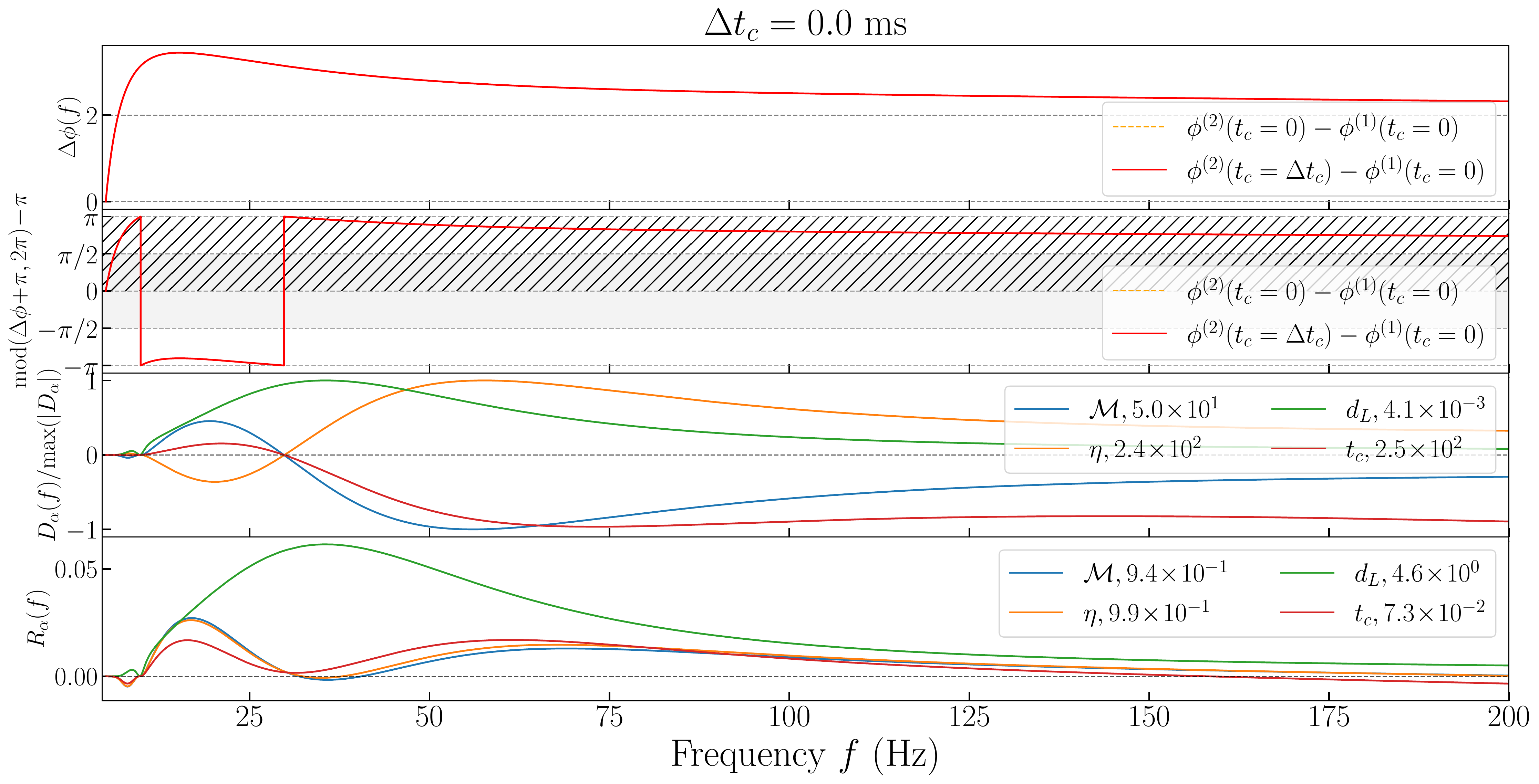}}
\\
\vspace{-0.4cm}
	\centering
\subfigure{\includegraphics[width=12cm]{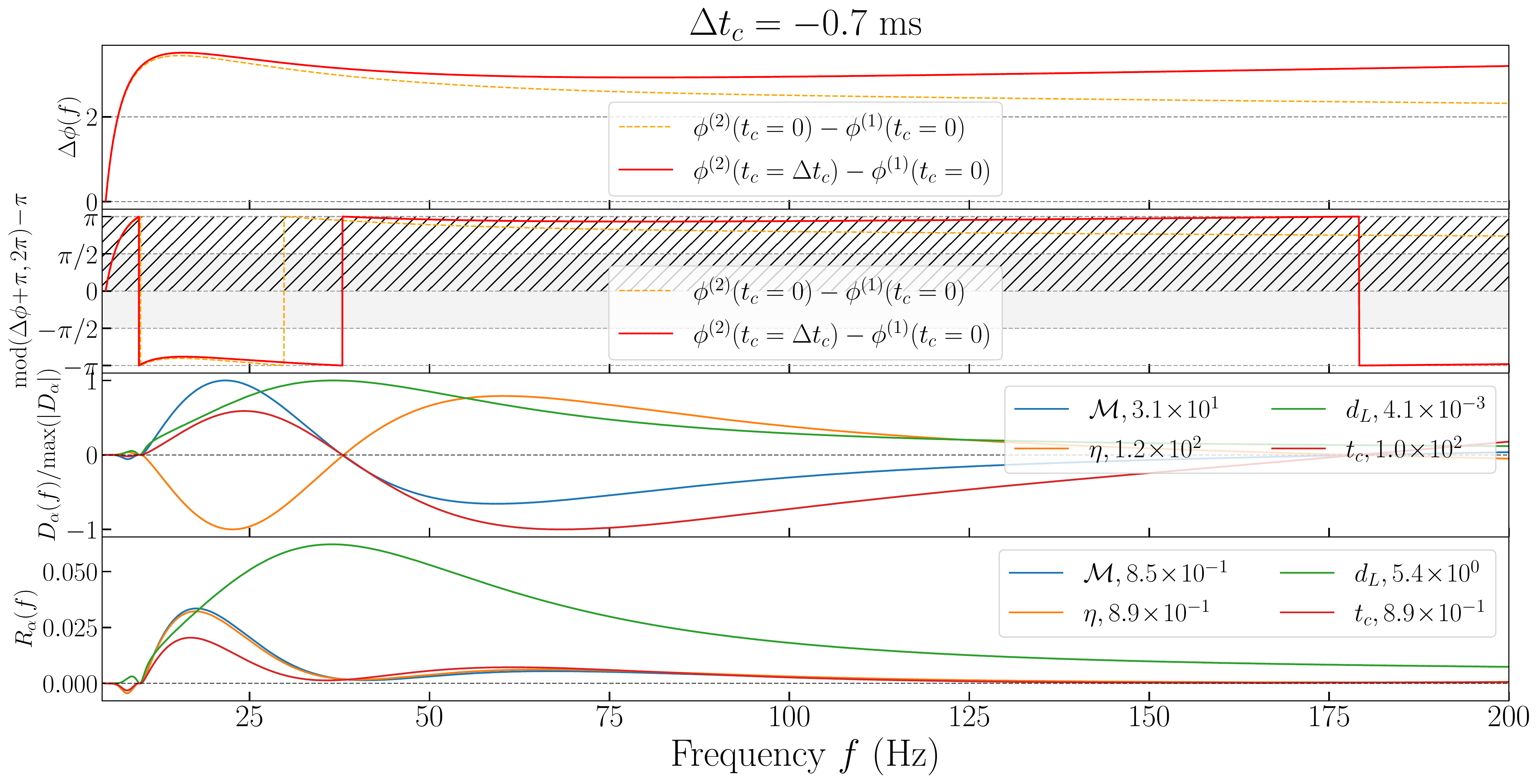}} 
\vspace{-0.3cm}
\caption{Same as Fig.~\ref{fig3}, but for the {\sc Symmetric and close}
configuration.} \label{fig6}
\end{figure*}

Next, we discuss the parameter configurations where $|\overline{\rm Asy}| \sim
0$. For the {\sc Symmetric and close} configuration, we choose $m_1\ud$ to be
slightly larger than $m_1\uf$, and then adjust $m_2\ud$ such that
$|\overline{\rm Asy}|$ is minimal.  It can be seen that $\bm B$ is approximately
symmetrical about $\dt$. This is very similar to the {\sc Equal} configuration,
except that all $B_\alpha(t)$ curves are slightly moved to the left.  Same as
before, we take some $\dt$ to calculate the frequency-domain phase, which is
shown in Fig.~\ref{fig6}. When $\dt = 0$, we find that the overall change of
$\Delta \phi_0$ under this configuration is not large, and it increases with $f$
first and then decreases.  The change of monotonicity leads to a long stable
stage, i.e., a large bias. This is significantly different from the {\sc
Asymmetric} configuration, where $\Delta \phi_0$ increases monotonically with
$f$, which requires additional non-zero $\dt$ to generate a significantly stable
stage. 

In the {\sc Symmetric and close} configuration, $\Delta \phi_0$ does not change
much with $f$, which is why $\bm B$ behaves similarly to the {\sc Equal}
configuration.  When $\dt$ significantly deviates from $0$, $|2\pi \dt f|$ will
be much larger than $|\Delta \phi_0|$ for most $f$, so $\Delta \phi = \Delta
\phi_0-2\pi \dt f \approx -2\pi \dt f$, which is exactly the phase difference in
the {\sc Equal} configuration.  Of course, on the basis of the relatively stable
$\Delta \phi_0$, we can slightly adjust $\dt$ to make $\Delta \phi$ more stable.
When $ \dt=-0.7  \mms $, $ \Delta  \phi $ is very close to $ \pm  \pi $ in a wide
frequency band, making the bias larger.  In other words, when $ \dt=-0.7  \mms $,
the behavior of $\Delta \phi$ is more like a constant, so the symmetry axis of
$\bm B$ will move to $ \dt=-0.7  \mms $. This explains the slight movement of the
bias curves in the {\sc Symmetric and close} configuration.

Physically, the process of adjusting $m_2\ud$ is equivalent to finding a $h\ud$
whose frequency evolution is most similar to $h\uf$.  As long as the frequency
evolution behavior of the two signals is similar, there will be a long
overlapping of the frequency evolutions no matter which binary merges first, and
the corresponding $|\overline{\rm Asy}|$ is small.  Therefore, if
$m_1\ud>m_1\uf$, to minimize $|\overline{\rm Asy}|$ one must have
$m_2\ud<m_2\uf$. Otherwise, $h\ud$ is generated by a heavier binary, leading to
similar results to that of the {\sc Asymmetric2} case.

\begin{figure*}[t]
	\centering
\subfigure{\includegraphics[width=12cm]{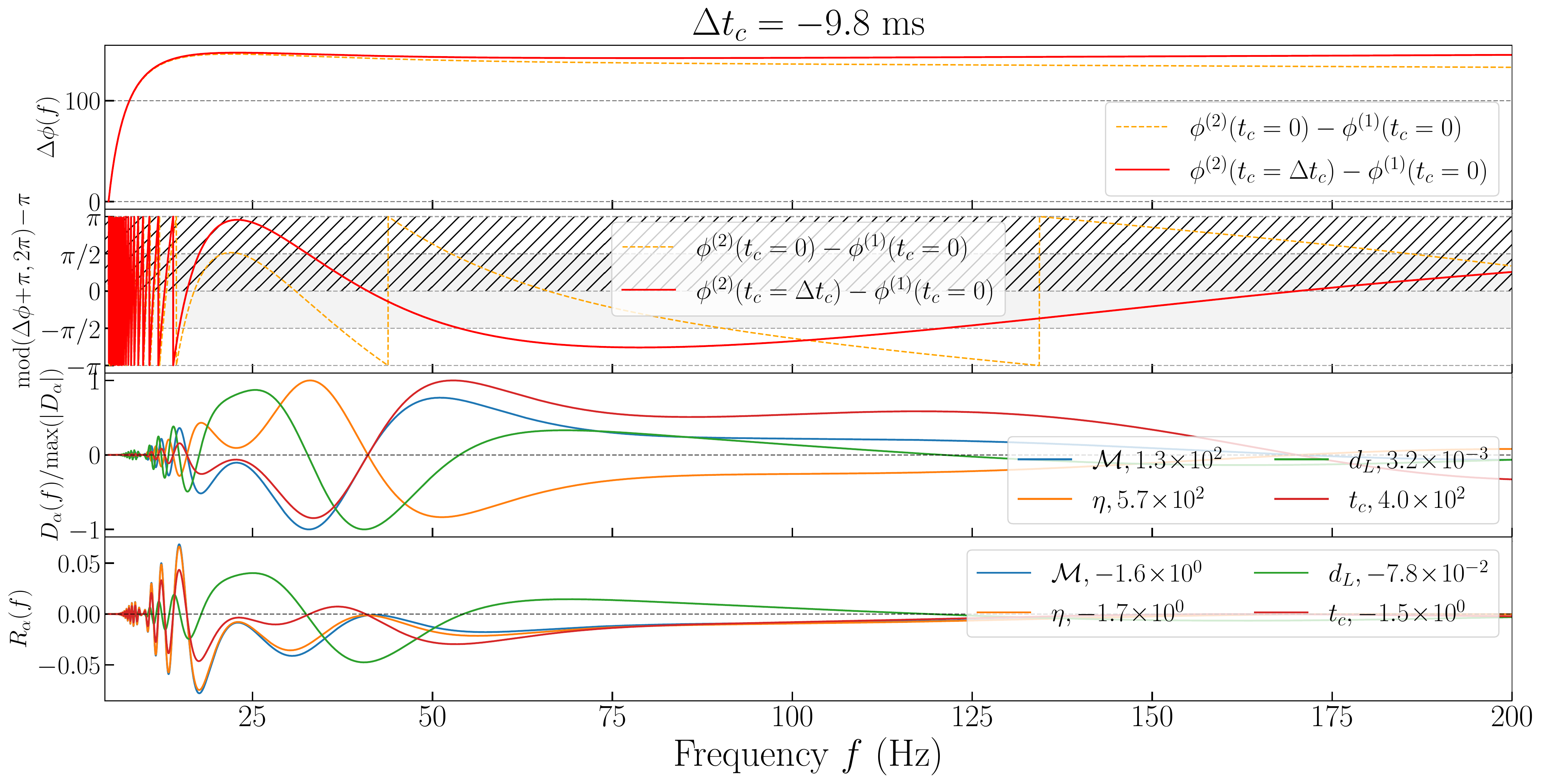}}
\\
\vspace{-0.4cm}
	\centering
\subfigure{\includegraphics[width=12cm]{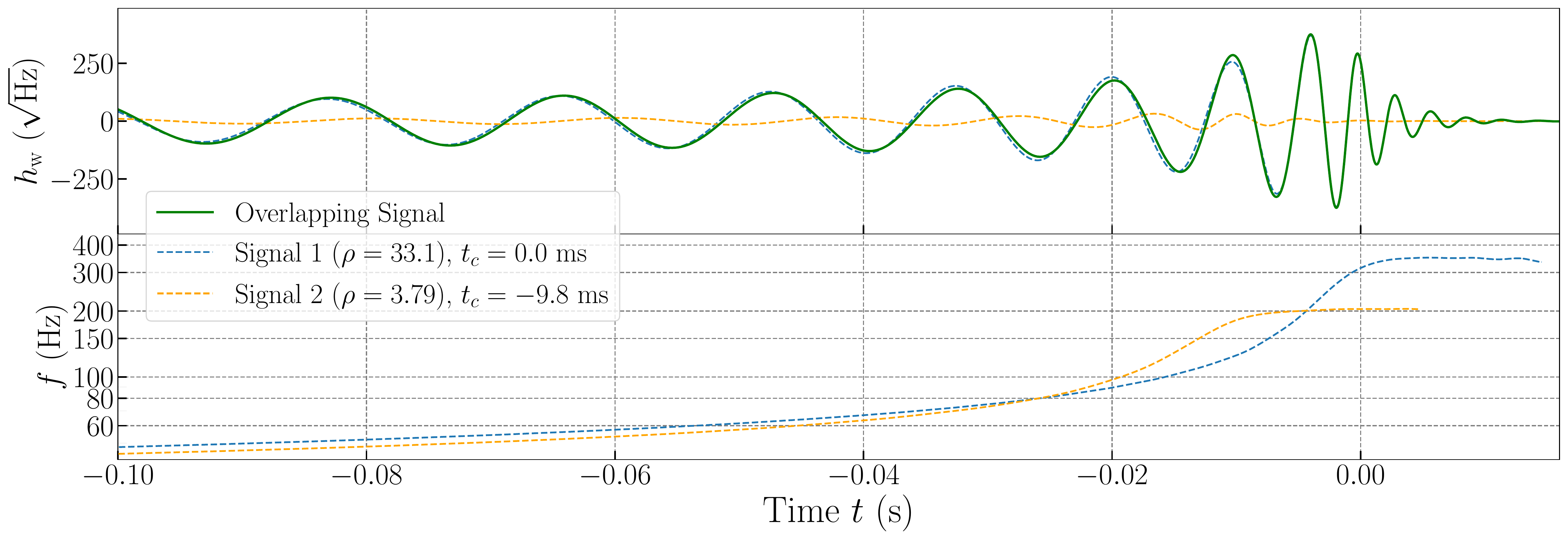}} 
\vspace{-0.3cm}
\caption{Same as Fig.~\ref{fig5}, but for the {\sc Symmetric and not close}
configuration and $\dt=-9.8\mms$.} \label{fig7}
\end{figure*}

Finally, we discuss the influence of $\eta$ on the bias or the bias behaviors of
binaries with extreme mass ratios. In general, the frequency evolutions are
mainly determined by the chirp mass $\cal M$.  This is why we take $\eta\ud =
\eta\uf$ in the {\sc Asymmetric} and {\sc Asymmetric2} configurations. In the
{\sc Symmetric and not close} configuration, we fix $m_1\ud = 2m_1\uf$ and
adjust $m_2\ud $ to minimize $|\overline {\rm Asy} | $.  Obviously, to obtain a
similar frequency evolution, $m_2\ud$ must be small. Interestingly, at this
time, although $M \ud>M  \uf $, there is ${\cal M}  \ud<{ \cal M}  \uf $.  This
phenomenon was mentioned in Sec.~\ref{sec3.3}, that is, when
$\big(m_1\uf,m_2\uf\big) = (30,20)$, the region where $|\overline {\rm Asy} | $
is small locates below the line where ${\cal M}\ud={\cal M}\uf$.  In
Fig.~\ref{fig7}, we draw $ \Delta \phi (f)$, $D_\alpha$, $R_\alpha$, whitened
waveforms and frequency evolutions of the two signals when $\dt = -9.8\mms$. 
Although ${\cal M}\ud<{\cal M}\uf$, $m_1\ud $ (also $M\ud $) is too large. In
general, the frequency of $h\ud$ is still lower than $h\uf$. In other words, the
frequency of a GW signal with an extreme mass ratio is lower, so smaller chirp
masses will be needed to generate a similar frequency evolution to $h\uf$. 
Therefore, in Fig.~\ref{fig2}, the region where $|\overline{\rm Asy}|\sim 0$
locates below the line where ${\cal M}\ud={\cal M}\uf$.  On the contrary, if
$\eta$ is small, such as GW190814, the region will be above (see
Fig.~\ref{figB1}).

Back to the {\sc Symmetric and not close} configuration, in fact, the biases are
not asymmetric about $\dt$, but just distributes on $ \dt<0$ and $ \dt>0 $
evenly.  In this configuration, although $ \Delta \phi_0$ also has a stable
stage similar to that in the {\sc Symmetric and close} configuration, the source
masses and the frequency evolutions of the two signals, are too different.  So
one cannot expect $ \bm B$ to have an obvious symmetric axis about $\dt$.
However, in this case, we can still choose an appropriate $\dt$ to increase the
length of the stable stage and cause a large bias.  The use of $\dt=-9.8 \, {\rm
ms}$ is an example, and its behavior will not be described repeatedly here. 

\section{Full Bayesian SPE results in other internal parameter configurations}
\label{appD:}

Figure~\ref{figD1} shows the marginalized distribution of parameters in the SPE
when the internal parameter configurations are taken as {\sc Asymmetric}, {\sc
Asymmetric2}, {\sc Random}, {\sc Symmetric and close}, and {\sc Symmetric and
not close}. Same as the {\sc Equal} configuration, we find that the results
forecasted by FM and the actual MLEs are in good agreement.

\begin{figure*}[htp]
\vspace{-0.4cm}
	\centering
\subfigure{\includegraphics[width=13cm]{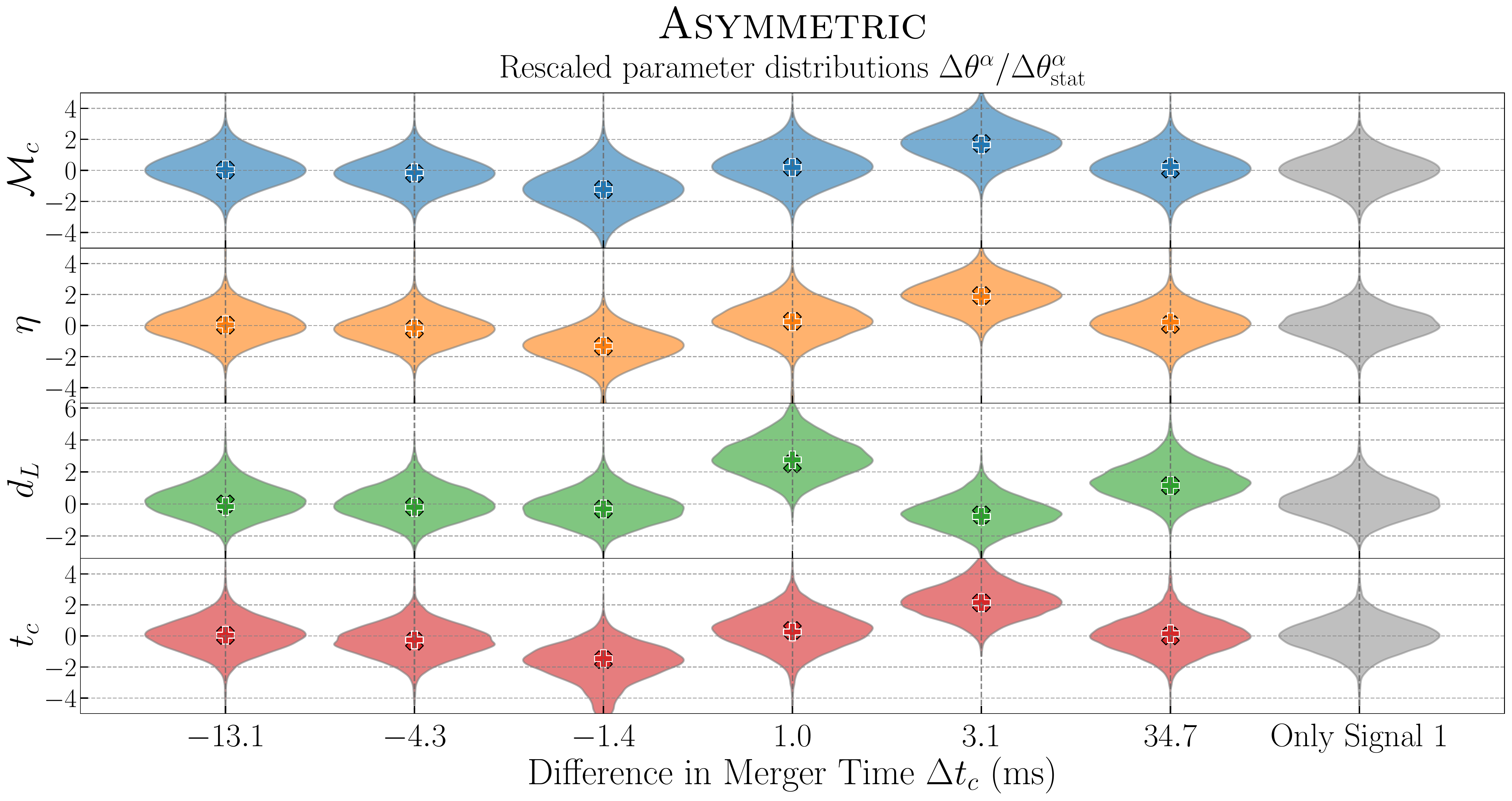}} 
\vspace{-0.3cm}
\\
	\centering
\subfigure{\includegraphics[width=13cm]{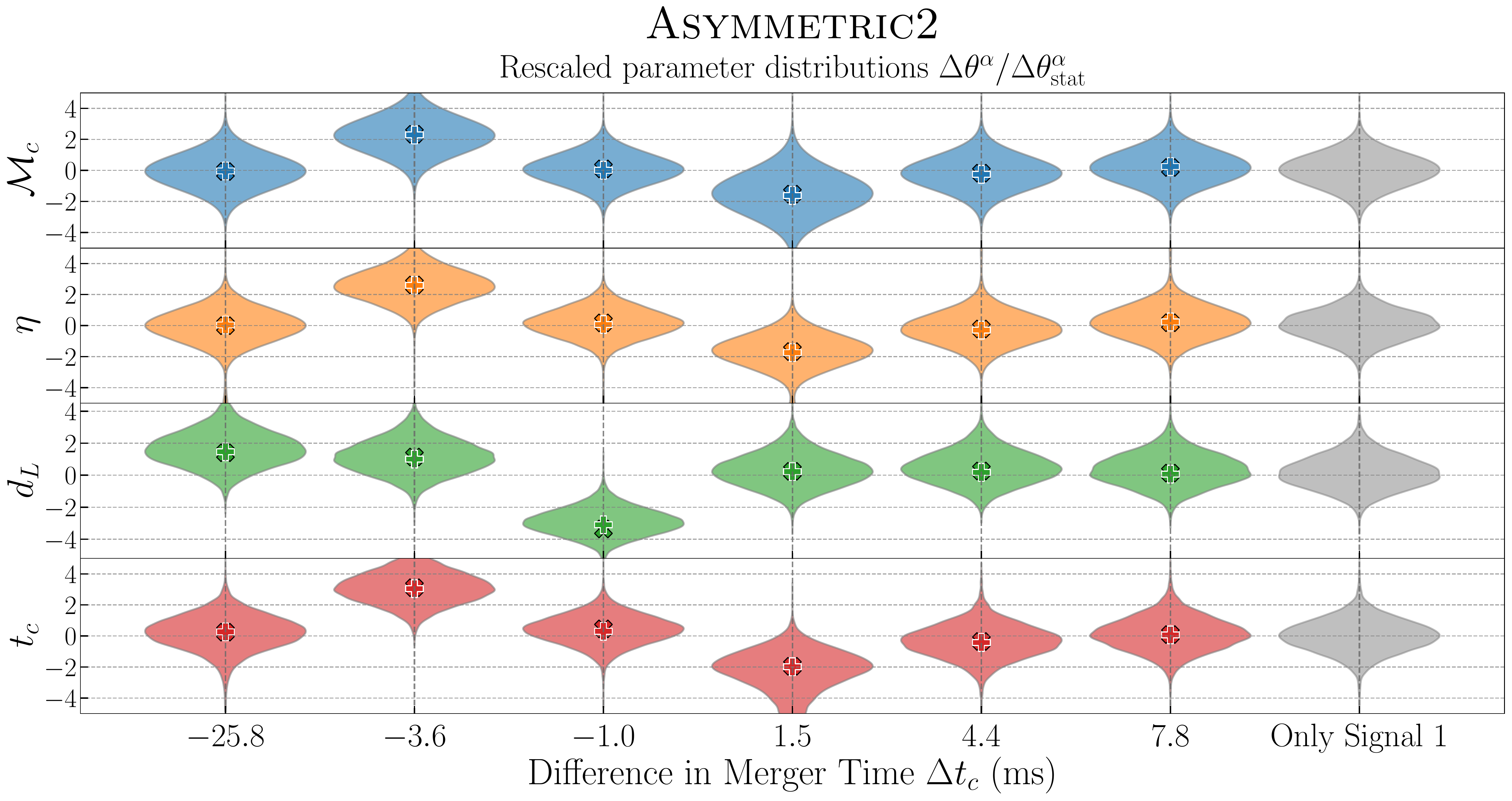}}
\vspace{-0.3cm}\\

	\centering
\subfigure{\includegraphics[width=13cm]{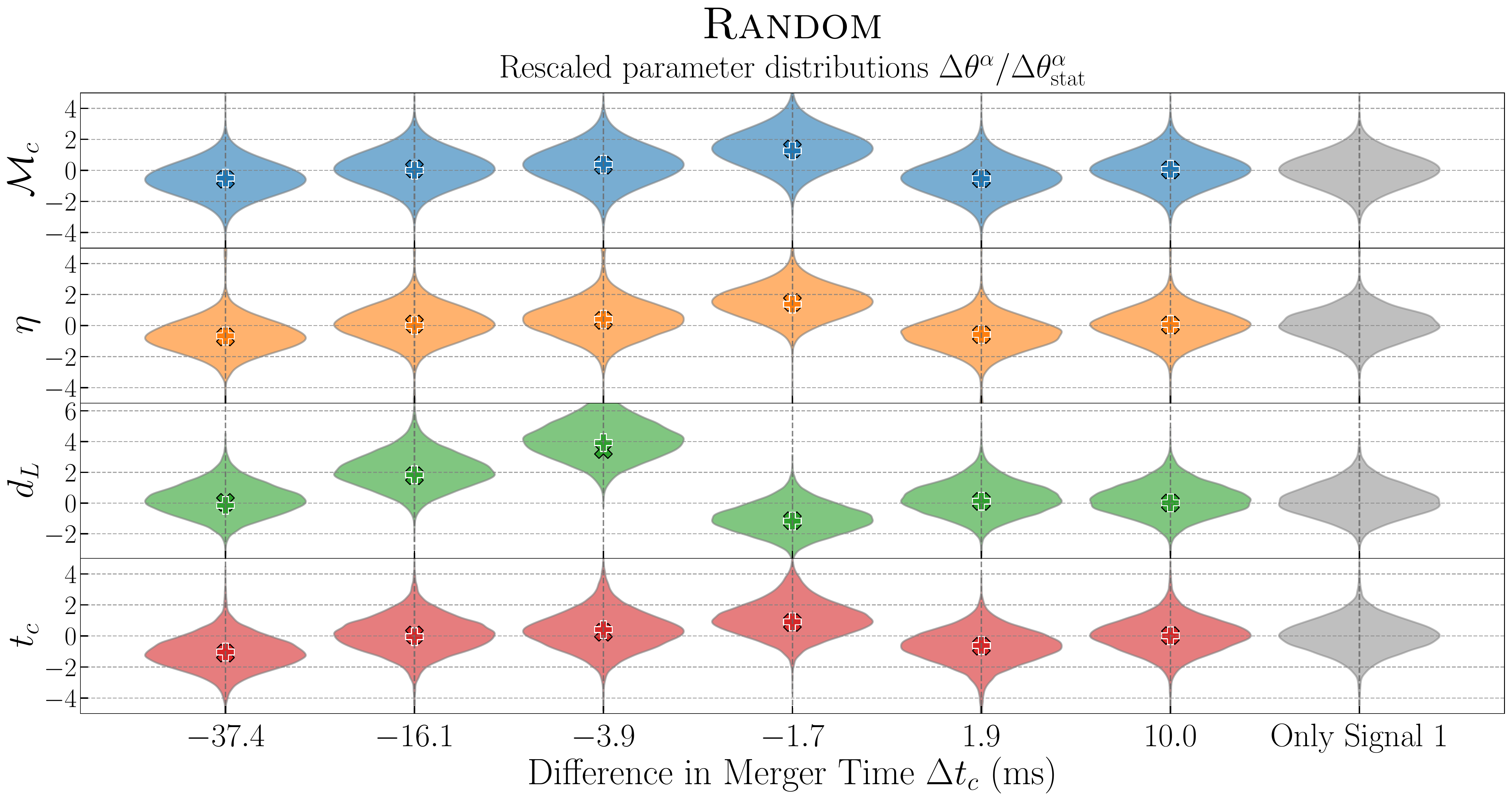}} 
\vspace{-0.4cm}\caption{Same as Fig.~\ref{fig9}, but for other parameter configurations. The
configuration names are marked on the top.} \label{figD1}
\end{figure*}

\begin{figure*}[htp]
\ContinuedFloat
	\centering
\subfigure{\includegraphics[width=13cm]{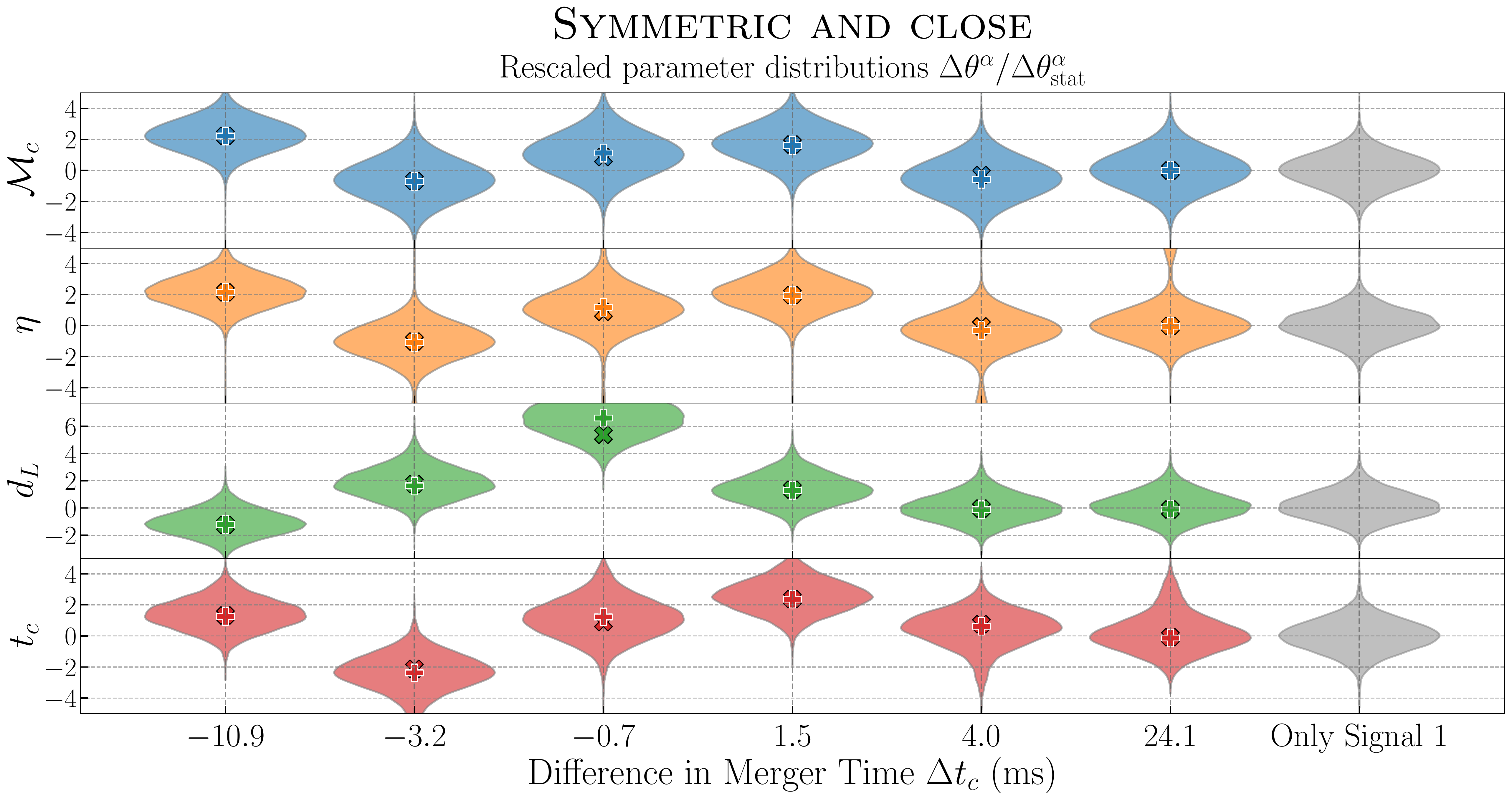}}
\vspace{-0.3cm}
\\
	\centering
\subfigure{\includegraphics[width=13cm]{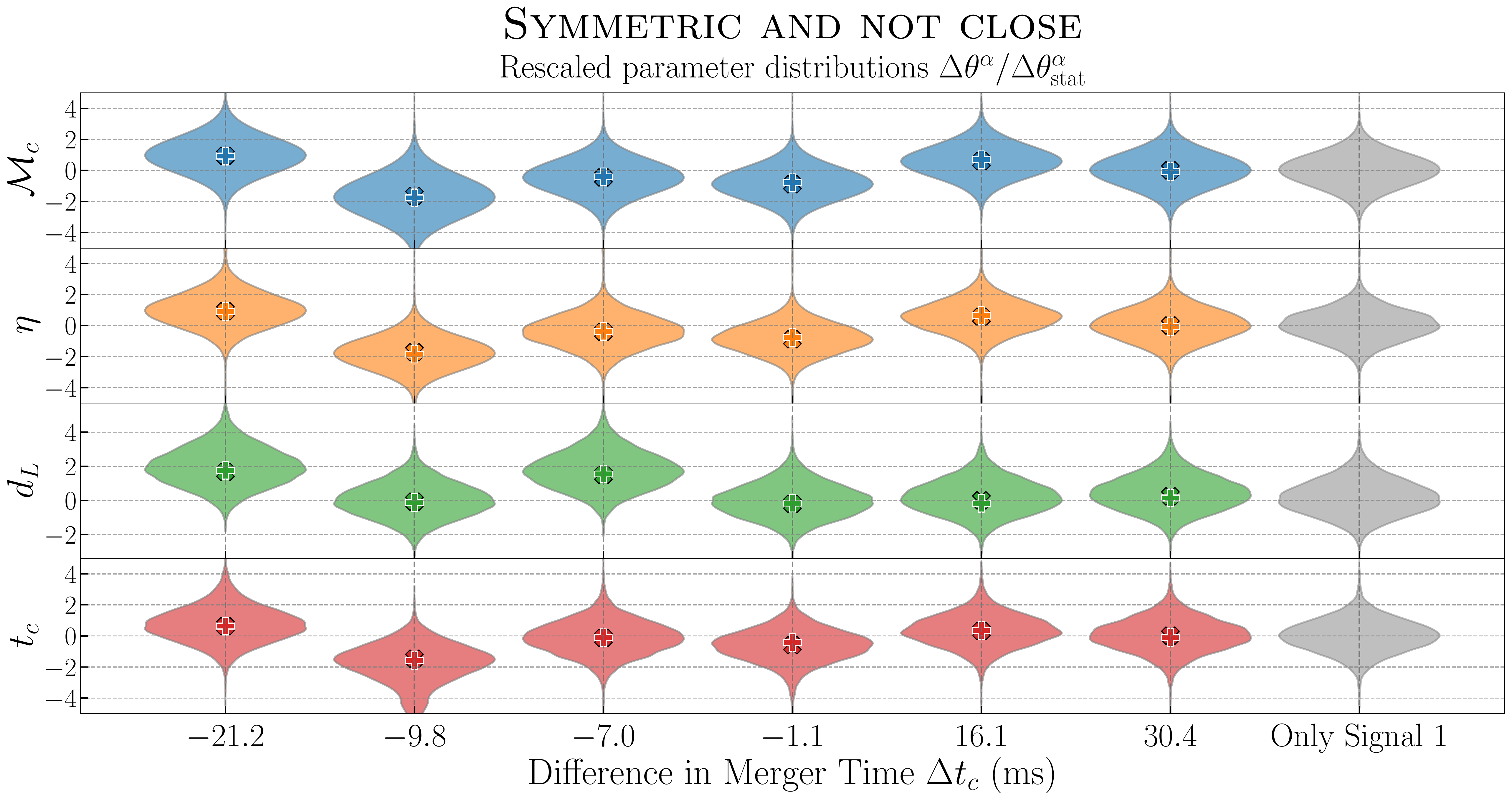}} 
\caption*{FIG D1 (continued).} 
\end{figure*}

\clearpage

\printbibliography{}

\end{document}